\documentclass[
    amsmath,
    amssymb,
    english,
    showkeys,
]{revtex4-2}
\usepackage{fix-cm}
\usepackage[utf8]{inputenc}
\usepackage[english]{babel}
\usepackage{graphicx}
\usepackage{latexsym,amsthm,amsfonts}
\usepackage{makeidx}   
\usepackage[T1]{fontenc}
\usepackage{subfigure}
\usepackage{dcolumn}
\usepackage{bm}
\usepackage{cancel}
\usepackage{xcolor}
\usepackage[hyperindex,colorlinks,linkcolor=blue,citecolor=red]{hyperref}

\usepackage{textgreek}
\usepackage{comment}

\allowdisplaybreaks[1]

\usepackage{upgreek}
\usepackage{graphicx}
\usepackage{tensor}

\begin{document}
\title{New sources of ghost fields in $k$-essence theories for black-bounce solutions}

\author{Carlos F. S. Pereira}
	\email{carlos.f.pereira@edu.ufes.br}
	\affiliation{Departamento de Física, Universidade Federal do Espírito Santo, Av. Fernando Ferrari, 514, Goiabeiras, 29060-900, Vit\'oria, ES, Brazil.}
	
	\author{Denis C. Rodrigues}
    \email[]{deniscr@gmail.com}
	\affiliation{Núcleo Cosmo-ufes \& Departamento de Física, Universidade Federal do Espírito Santo, Av. Fernando Ferrari, 514, Goiabeiras, 29060-900, Vit\'oria, ES, Brazil.}
		
	\author{Júlio C. Fabris}
    \email[]{julio.fabris@cosmo-ufes.org}
	\affiliation{Núcleo Cosmo-ufes \& Departamento de Física, Universidade Federal do Espírito Santo, Av. Fernando Ferrari, 514, Goiabeiras, 29060-900, Vit\'oria, ES, Brazil}
    \affiliation{National Research Nuclear University MEPhI (Moscow Engineering Physics Institute), 115409, Kashirskoe shosse 31, Moscow, Russia.}
	
	\author{Manuel E. Rodrigues}
    \email[]{esialg@gmail.com}
	\affiliation{Faculdade de Ci\^encias Exatas e Tecnologia, Universidade Federal do Par\'a Campus Universit\'ario de Abaetetuba, 68440-000, Abaetetuba, Par\'a, Brazil and Faculdade de F\'isica, Programa de P\'os-Gradua\c{c}\~ao em F\'isica, Universidade Federal do Par\'a, 66075-110, Bel\'em, Par\'a, Brazil.}

\author{Ébano L. Martins}
	\email{ebano.lm@gmail.com}
	\affiliation{Departamento de Física, Universidade Federal do Espírito Santo, Av. Fernando Ferrari, 514, Goiabeiras, 29060-900, Vit\'oria, ES, Brazil.}	
	
\begin{abstract}

\noindent  In the present study, we generalize the possible ghost field configurations within the framework of $k$-essence theory to the Simpson-Visser metric area function $\Sigma^2=x^2+a^2$. Our analysis encompasses field configurations for the region-defined metric function $dA_\pm$ as well as the general solution that asymptotically behaves as Schwarzschild-de Sitter for $x\to-\infty$. Specifically, we investigate two scalar field configurations and define the associated potential for each one. Through rigorous calculations, we verify that all equations of motion are satisfied. Notably, our findings indicate that even when proposing new configurations of ghost scalar fields, the energy conditions remain unchanged. This result serves to validate the wormhole solutions obtained in previous studies. 
\end{abstract}
	
	
\maketitle
	
\section{Introduction}\label{sec1}

The quest for geometrically regular black hole solutions has been a topic of significant interest in theoretical physics. The pioneering work in this domain was undertaken by Bardeen in 1968 \cite{regular1}, who proposed the first solution for a geometrically regular black hole. This solution is characterized by the absence of a singularity, possessing only the so-called event horizon. Notably, Bardeen's solution introduces a parameter that distinguishes it from the Schwarzschild metric; however, it does not satisfy Einstein's field equations. 
It was not until 2000 that Beato and Garcia \cite{regular2} obtained the matter content that solved Einstein's equations through the framework of nonlinear electrodynamics. In their work \cite{regular2}, the parameter introduced in Bardeen's metric gained a physical interpretation as a magnetic charge. Following this breakthrough, numerous other regular black hole solutions have been proposed \cite{regular3,regular4,regular5,regular6,regular7,regular8,regular9,regular10,regular11,regular12,regular13,regular14,regular15}, enriching the understanding of these intriguing objects in the realm of general relativity.

From this perspective, the so-called Black-bounce solutions were recently proposed by Simpson and Visser \cite{matt}. These are regular solutions \cite{regular1,regular15} with possibilities to describe different wormhole configurations, including the case of traversability in both directions ($a>2m$), the unidirectional case ($a=2m$), and the scenario where the wormhole presents two symmetric horizons ($a<2m$). This class of solutions is characterized by the presence of the area function $\Sigma^2=x^2+a^2$, where the parameter $a$ describes the radius of the wormhole throat. Logically, the Schwarzschild metric must be recovered in the limit of $a\rightarrow0$ \cite{intro1,intro2,intro3,intro4,intro5}. The material content regarding the Simpson-Visser metric was proposed in Refs. \cite{intro6,intro7}, requiring a combination of nonlinear electrodynamics and a scalar field. This solution class has also been explored in the context of black strings \cite{geo1,geo2,geo3}.

Subsequently, several other works have emerged in various contexts related to the Simpson-Visser black-bounce solutions. For instance, modifications have been made to the mass function \cite{intro8}, alternative throat forms for this wormhole have been proposed \cite{intro9}, and examinations have been undertaken in modified gravity theories such as $f(T)$ and $f(R)$ gravities \cite{intro10,intro11}. Additionally, these solutions have been extended to stationary and spherically symmetric scenarios \cite{intro12,intro13}. Studies of light deflection and gravitational lensing effects associated with these wormhole geometries were made \cite{intro14,intro15,intro16,intro17,intro18}.  Notably, in the zero mass limit, this solution reduces to the Ellis-Bronnikov charged wormhole. Furthermore, the quantum dynamics of these configurations have also been investigated using the Ellis-Bronnikov metric \cite{intro19,intro20,intro21,intro22}, providing insights into the quantum aspects of these wormhole solutions.



$k$-Essence theories \cite{mukha1} constitute a class of models that involve fundamental fields coupled to gravity, characterized by a non-standard kinetic term for a scalar field. This concept was initially proposed in \cite{mukha2} to achieve inflation driven by the kinetic energy of the field, rather than by its potential. Subsequently, this idea was also employed to account for the current accelerated expansion of the Universe \cite{mukha3}. Interestingly, $k$-essence structures arise naturally in string theories, such as in the Dirac-Born-Infeld action, where the kinetic term of the scalar field resembles the Maxwell-like term in Born-Infeld electrodynamics \cite{dbi}. $k$-Essence theories have also been shown to admit static and spherically symmetric solutions, such as black holes \cite{bronnikov1}. Bronnikov et al. investigated the conditions under which the Ellis-Bronnikov wormhole metric can be a solution within the $k$-essence framework. Previously, regular black-bounce solutions were constructed in $k$-essence theory \cite{simple}. These solutions featured a regular metric function at the origin and a Simpson-Visser black-bounce area function. They exhibited a phantom scalar field and potential with properties akin to energy condition violations \cite{CDJM}. In this sense, in the present work we generalize the possible configurations of ghost fields investigated in the first article \cite{CDJM}. Thus, the energy conditions were analyzed for the scenario in which the metric function represents a wormhole as well as the case in which we have behavior similar to the internal region of a regular black hole. Intuitively we are led to conclude that the energy conditions for all new field configurations $\bar{n}=2,3,4,5,\dots$, remain unchanged. However, to verify this it is necessary to investigate case by case.

The paper is structured as follows: Section \ref{sec2} establishes the theoretical background of the $k$-essence model, including the key relationships and equations. In Section \ref{sec3}, we introduce the generic expression for the scalar field and verify that a class of configurations can be obtained. Section \ref{sec4} is devoted to developing and investigating the properties of the scalar field for the parameter $\bar{n}=2$, along with its associated energy conditions. In Section \ref{sec5}, we explore the second configuration of the scalar field with $\bar{n}=3$ and analyze its energy conditions. Section \ref{sec6} focuses on analyzing the physical quantities of interest for the general solution. Finally, the conclusions are presented in Section \ref{sec7}.


\section{General relations}\label{sec2}

$k$-Essence theories are characterized by a non-canonical kinetic term for the scalar field, represented by the Lagrangian
\begin{equation}\label{Lagran}
    \mathcal{L} = \sqrt{-g}[R-F(X,\phi)]\,,
\end{equation}
where $R$ is the Ricci scalar and $X=\eta\phi_{;\rho}\phi^{;\rho}$ denotes the kinetic term. Though $k$-essence models can include a potential term and non-trivial couplings, the scalar sector is typically minimally coupled to gravity. The parameter $\eta=\pm 1$ is introduced to avoid imaginary terms in the kinetic expression $X$. Through selecting different forms for the function $F(X,\phi)$, $k$-essence theories can describe both phantom \cite{phan1,phan2,phan3,phan4} and standard scalar fields.

The variation of the Lagrangian \eqref{Lagran} with respect to the metric tensor and the scalar field yields the field equations.
\begin{eqnarray}\label{eq1}
G_\mu^{\nu}=-T_\mu^{\nu}\left(\phi\right)=-\eta{F_X}\phi_\mu{\phi^{\nu}} + \frac{1}{2}\delta_\mu^{\nu}F, \\\label{eq2}
\eta\nabla_\alpha\left(F_X\phi^{\alpha}\right)-\frac{1}{2}F_\phi =0,
\end{eqnarray} 
where $G_\mu^{\nu}$ is the Einstein tensor, $T_\mu^{\nu}$ the stress-energy tensor, $F_X=\frac{\partial{F}}{\partial{ X}}$, $F_\phi=\frac{\partial{F}}{\partial\phi}$ and $\phi_\mu=\partial_\mu\phi$.

The line element representing the most general spherically symmetric and static spacetime takes the form:
\begin{eqnarray}\label{eq3}
ds^2=e^{2\gamma\left(u\right)}dt^2-e^{2\alpha\left(u\right)}du^2-e^{2\beta\left(u\right)}d\Omega^2,
\end{eqnarray} where $u$ is an arbitrary radial coordinate, $d\Omega^2=d\theta^2+\sin^2\theta{d\varphi^2}$ the volume element, and $\phi=\phi\left (u\right)$. 

The non-zero components of the stress-energy tensor are,
\begin{eqnarray}\label{eq4}
T_0^{0}= T_2^{2} =T_3^{3}= -\frac{F}{2}, \\\label{eq5}
T_1^{1}=-\frac{F}{2} -\eta{F_X}e^{-2\alpha}{\phi}'^2,
\end{eqnarray} with $\phi'=\frac{d\phi}{du}$.

It is assumed that the function $X=-{\eta}e^{-2\alpha}{\phi}'^2$ is positive, which implies that $\eta=-1$. As a result, the equations of motion take the form:
\begin{eqnarray}\label{eq6}
2\left(F_X{e^{-\alpha+2\beta+\gamma}}\phi'\right)' - {e^{\alpha+2\beta+\gamma}}F_\phi=0, \\\label{eq7}
{\gamma}'' + {\gamma}'\left(2{\beta}'+ {\gamma}'-{\alpha}'\right)-\frac{e^{2\alpha}}{2}\left(F-XF_X\right)=0, \\ \label{eq8}
-e^{2\alpha-2\beta} + {\beta}'' +{\beta}'\left(2{\beta}'+ {\gamma}'-{\alpha}'\right) -\frac{e^{2\alpha}}{2}\left(F-XF_X\right)=0, \\ \label{eq9}
-e^{-2\beta} + e^{-2\alpha}{\beta}'\left({\beta}'+2{\gamma}'\right) -\frac{F}{2} + XF_X=0.
\end{eqnarray}

The notation used here follows the same as used in the reference \cite{bronnikov1}. The following coordinate transformation is defined: $u=:x$, and the \textit{quasi-global} gauge $\alpha\left(u\right)+\gamma\left(u\right)=0$ is employed. As a result, the line element in Eq. (\ref{eq3}) can be expressed in the following form:
\begin{eqnarray}\label{eq10}
ds^2= A\left(x\right)dt^2- \frac{dx^2}{A\left(x\right)} - \Sigma^2\left(x\right)d\Omega^2,
\end{eqnarray}
where the metric functions are defined as $A(x) = e^{2\gamma} = e^{-2\alpha}$ and $e^\beta = \Sigma(x)$. The equations of motion defined in Eqs. (\ref{eq6}-\ref{eq9}) can then be rewritten in the new coordinates. Combining Eqs. (\ref{eq7}-\ref{eq9}) yields the expressions:
\begin{eqnarray}\label{eq11}
2A\frac{{\Sigma}''}{\Sigma} - XF_X =0, \\\label{eq12}
{A}''\Sigma^2 - A\left(\Sigma^2\right)''+ 2 =0,
\end{eqnarray} 
where the primes now represent derivatives with respect to $x$.

The two remaining equations, Eq. (\ref{eq6}) and Eq. (\ref{eq9}), are rewritten in the new coordinates as
\begin{eqnarray}\label{eq13}
2\left(F_X{A\Sigma^2}\phi'\right)' - \Sigma^2F_\phi = 0, \\\label{eq14}
\frac{1}{\Sigma^2}\left(-1 + A'\Sigma'\Sigma + A{\Sigma'}^2\right) -\frac{F}{2} + XF_X = 0.
\end{eqnarray}

It was verified in the previous work \cite{CDJM} that it is not mathematically consistent to pursue black-bounce solutions with only the kinetic term of the $k$-essence function. Therefore, when constructing these new solutions, we must adopt the presence of a scalar potential $F(X)=F_0X^n-2V(\phi)$, where $F_0$ is a constant, $n$ is a real number, and $V(\phi)$ is the potential.

The previous work also discussed the process of constructing the metric function $A\left(x\right)$ by solving the differential equation Eq.(\ref{eq12}), for a given Simpson-Visser type area function $\Sigma^2(x) = x^2 + a^2$ \cite{matt}. The solution is given by

\begin{eqnarray}\label{GN}
    A\left(x\right)= 1+  C_1\left[(x^2+a^2)\arctan\left(\frac{x}{a}\right)+xa\right] +C_2 (x^2+a^2),
\end{eqnarray} where $C_1$ and $C_2$ are arbitrary integration constants that can be adjusted for the case of interest. As already discussed in the first work \cite{CDJM} we impose that the general solution should be asymptotically flat and at the origin recover the Simpson-Visser limit. In this way, the metric function to represent a wormhole region is defined and its properties were examined in detail in that work. Its shape is given by

\begin{eqnarray}\label{eq15}
A_+\left(x\right)=1 + \left(\frac{4m}{\pi{a^3}}\right)\left[xa + \left(x^2+a^2\right)\left(\arctan\left(\frac{x}{a}\right)-\frac{\pi}{2}\right)\right] \qquad x \geq 0, \nonumber \\ 
A_-\left(x\right)=1 - \left(\frac{4m}{\pi{a^3}}\right)\left[xa + \left(x^2+a^2\right)\left(\arctan\left(\frac{x}{a}\right)+\frac{\pi}{2}\right)\right] \qquad x \leq 0.
\end{eqnarray}

We emphasize that we can recover the same metric function found in the article \cite{PRL} considering the following redefinition of the constants $\rho_0=\frac{4m}{\pi}$ and $c=-\frac{2m}{a}$. The coincidence of these results occurs due to the requirement in solving the differential equation Eq. (\ref{eq12}) that the metric function must be asymptotically flat in the limit of $x\to{\infty}$ and recover Simpson-Visser in the origin. Thus, we can qualitatively reproduce the wormhole solutions for the region-defined metric function Eq. (\ref{eq15}). For the general solution, the system starts in a static universe as $x\to{+\infty}$ and ends in a solution characteristic of a cosmological black hole as $x\to{-\infty}$.

\section{New field configurations}\label{sec3}

As demonstrated in the aforementioned work \cite{CDJM}, the equation of motion (Eq. \ref{eq12}) is employed to construct the metric function (Eq. \ref{eq15}). Subsequently, the equation of motion (Eq. \ref{eq11}) is utilized to obtain the scalar field. As stated at the outset of the preceding section, the form of the $k$-essence field is adopted as defined by $F(X)=F_0X^n-2V(\phi)$. For the sake of simplicity, we introduce the following change in the power of the $k$-essence field: $\bar{n}=\frac{1-n}{2n}$. Consequently, the scalar field can be expressed in a general form in terms of the previously defined metric functions as

\begin{eqnarray}\label{eq16}
\phi_\pm\left(x\right)=\Omega\int{dx}\left[\frac{A^{\bar{n}}_{\pm}}{\left(x^2+a^2\right)^{2\bar{n}+1}}\right],
\end{eqnarray} where $\Omega=\sqrt{-\frac{1}{\eta}}\left[\frac{2\left(2\bar{n}+1\right)a^2}{F_0}\right]^{\frac{2\bar{n}+1}{2}}$ is a constant that depends on the power law of the $k$-essence field.

The scalar field expression defined above allows for the construction of a class of configurations that exhibit a power-law dependence on the metric function $A_{\pm}$ within the integrand of Eq. (\ref{eq16}). For illustrative purposes, consider the following examples with $\eta=-1$ for the scalar field:


\begin{eqnarray}\label{eq17}
\bar{n}=1 \qquad \rightarrow \qquad n=\frac{1}{3} \qquad \rightarrow \qquad \phi_\pm\left(x\right)&=&\int{dx}\left(\frac{6a^2}{F_0}\right)^\frac{3}{2}\frac{A_{\pm}}{\left(x^2+a^2\right)^3}, \label{eq18} \\
\bar{n}=2 \qquad \rightarrow \qquad n=\frac{1}{5} \qquad \rightarrow \qquad \phi_\pm\left(x\right)&=&\int{dx}\left(\frac{10a^2}{F_0}\right)^\frac{5}{2}\frac{A^2_{\pm}}{\left(x^2+a^2\right)
^5}, \label{eq19} \\
\bar{n}=3 \qquad \rightarrow \qquad n=\frac{1}{7} \qquad \rightarrow \qquad \phi_\pm\left(x\right)&=&\int{dx}\left(\frac{14a^2}{F_0}\right)^\frac{7}{2}\frac{A^3_{\pm}}{\left(x^2+a^2\right)^7}, \label{eq19a}
\end{eqnarray} where

\begin{eqnarray}\label{eqjunção1}
\phi\left(x\right)= 
\begin{cases}
\phi_{+}, & x \geq {0} \\
\phi_{-}, & x \leq {0}.
\end{cases}
\end{eqnarray}

In the expressions for the scalar fields, a class of configurations with $\bar{n}=0,1,2,3,4,\dots$ allows for straightforward integration. The case of $\bar{n}=0$ corresponds to the ghost field configuration commonly found in the literature. However, for the present purposes, the simplest case is $\bar{n}=1$ \cite{PRL,KJD}, as investigated in \cite{CDJM}. The geometric properties of the metric function $A_\pm$ were examined in \cite{CDJM}, where it was found to be regular throughout spacetime. This admits solutions for two-way traversable wormholes with $a>2m$ and wormholes with symmetric horizons when $a<2m$.

\section{Special Configuration I}\label{sec4}

In this section, the special case for $\bar{n}=2$ is analyzed and the properties of the scalar field and potential inside and outside the event horizon are examined. The scalar field can be obtained by direct integration of Eq. (\ref{eq19}) and its explicit form is defined in (\ref{apen11}). To characterize some properties that appear in the behavior of the scalar field, it is useful to evaluate its asymptotic behavior $\phi\left(x\rightarrow \pm\infty\right)$ and at the origin. Therefore, one must

\begin{eqnarray}\label{eq20}
\phi_{+}\left(x\rightarrow\infty\right)=-\phi_{-}\left(x\rightarrow{-}\infty\right)&=&\frac{25\sqrt{10}\left[15 \pi ^2 a (7 a-16 m)+16m^2\left(12\pi^2-35\right)\right]}{192\pi{a^6}},\\\label{eq22}
\phi_{+}\left(x\rightarrow{0}\right)=-\phi_{-}\left(x\rightarrow{0}\right)&=&\frac{400m\sqrt{10}\left(2a-3m\right)}{9\pi{a^6}}.
\end{eqnarray}



\begin{figure}[htb!]
		\centering
	    \subfigure[]
			{\label{campofora}
			{\includegraphics[scale=0.4]{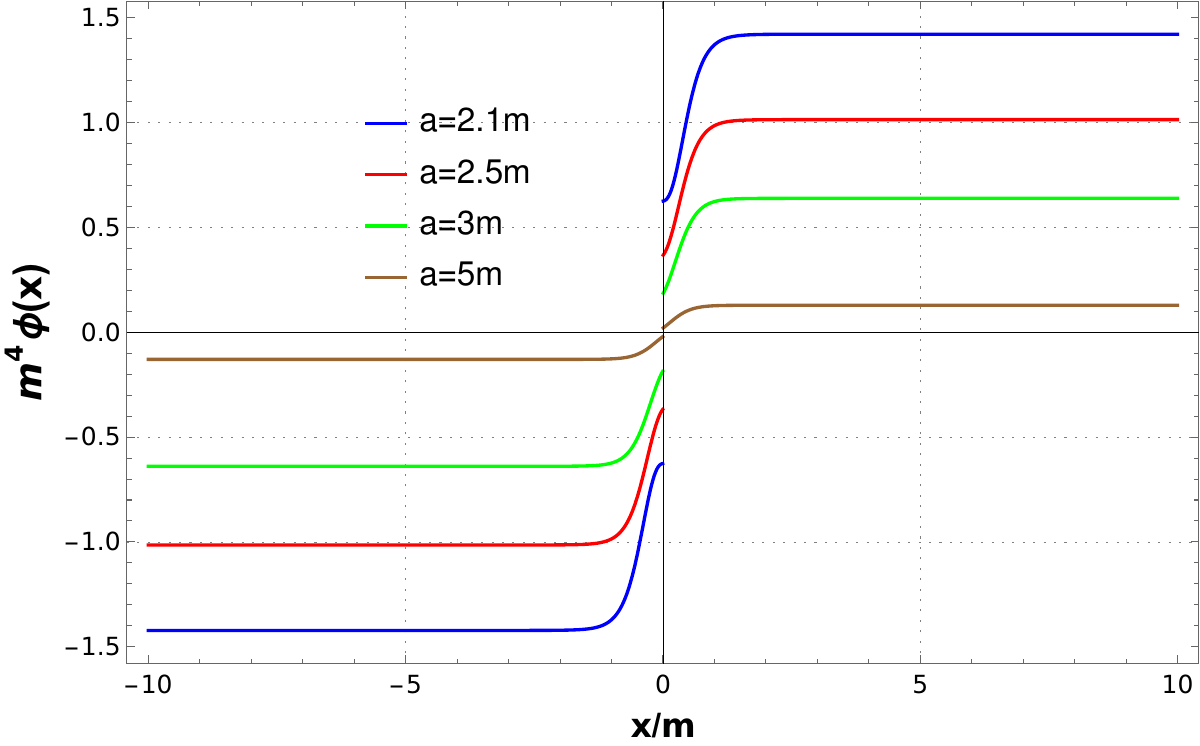}}}\quad
			\subfigure[]
			{\label{campodentro}
				\includegraphics[scale=0.4]{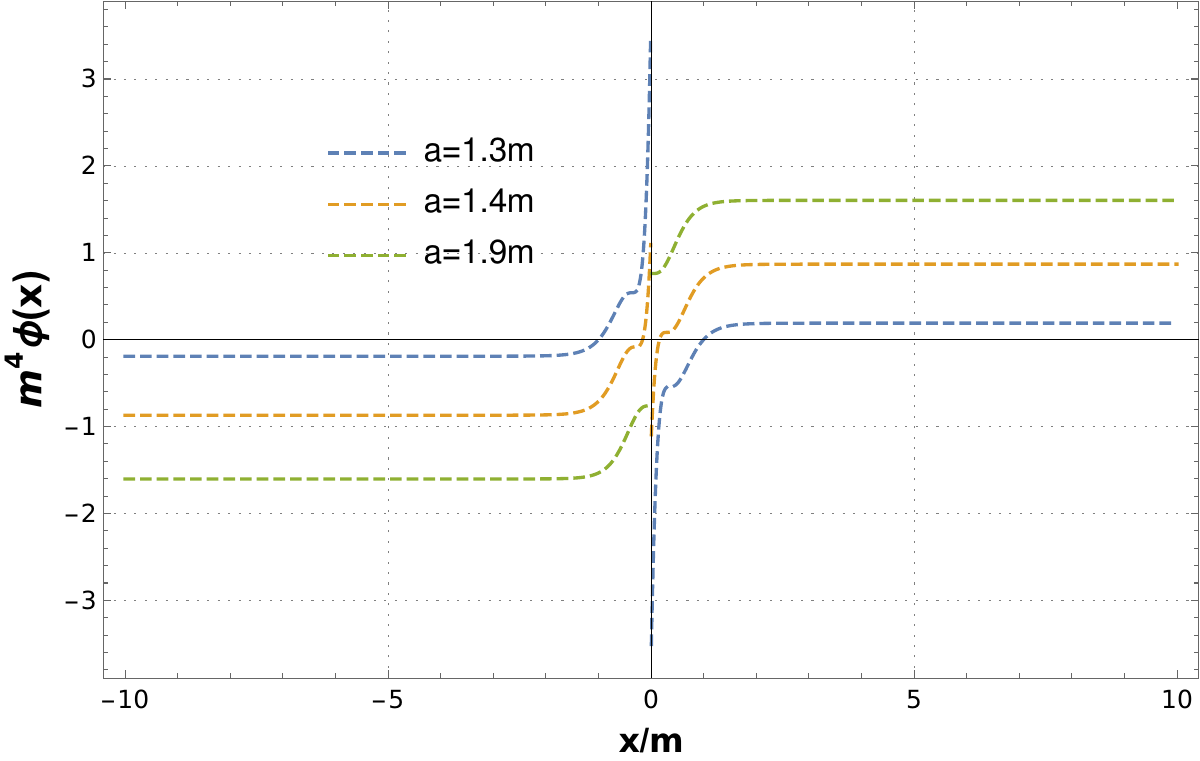}}
			\subfigure[]
			{\label{fora}
				\includegraphics[scale=0.4]{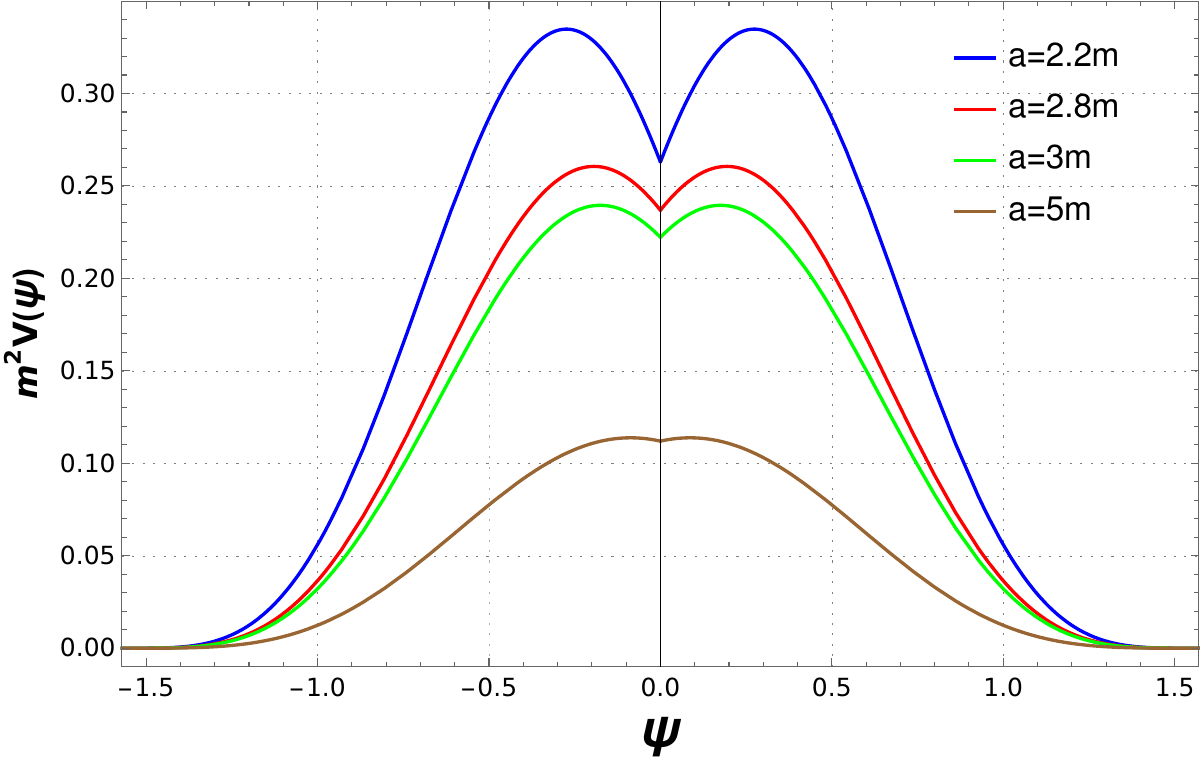}}\qquad
			\subfigure[]
			{\label{dentro}
				\includegraphics[scale=0.4]{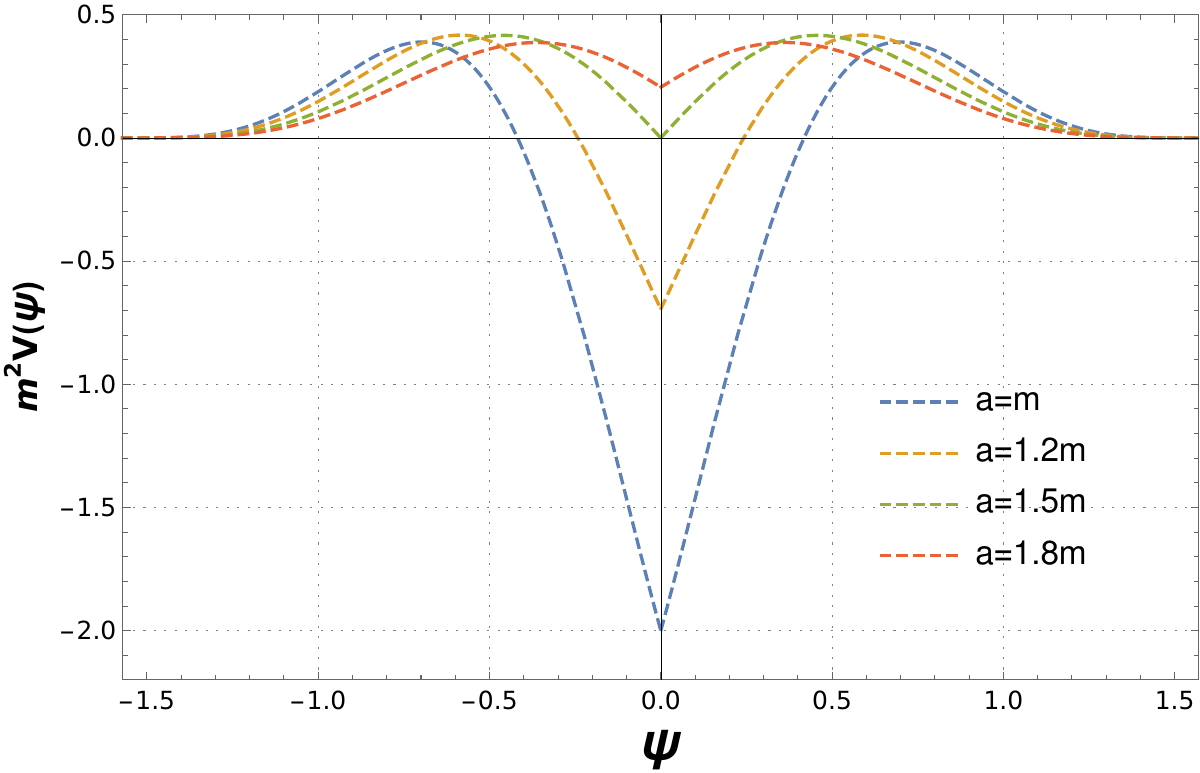}}

    \caption{(a) The scalar field from Eq. (\ref{eq19}) for radius values outside the event horizon. (b) The scalar field from Eq. (\ref{eq19}) for radius values inside the event horizon. In both cases, the constant $F_0=1$ is fixed. (c) The potential from Eq. (\ref{eq24}) for throat radii outside the event horizon. (d) The potential from Eq. (\ref{eq24}) for throat radii inside the event horizon.
    }
\label{ESCALARCASOI}
\end{figure}


Similarly, the potential for this novel configuration can be defined. This is achieved by utilizing the equation of motion, Eq. (\ref{eq14}), in conjunction with the equation of motion defining the scalar field, Eq. (\ref{eq11}). Consequently, the potential can be expressed as:

\begin{eqnarray}\label{eq23}
V_{\pm}\left(\phi\left(x\right)\right)= \frac{1}{\Sigma^2} -\frac{{{\Sigma}'A'_{\pm}}}{\Sigma}-\frac{A_{\pm}{\Sigma'}^2}{\Sigma^2} + \frac{3A_{\pm}{\Sigma}''}{\Sigma},
\end{eqnarray} where


\begin{eqnarray}\label{eqjunção2}
V\left(\phi\left(x\right)\right)= 
\begin{cases}
V_{+}, & x \geq {0} \\
V_{-}, & x \leq {0}.
\end{cases}
\end{eqnarray}

The potential has been analyzed within the context of an auxiliary variable, as discussed in \cite{CDJM}. Employing the transformation $\psi=\arctan\left(\frac{x}{a}\right)$, it is important to observe that as $x\rightarrow{\pm}\infty$, this transformation corresponds to $\psi\rightarrow{\pm}\frac{\pi}{2}$. Consequently, the potential can be reformulated in terms of the new variable as follows:


\begin{eqnarray}\label{eq24}
V_{\pm}\left(\psi\right)=\frac{\left[3+ \left(4-3c \pi  a^2\right) \cos (2 \psi )+\cos (4 \psi ) +ca^2 (\mp\sin (2 \psi )\pm\sin (4 \psi )\pm 6 \psi  \cos (2 \psi ))\right]}{2 a^2},
\end{eqnarray} where $c=\frac{4m}{\pi{a^3}}$ is a combination of constants.


The upper section of Figure \ref{ESCALARCASOI} presents a graphical depiction of the scalar field given by Eq. (\ref{eq19}) for throat radius values both exterior to (cf. \ref{campofora}) and within (cf. \ref{campodentro}) the event horizon. A notable aspect in the behavior of the scalar field is the presence of discontinuity at the origin, for both radii inside and outside the horizon. This characteristic suggests dependence on the odd derivatives of the metric function $A_{\pm}$, indicating the scalar field feels the presence of the thin shell at the junction point.

The behavior of the scalar field at the origin can also be observed through the limit of the configuration in Eq. (\ref{eq23}). A limit between the throat radius parameters and mass can be noted, wherein the sign inversion occurs when $a>\frac{3m}{2}$. This behavior extends to the analysis of the scalar field in the asymptotic limit $x\to\infty$, where it assumes positive values at upper throat radii or on the order of $a\approx{1.28m}$ and negative for smaller values (see Figures \ref{campodentro} and \ref{campofora}). In the asymptotic limit where, $x\to{-\infty}$ for values of throat radii increasingly distant from the horizon, the scalar field approaches zero.

The lower section of Figure \ref{ESCALARCASOI} presents a graphical representation of the potential described in Eq. (\ref{eq24}). Figure \ref{fora} on the left illustrates the potential's behavior for throat values located outside the event horizon. In this case, the potential approaches zero as $\psi$ tends to $\pm{\frac{\pi}{2}}$ and exhibits a barrier-like shape near the origin. For radius values within the event horizon, as depicted in Figure \ref{dentro}, the potential progressively resembles the Pöschl-Teller potential \cite{teller1,teller2,teller3,teller4,molina} as the radius decreases, while still asymptotically approaching zero as $\psi$ tends to $\pm{\frac{\pi}{2}}$.

As previously established in \cite{CDJM}, only three of the equations of motion were employed in the construction. Therefore, it is imperative to verify that the fourth equation, Eq. (\ref{eq13}), is also satisfied. This condition can be expressed as follows:

\begin{eqnarray}\label{eq25}
\frac{dV_{\pm}}{dx} + \frac{F_{0}}{5}\left(\frac{F_{0}}{10a^2}\right)^\frac{3}{2}\left(\frac{{\phi'_\pm}}{\Sigma^2}\right)\left(\frac{\Sigma^8}{A_{\pm}}\right)'=0.
\end{eqnarray}

\subsection{Energy conditions}\label{sec41}

The analysis of null energy conditions, as presented in Ref. \cite{intro2}, begins with Einstein's field equations, given by Eq. (\ref{eq1}). Subsequent calculations yield the following non-vanishing components of the stress-energy tensor \cite{intro1}:
 
\begin{eqnarray}\label{eq26}
\tensor{T}{^\mu}_{\nu}= {\rm diag}\left[\rho^\phi,-p^\phi_1,-p^\phi_2,-p^\phi_2\right],
\end{eqnarray}
where the scalar field energy density is denoted by $\rho^\phi$, while $p^\phi_1$ and $p^\phi_2$ represent the radial and tangential pressures, respectively. By employing the diagonal components of the stress-energy tensor, as expressed in Eqs. (\ref{eq4}-\ref{eq5}), for the novel $k$-essence configuration with $\bar{n}=2$ (Eq. \ref{eq19}) and its corresponding potential (Eq. \ref{eq23}),

\begin{eqnarray}\label{eq27}
\rho^\phi_{\pm}&=& -\frac{F_0}{2}\left[-\eta{A_{\pm}{\left(\phi'_{\pm}\right)}^2}\right]^\frac{1}{5} + V_{\pm}\left(x\right)= -\frac{5{A_{\pm}{\Sigma}''}}{\Sigma} +V_{\pm}\left(x\right), \\\label{eq28} 
p^\phi_{1\pm}&=& - T^{1}_{1} =  \frac{3{A_{\pm}{\Sigma}''}}{\Sigma} -V_{\pm}\left(x\right), \\\label{eq29} 
p^\phi_{2\pm}&=&  - T^{2}_{2}= - T^{0}_{0}=- \rho^\phi_{\pm}= \frac{5{A_{\pm}{\Sigma}''}}{\Sigma} - V_{\pm}\left(x\right).
\end{eqnarray}

The diagonal components of the stress-energy tensor, as defined above, are applicable solely outside the horizon where $A_\pm>0$, with the metric signature $(+,-,-,-)$. In this domain, the coordinate $t$ is timelike, and the coordinate $x$ is spacelike.

Within the event horizon, where $A_\pm<0$, the coordinate $t$ becomes spacelike and $x$ becomes timelike. Consequently, the metric signature transforms to $(-,+,-,-)$, thereby reversing the roles of the coordinates. The stress-energy tensor components must then be reformulated as:



\begin{eqnarray}\label{eq30}
\tensor{T}{^\mu}_{\nu}= {\rm diag}\left[-p^\phi_1,\rho^\phi,-p^\phi_2,-p^\phi_2\right],
\end{eqnarray} and therefore, the equations for energy density, radial pressure, and tangential pressure must be rewritten as:

\begin{eqnarray}\label{eq31}
\rho^\phi_{\pm}&=& -\frac{{3A_{\pm}{\Sigma}''}}{\Sigma} +V_{\pm}\left(x\right),\\\label{eq32} 
p^\phi_{1\pm}&=&  \frac{5{A_{\pm}{\Sigma}''}}{\Sigma} - V_{\pm}\left(x\right),\\\label{eq33}
p^\phi_{2\pm}&=& - T^{2}_{2}= - T^{0}_{0}=-\rho^\phi_{\pm}=-\left(-p^\phi_{1\pm}\right)=\frac{5{A_{\pm}{\Sigma}''}}{\Sigma} -V_{\pm}\left(x\right).
\end{eqnarray} 

The dependence of geometric quantities on the piecewise-defined metric function $A_\pm(x)$ naturally extends to any physical quantities derived from it. Consequently, the energy density and pressure components inherit this piecewise characteristic. These components play a crucial role in defining the energy conditions for Black-Bounce solutions, as outlined in Ref. \cite{intro5}.

Energy conditions are typically expressed as inequalities involving the energy density and pressure components \cite{intro5}:

%
\begin{eqnarray}\label{eq34}
NEC_{1,2}&=&WEC_{1,2}=SEC_{1,2} \Longleftrightarrow \rho^\phi_{\pm} + p^\phi_{\left(1,2\right)\pm} \geq 0, \\\label{eq35}
SEC_3 &\Longleftrightarrow & \rho^\phi_{\pm} + p^\phi_{1\pm} + 2p^\phi_{2\pm} \geq 0, \\\label{eq36}
DEC_{1,2} &\Longleftrightarrow &  \rho^\phi_{\pm} + p^\phi_{\left(1,2\right)\pm} \geq 0  \qquad    \mbox{and} \qquad \rho^\phi_{\pm} - p^\phi_{\left(1,2\right)\pm} \geq 0 , \\\label{eq37}
DEC_3&=&WEC_{3} \Longleftrightarrow   \rho^\phi_{\pm}  \geq 0 .
\end{eqnarray}

The explicit forms of the energy conditions can be obtained by substituting the stress-energy tensor components, as expressed in Eqs. (\ref{eq27}-\ref{eq29}), into the inequalities defining the energy conditions in Eqs. (\ref{eq34}-\ref{eq37}).

This procedure yields the following expressions for the energy conditions within the timelike region located outside the event horizon, characterized by $A_\pm > 0$:


\begin{eqnarray}\label{eq38}
NEC^\phi_{1}&=&WEC^\phi_{1}=SEC^\phi_{1} \Longleftrightarrow  -\frac{2A_{\pm}\Sigma''}{\Sigma} \geq 0, \\\label{eq39}
NEC^\phi_{2}&=&WEC^\phi_{2}=SEC^\phi_{2} \Longleftrightarrow  0, \\\label{eq40}
SEC^\phi_3 & \Longleftrightarrow &  \frac{8{\Sigma}''{A_{\pm}}}{\Sigma} -2V_{\pm}\left(x\right)  \geq 0, \\\label{eq41}
DEC^\phi_{1} & \Longleftrightarrow & -\frac{8{\Sigma}''{A_{\pm}}}{\Sigma}                                + 2V_{\pm}\left(x\right) \geq 0, \\\label{eq42}
DEC^\phi_{2} & \Longleftrightarrow & -\frac{10{\Sigma}''{A_{\pm}}}{\Sigma}                                + 2V_{\pm}\left(x\right) \geq 0, \\\label{eq43}
DEC^\phi_{3}&=&WEC^\phi_{3}  \Longleftrightarrow   -\frac{5{A_{\pm}{\Sigma}''}}{\Sigma} +V_{\pm}\left(x\right) \geq 0.
\end{eqnarray}

Similarly, the energy conditions within the horizon, where $t$ is spacelike, are derived by substituting the stress-energy tensor components from Eqs. (\ref{eq31}-\ref{eq33}) into the inequalities in Eqs. (\ref{eq34}-\ref{eq37}). This results in the energy conditions for $A_\pm < 0$ as:

\begin{eqnarray}\label{eq44}
NEC^\phi_{1}&=&WEC^\phi_{1}=SEC^\phi_{1} \Longleftrightarrow  \frac{2A_{\pm}\Sigma''}{\Sigma} \geq 0, \\\label{eq45}
NEC^\phi_{2}&=&WEC^\phi_{2}=SEC^\phi_{2} \Longleftrightarrow   \frac{2A_{\pm}\Sigma''}{\Sigma} \geq 0, \\\label{eq46}
SEC^\phi_3 & \Longleftrightarrow &  \frac{12A_{\pm}\Sigma''}{\Sigma} -2V_{\pm}\left(x\right)\geq 0, \\\label{eq47}
DEC^\phi_{1} & \Longleftrightarrow & -\frac{8A_{\pm}\Sigma''}{\Sigma} +2V_{\pm}\left(x\right) \geq 0, \\\label{eq48}
DEC^\phi_{2} & \Longleftrightarrow & -\frac{8A_{\pm}\Sigma''}{\Sigma} +2V_{\pm}\left(x\right) \geq 0, \\\label{eq49}
DEC^\phi_{3}&=&WEC^\phi_{3} \Longleftrightarrow   -\frac{3A_{\pm}\Sigma''}{\Sigma} +V_{\pm}\left(x\right) \geq 0.
\end{eqnarray}

Equations (\ref{eq38}) and (\ref{eq44}) demonstrate a violation of the null energy condition, $NEC^\phi_1$, both within and beyond the event horizon in this novel configuration. Due to the association between the dominant energy condition, $DEC^\phi_1$, and the violated null energy condition, a violation of $DEC^\phi_1$ is also observed.

Beyond the horizon, the secondary null energy condition $NEC^\phi_2$ remains unviolated, as indicated by Eq. (\ref{eq39}). Consequently, Eqs. (\ref{eq40}), (\ref{eq42}), and (\ref{eq43}) remain open to interpretation. However, within the event horizon, the secondary null energy condition is violated, as observed in Eq. (\ref{eq45}). This leads to the conclusion that the dominant energy condition $DEC^\phi_2$ is also violated, as expressed in Eq. (\ref{eq48}). Since no definitive conclusions can be drawn about the energy conditions in Eqs. (\ref{eq46}) and (\ref{eq49}), a graphical representation is necessary for further analysis.

\begin{figure}[htb!]
		\subfigure[]
			{\label{DEC2a}
				\includegraphics[scale=0.54]{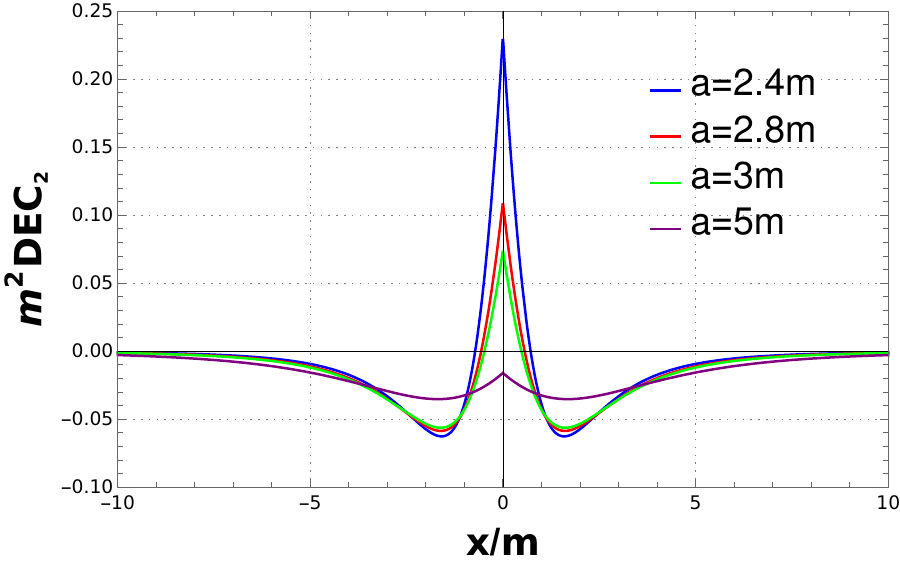}}\qquad
			\subfigure[]
			{\label{DEC3}
				\includegraphics[scale=0.54]{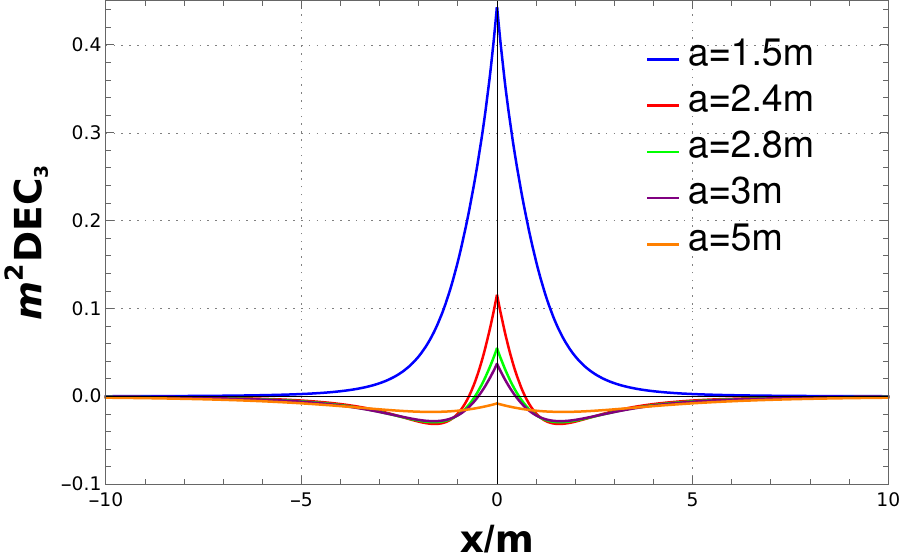}} 
    \caption{The DEC relating energy density and tangential pressure is plotted for radii inside and outside the event horizon.
    }
\label{DEC2}
\end{figure}
\begin{figure}[htb!]
				{\includegraphics[scale=0.7]{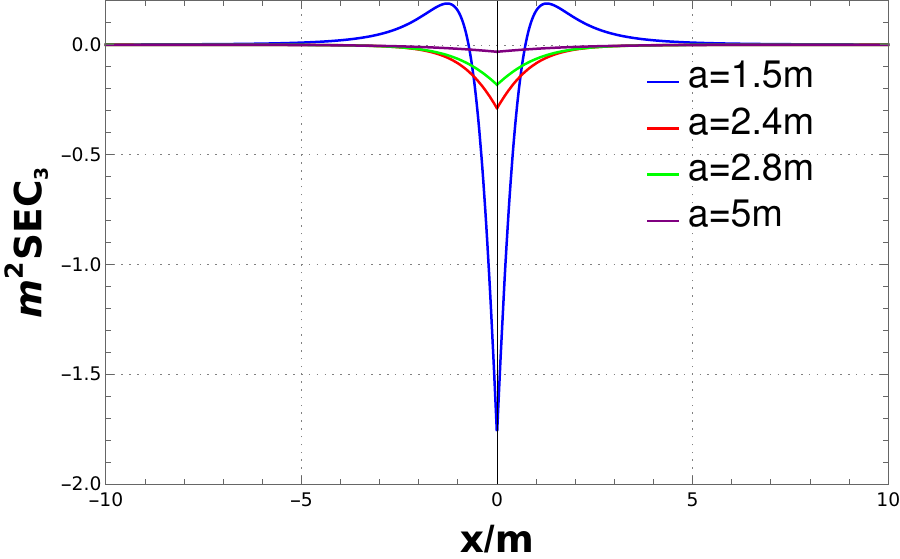}}
    \caption{The strong energy condition (SEC) combining energy density and all pressure components is plotted for various radii inside and outside the horizon.
    }
\label{SEC3}
\end{figure}

This complementary analysis is presented in Figures \ref{DEC2} and \ref{SEC3}. As depicted in Figure \ref{DEC2a}, the dominant energy condition $DEC^{\phi}_2$ is violated both outside and inside the event horizon. The same violation is observed for the strong energy condition $SEC^{\phi}_3$, as illustrated in Figure \ref{SEC3}. Figure \ref{DEC3} reveals that the energy density remains positive within the event horizon but is violated for throat radii outside it. This analysis aligns with the potential solutions previously obtained in Ref. \cite{CDJM}, where violating $DEC^{\phi}_3$ allows for two-way traversable wormholes, while maintaining positivity for $A_{\pm}<2m$ yields wormholes with symmetric horizons.



\newpage

\section{Special Configuration II}\label{sec5}

In this section, $k$-essence configurations with $\bar{n}=3$ are investigated. The scalar field solution, obtained via direct integration of Eq.~(\ref{eq19a}), is presented in (\ref{apen21}). An asymptotic expansion proves useful for evaluating the characteristics of this solution:

\begin{eqnarray}\label{SC1}
\phi_{+}\left(x\rightarrow\infty\right)=-\phi_{-}\left(x\rightarrow{-}\infty\right)=\frac{343\sqrt{14}\left[315 \pi ^2 a^2 (11 a-36 m)+160m^3\left(539-60\pi^2\right)+28am^2\left(600\pi^2-3289\right)\right]}{3840\pi{a}^9}, \nonumber  \\
\end{eqnarray}


\begin{eqnarray}\label{SC3}
\phi_{+}\left(x\rightarrow{0}\right)=-\phi_{-}\left(x\rightarrow{0}\right)=\frac{5488\sqrt{14}m\left[8m^2\left(225\pi^2-446\right)+9\pi^2{a}\left(57a-220m\right)\right]}{675\pi^3{a^9}},
\end{eqnarray} where

\begin{eqnarray}\label{eqjunção3}
\phi\left(x\right)= 
\begin{cases}
\phi_{+}, & x \geq {0} \\
\phi_{-}, & x \leq {0}.
\end{cases}
\end{eqnarray}

\begin{figure}[htb!]
		\centering
		\subfigure[]
			{\label{campoforaSC2}
				\includegraphics[scale=0.4]{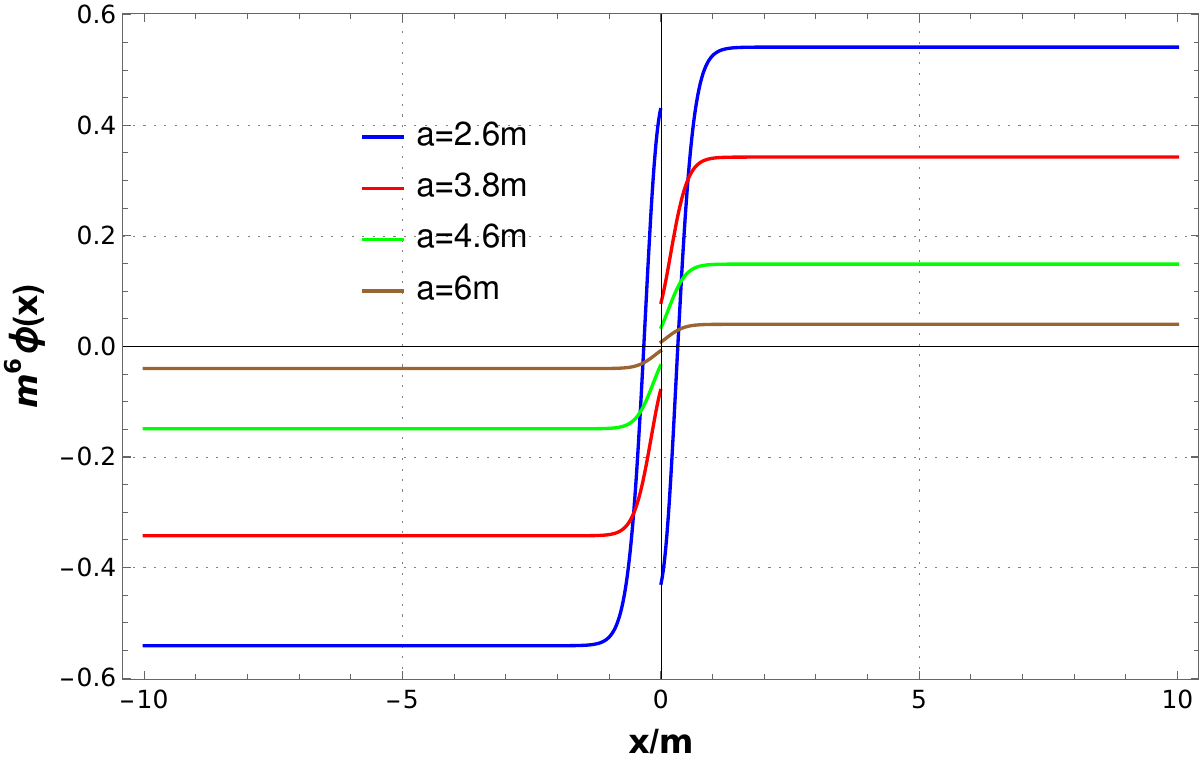}}\qquad
			\subfigure[]
			{\label{campodentroSC2}
				\includegraphics[scale=0.4]{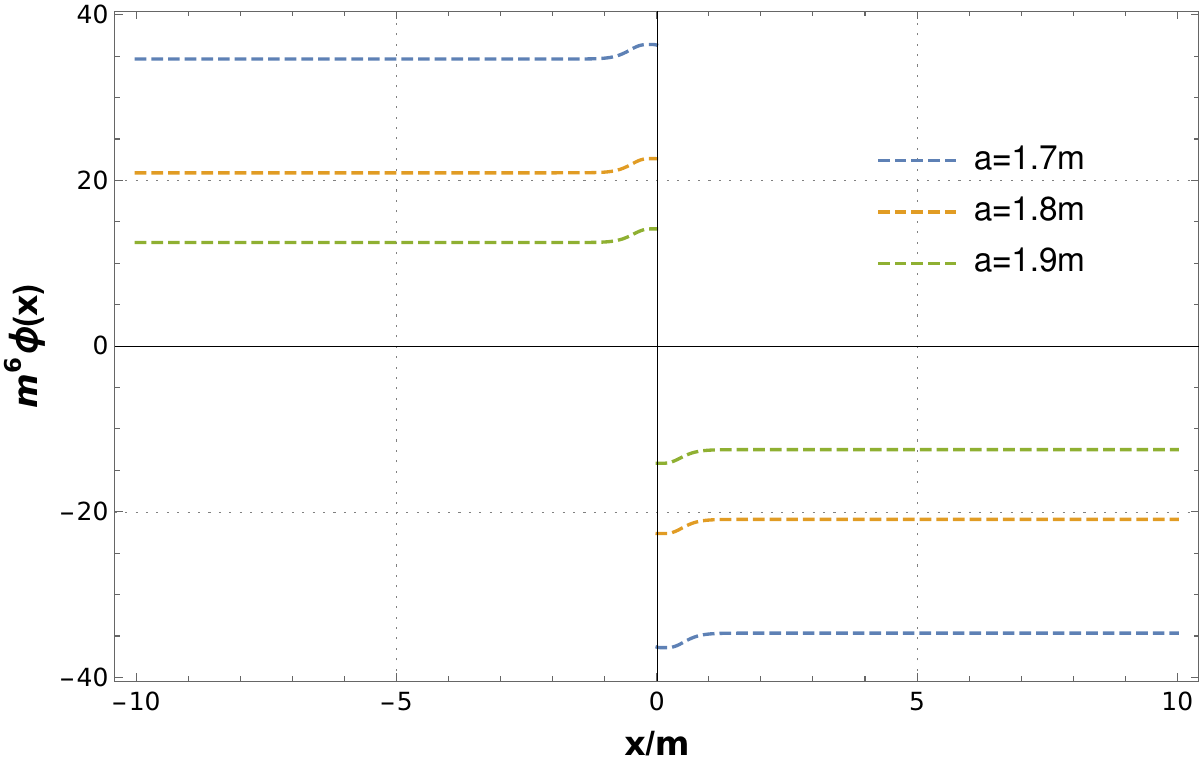}}
			\subfigure[]
			{\label{foraSC2}
				\includegraphics[scale=0.4]{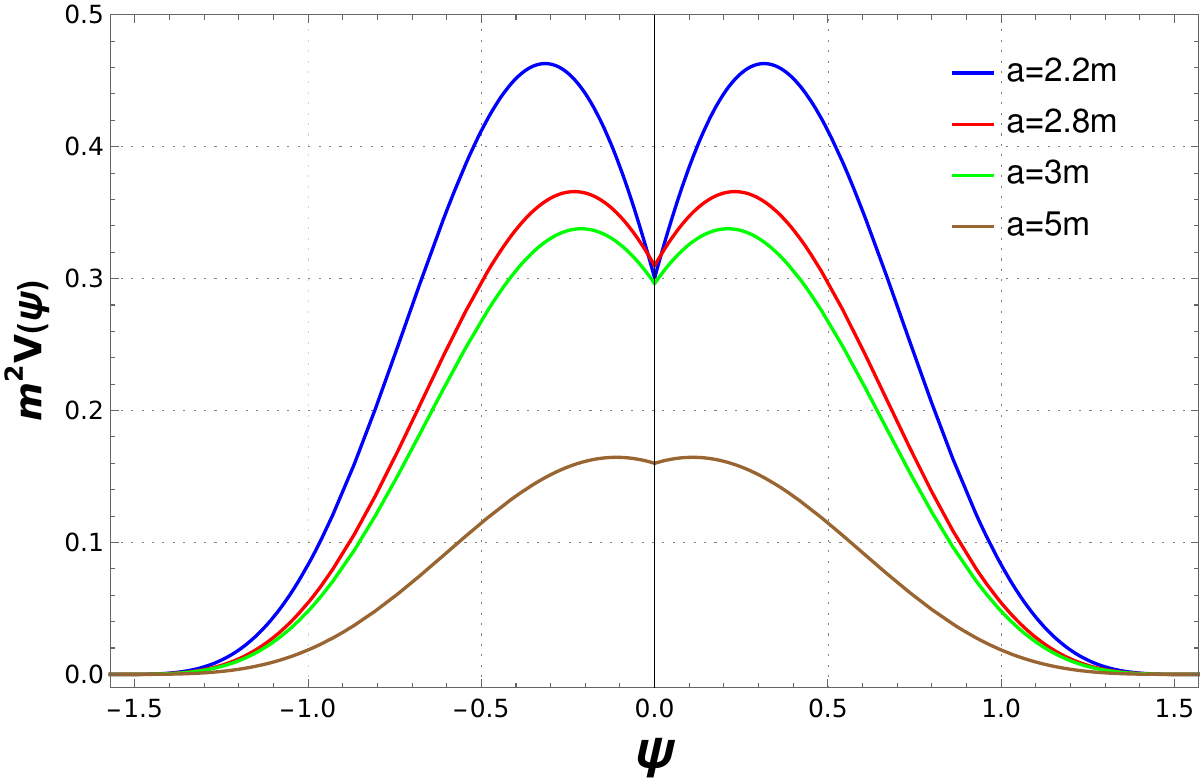}}\qquad
			\subfigure[]
			{\label{dentroSC2}
				\includegraphics[scale=0.4]{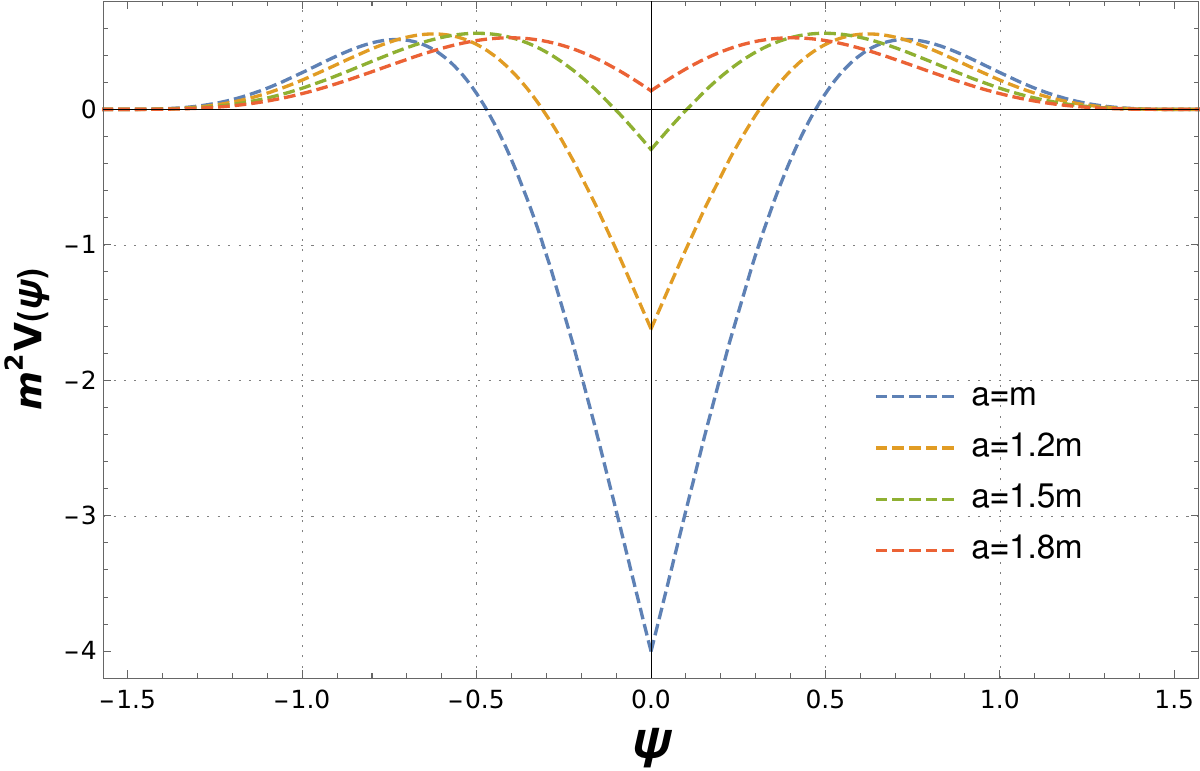}}
				
    \caption{(a) The scalar field from Eq. (\ref{eq19a}) for radius values outside the event horizon. (b) The scalar field from Eq. (\ref{eq19a}) for radius values inside the event horizon. In both cases, the constant $F_0=1$ is fixed. (c) The potential in Eq. (\ref{SC5}) for throat radii outside the event horizon. (d) The potential in Eq. (\ref{SC5}) for throat radii inside the event horizon.}
\label{ESCALARCASOII}
\end{figure}

Figure \ref{ESCALARCASOII} provides a graphical representation of the scalar field configuration given by Eq. (\ref{eq19a}) for radius values inside and outside the event horizon. Similar to the configuration examined in Section \ref{sec4}, the scalar field exhibits a discontinuity at the origin for radii inside \ref{campodentroSC2} and outside \ref{campoforaSC2} the horizon. This behavior suggests a dependence on the odd derivatives of the metric function, as described by Eq. (\ref{eq15}).


Furthermore, this effect can be visualized by examining the asymptotic values of the scalar field in Eq. (\ref{SC3}). The scalar field tends to assume progressively smaller values for throat radii further from the event horizon (Fig. \ref{campoforaSC2}). However, sign inversion may occur in the limit as $x$ approaches infinity for smaller throat radii or those on the order of $a \approx 2.43m$ (see Figs. \ref{campoforaSC2} and \ref{campodentroSC2}). Similarly, as the scalar field approaches the origin, it takes on positive values for larger throat radii or those on the order of $a \approx 2.89m$, while exhibiting negative values for smaller radii (see Figs. \ref{campoforaSC2} and \ref{campodentroSC2}). In the asymptotic limit where $x$ approaches negative infinity for throat radii far from the event horizon, the scalar field tends towards zero (see Fig. \ref{campoforaSC2}). However, a sign reversal may occur for throat radius values on the order of $a \approx 2.42m$ or smaller (see Figs. \ref{campoforaSC2} and \ref{campodentroSC2}).

The potential for such a configuration is obtained through Eq. (\ref{eq11}) using the scalar field profile in Eq. (\ref{eq19a}). The potential must therefore be expressed as:

\begin{eqnarray}\label{SC4}
V_{\pm}\left(\phi\left(x\right)\right)= \frac{1}{\Sigma^2} -\frac{{{\Sigma}'A'_{\pm}}}{\Sigma}-\frac{A_{\pm}{\Sigma'}^2}{\Sigma^2} + \frac{5a^2A_{\pm}}{\Sigma^4},
\end{eqnarray} where


\begin{eqnarray}\label{eqjunção4}
V\left(\phi\left(x\right)\right)= 
\begin{cases}
V_{+}, & x \geq {0} \\
V_{-}, & x \leq {0}.
\end{cases}
\end{eqnarray}

As discussed in Section \ref{sec4} and Ref. \cite{CDJM}, to better visualize the behavior of the potential in Eq. (\ref{SC4}), the variable transformation $\psi=\arctan\left( \frac{x}{a}\right)$ can be employed. Taking the asymptotic limits as $x$ approaches $\pm\infty$ is equivalent to considering $\psi\rightarrow\pm\frac{\pi}{2}$. Consequently, the potential can be rewritten in terms of the new variables as:


\begin{eqnarray}\label{SC5}
V_{\pm}\left(\psi\right)= \frac{9-2\pi{ca^2}+12\cos\left(2\psi\right)+3\cos\left(4\psi\right)\pm{ca^2}\left[4\psi\mp{8}\left(\pi\mp{2\psi}\right)\cos\left(2\psi\right)+3\sin\left(4\psi\right)\right]}{4a^2},
\end{eqnarray} where $c=\frac{4m}{\pi{a^3}}$ is a combination of constants.


At the bottom of the Figure \ref{ESCALARCASOII}, a graphical representation of the potential for the new configuration of interest is provided, with throat radii lying outside \ref{foraSC2} and inside \ref{dentroSC2} the horizon. The behavior is similar to the previous case discussed in Section \ref{sec4}, although numerically displaced. Thus, the same analysis performed in Section \ref{sec4} is applicable to this configuration.

Thus far, three of the four equations of motion have been utilized. It remains to ensure that the fourth equation, Eq. (\ref{eq13}), is also satisfied for this configuration. This can be verified through the following expression:

\begin{eqnarray}\label{SC6}
\frac{dV_{\pm}}{dx} + \frac{F_{0}}{7}\left(\frac{F_{0}}{14a^2}\right)^\frac{5}{2}\left(\frac{{\phi'_\pm}}{\Sigma^2}\right)\left(\frac{\Sigma^{12}}{A^2_{\pm}}\right)'=0.
\end{eqnarray}

\subsection{Energy conditions}\label{sec51}

As demonstrated in the preceding section \ref{sec41} and in the work by \cite{CDJM}, for this configuration with $\bar{n}=3$, the non-zero components of the energy-momentum tensor can be defined through the scalar field in Eq. (\ref{eq19a}) and the potential in Eq. (\ref{SC4}). The energy conditions outside the event horizon, where $A_\pm > 2m$, can then be constructed as:

\begin{eqnarray}\label{SC7}
NEC^\phi_{1}&=&WEC^\phi_{1}=SEC^\phi_{1} \Longleftrightarrow  -\frac{2A_{\pm}\Sigma''}{\Sigma} \geq 0, \\\label{SC8}
NEC^\phi_{2}&=&WEC^\phi_{2}=SEC^\phi_{2} \Longleftrightarrow  0, \\\label{SC9}
SEC^\phi_3 & \Longleftrightarrow &  \frac{12{\Sigma}''{A_{\pm}}}{\Sigma} -2V_{\pm}\left(x\right)  \geq 0, \\\label{SC10}
DEC^\phi_{1} & \Longleftrightarrow & -\frac{12{\Sigma}''{A_{\pm}}}{\Sigma}                                + 2V_{\pm}\left(x\right) \geq 0, \\\label{SC11}
DEC^\phi_{2} & \Longleftrightarrow & -\frac{14{\Sigma}''{A_{\pm}}}{\Sigma}                                + 2V_{\pm}\left(x\right) \geq 0, \\\label{SC12}
DEC^\phi_{3}&=&WEC^\phi_{3}  \Longleftrightarrow   -\frac{7{A_{\pm}{\Sigma}''}}{\Sigma} +V_{\pm}\left(x\right) \geq 0.
\end{eqnarray}

Inside the event horizon, where $A_\pm < 2m$, the energy conditions are expressed as:

\begin{eqnarray}\label{SC13}
NEC^\phi_{1}&=&WEC^\phi_{1}=SEC^\phi_{1} \Longleftrightarrow  \frac{2A_{\pm}\Sigma''}{\Sigma} \geq 0, \\\label{SC14}
NEC^\phi_{2}&=&WEC^\phi_{2}=SEC^\phi_{2} \Longleftrightarrow   \frac{2A_{\pm}\Sigma''}{\Sigma} \geq 0, \\\label{SC15}
SEC^\phi_3 & \Longleftrightarrow &  \frac{16A_{\pm}\Sigma''}{\Sigma} -2V_{\pm}\left(x\right)\geq 0, \\\label{SC16}
DEC^\phi_{1} & \Longleftrightarrow & -\frac{12A_{\pm}\Sigma''}{\Sigma} +2V_{\pm}\left(x\right) \geq 0, \\\label{SC17}
DEC^\phi_{2} & \Longleftrightarrow & -\frac{12A_{\pm}\Sigma''}{\Sigma} +2V_{\pm}\left(x\right) \geq 0, \\\label{SC18}
DEC^\phi_{3}&=&WEC^\phi_{3} \Longleftrightarrow   -\frac{5A_{\pm}\Sigma''}{\Sigma} +V_{\pm}\left(x\right) \geq 0.
\end{eqnarray}

Despite constructing the expressions for the energy conditions above for this new configuration, they are found to be equivalent to the scenario analyzed previously in section \ref{sec41}. Thus, the physical interpretation must remain the same.

\section{Analyzing the General Solution} \label{sec6}

In this section, the physical quantities described above will be investigated for the general solution of the equation of motion (\ref{eq12}) for the Simpson-Visser area function $\Sigma^2=x^2 +a^2$. In this solution Eq. (\ref{GN}), boundary conditions were imposed such that the solution is asymptotically flat as $x\to\infty$ and exhibits the Simpson-Visser limit at the origin. Therefore, its form is given by the positive sign of the expression (\ref{eq15}), now defined for all $x$. It was explored in \cite{CDJM} that this solution (\ref{eq15}) is asymptotically flat when $x\to\infty$ and behaves in a similar way to Schwarzschild-de Sitter when $x\to{-\infty}$ \cite{desitter}. And regarding its regularity, it is shown that the Kretschmann scalar tends to zero when $x\to\infty$ (Minkowski limit) and tends to a positive constant as $x\to{-\infty}$.

\subsection{Case $\bar{n}=2$}\label{sec61}

The analysis initially focuses on the scalar field and the potential based on the general metric function (\ref{eq15}) (positive sign), which is equivalent to considering the positive sign of the respective equations (\ref{apen11}) and (\ref{eq23}) defined for all values of $x$. Therefore, the behavior of the scalar and potential fields in the asymptotic limit where $x\to\infty$ is the same as described in section \ref{sec4}. On the other hand, as the analysis is conducted on the general metric function without any type of match, the behavior of the scalar and potential fields in the asymptotic limit of $x\to-\infty$ differs.

Consequently, one obtains the asymptotic expression for the scalar field as follows:

\begin{eqnarray}\label{Asscaso1}
\phi\left(x\rightarrow{-\infty}\right)= -\frac{25 \sqrt{\frac{5}{2}} \left(105 \pi ^2 a^2-720 \pi ^2 a m+112 \left(12 \pi ^2-5\right) m^2\right)}{96 \pi  a^6}.
\end{eqnarray}

\begin{figure}[htb!]
		\centering
		\subfigure[]
			{\label{cpforasemcolmod2}
				\includegraphics[scale=0.4]{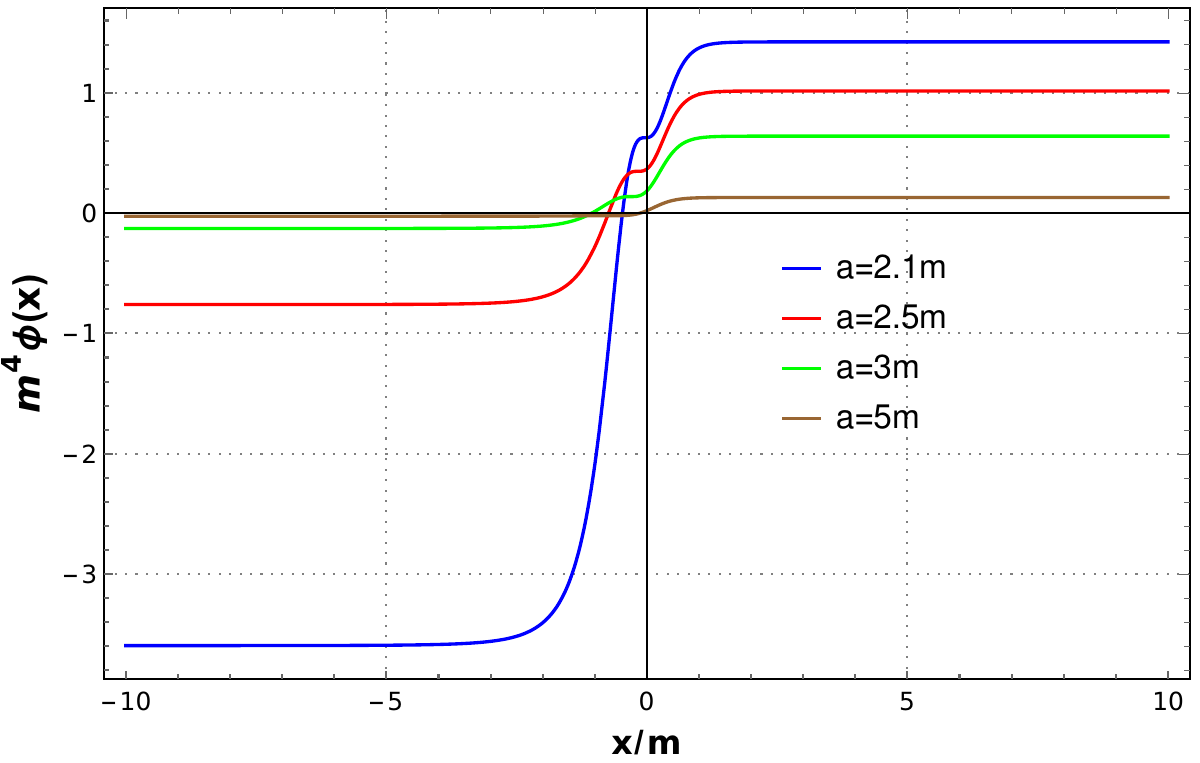}}\qquad
			\subfigure[]
			{\label{cpdentrosemcolmod2}
				\includegraphics[scale=0.4]{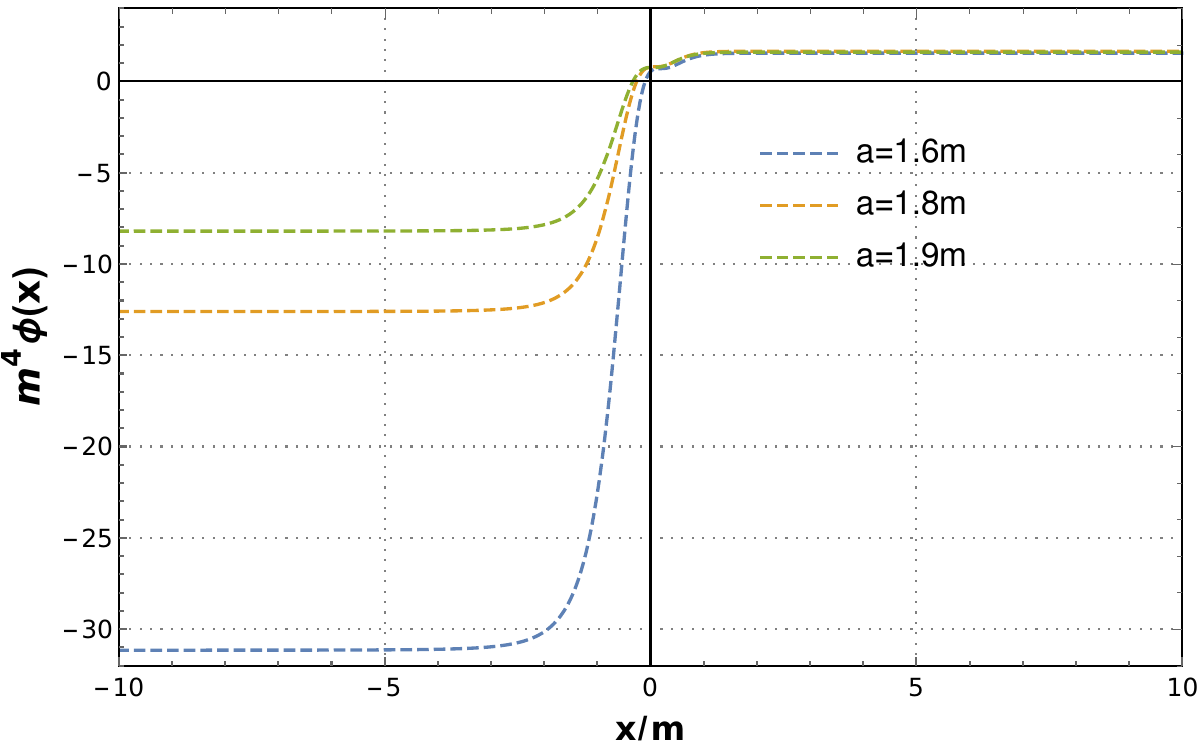}}
			\subfigure[]
			{\label{potforasemcolmod2}
				\includegraphics[scale=0.4]{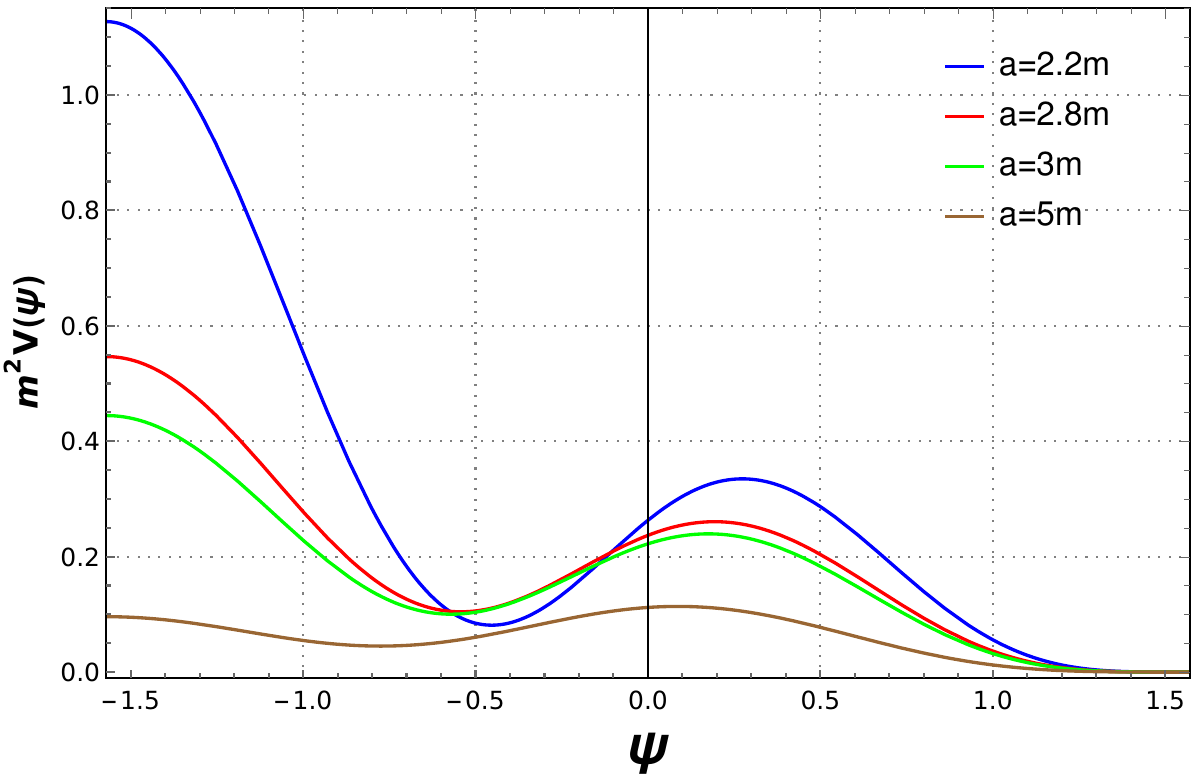}}\qquad
			\subfigure[]
			{\label{potdentrosemcolmod2}
				\includegraphics[scale=0.4]{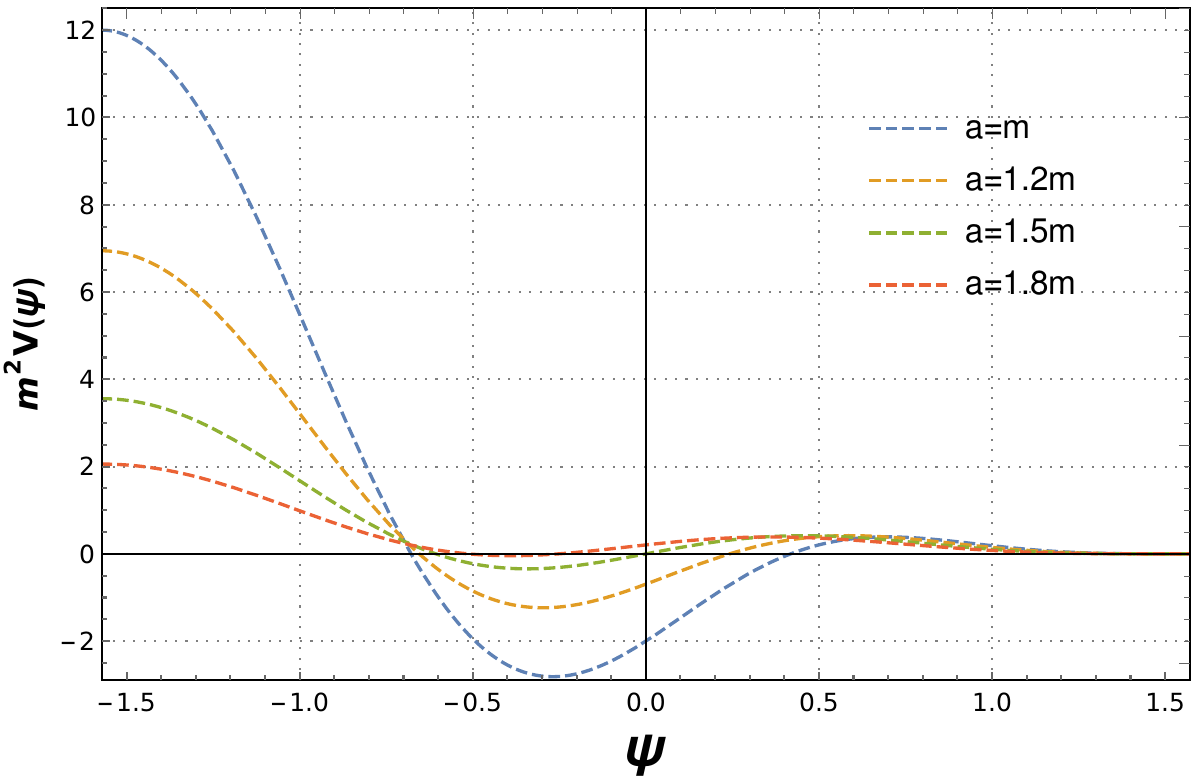}}

    \caption{Scalar and potential field for throat radius values inside and outside the event horizon. In both cases, the constant $F_0=1$ is fixed.}
\label{CASOISEMCOLAGEM}
\end{figure}

As in the scenario investigated in Section \ref{sec4}, for values of throat radii far from the event horizon, the scalar field tends to zero (Eq. (\ref{Asscaso1}), see Figure \ref{cpforasemcolmod2}). And for values of throat radii increasingly internal to the horizon, the field tends to assume increasingly negative values (see Figure \ref{cpdentrosemcolmod2}).

Likewise, the analysis of the potential in the asymptotic limit of $x\to\infty$ is identical to that investigated in Section \ref{sec4}. Regarding the left side of the potential, which refers to the asymptotic of $x\to{-\infty}$, it is observed that it tends to assume increasingly greater values for throat radii increasingly internal to the event horizon (see Figures \ref{potforasemcolmod2} and \ref{potdentrosemcolmod2}). A very interesting aspect that also appears in this scenario is that the potential for throat radius values in the order of $a\approx{1.8m}$ tends to start creating a negative minimum that grows towards more internal radii (see Figure \ref{potdentrosemcolmod2}). This fact may indicate the possibility of obtaining normal and/or quasi-normal modes for linear perturbations of the scalar field for some ray regimes \cite{teller1,teller2,teller3,teller4,molina}.



\subsubsection{Energy conditions}

\begin{figure}[htb!]
		\centering
		\subfigure[]
			{\label{DEC2SEMCOL}
				\includegraphics[scale=0.52]{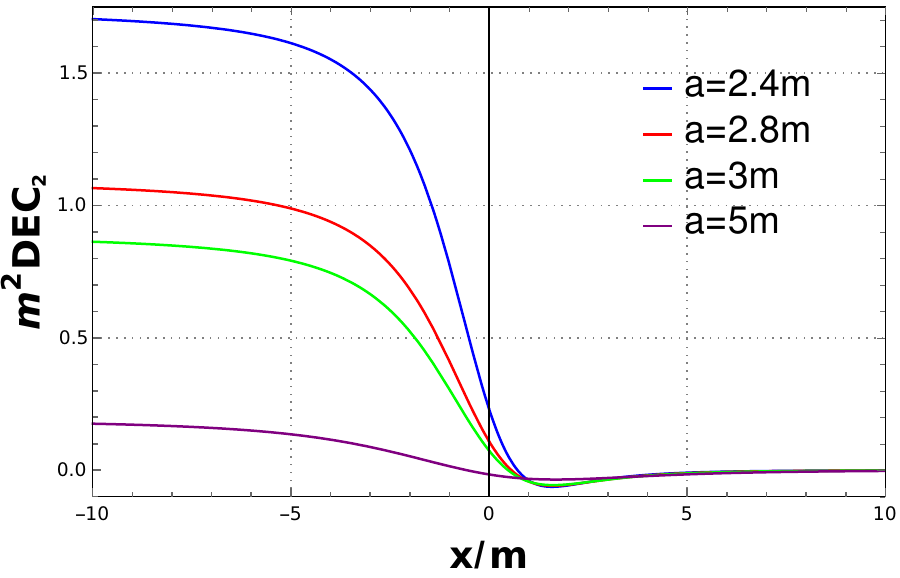}}\qquad
			\subfigure[]
			{\label{SEC3SEMCOL}
				\includegraphics[scale=0.52]{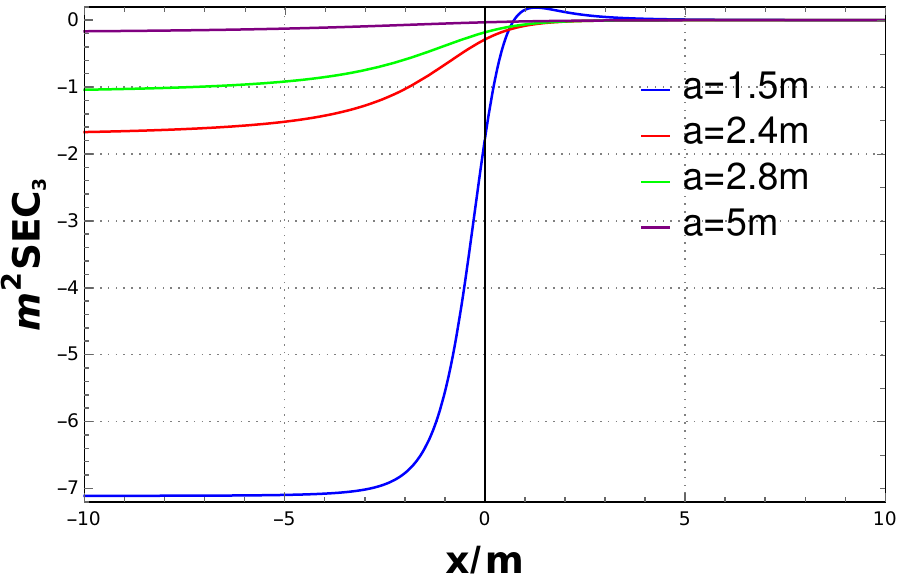}}
			\subfigure[]
			{\label{DEC3ASEMCOL}
				\includegraphics[scale=0.4]{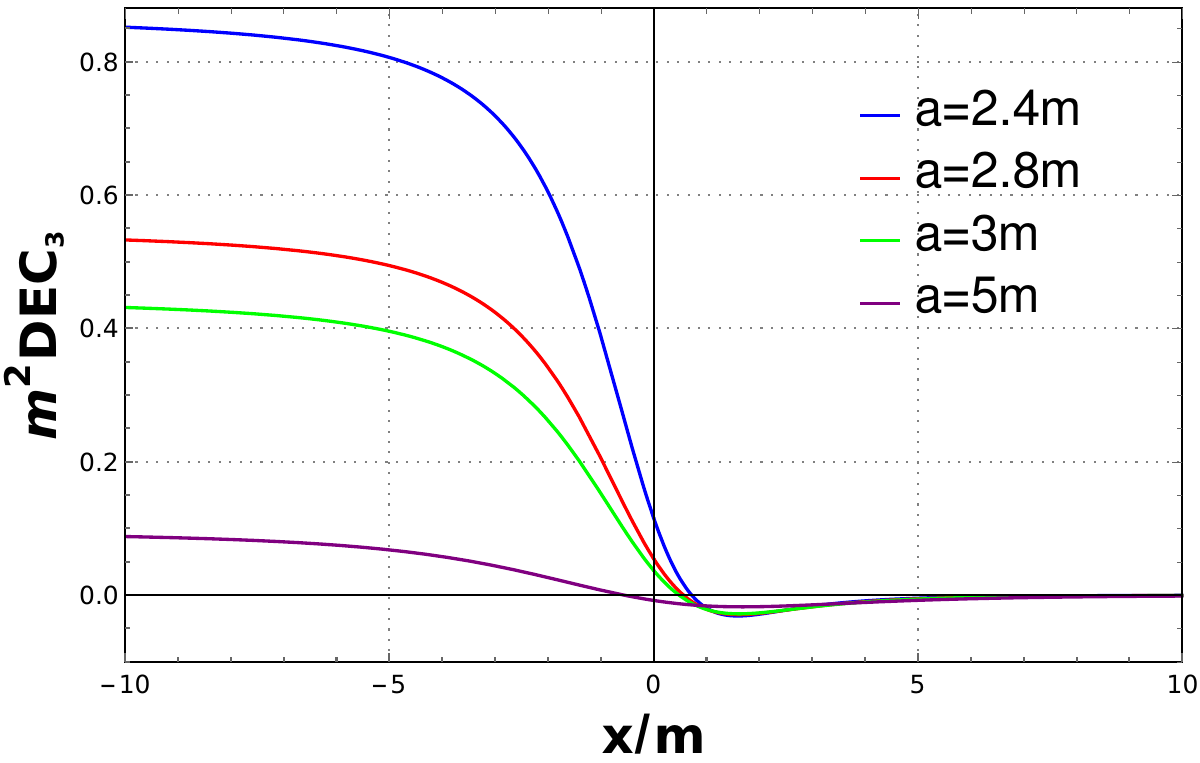}}\qquad
			\subfigure[]
			{\label{DEC3BSEMCOL}
				\includegraphics[scale=0.4]{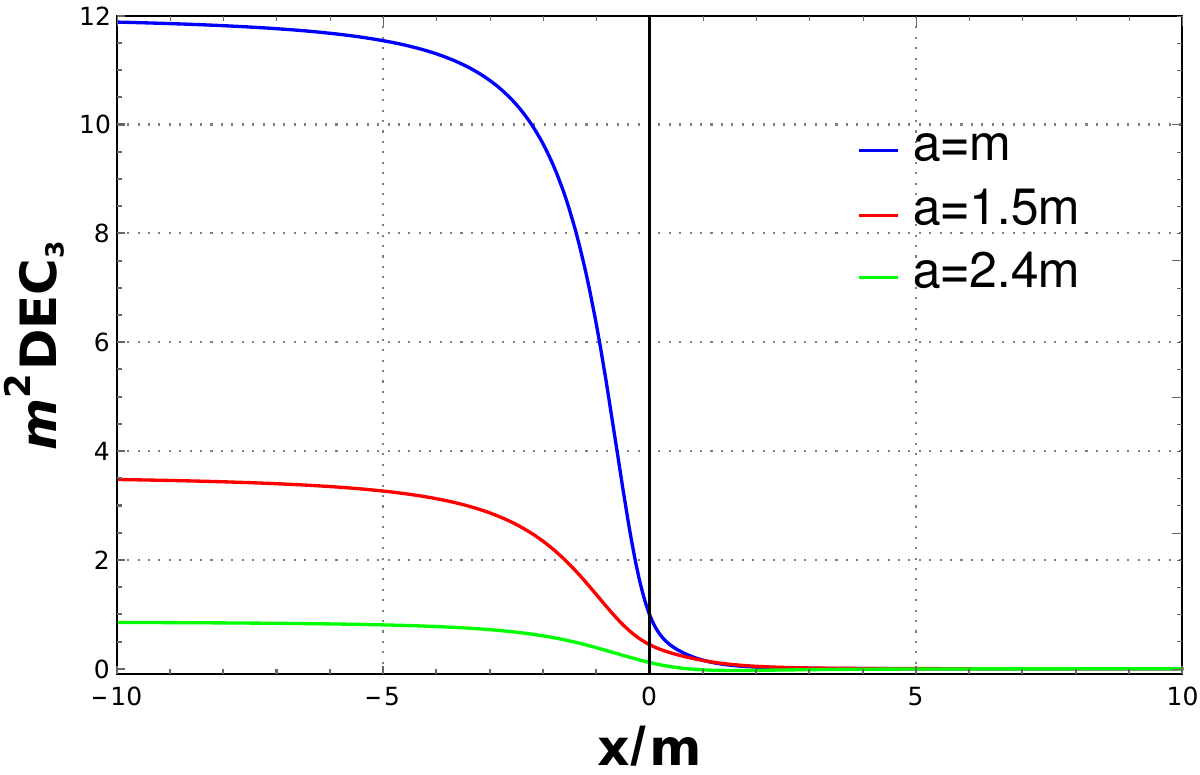}}

    \caption{Complementary energy conditions for throat lightning inside and outside the event horizon.}
\label{CDSEMCOL}
\end{figure}

Regarding the analysis of energy conditions, the procedure mirrors that of Section \ref{sec41}, but without considering the combination of the metric function. The conclusions for the model regarding energy condition violations remain consistent with those analyzed in Section \ref{sec41}, except for the energy conditions not found in Figure \ref{CDSEMCOL}, which will be investigated below.

Therefore, as in the analysis conducted in Section \ref{sec41}, the dominant energy condition $DEC^{\phi}_{2}$ is violated both inside and outside the event horizon (Fig. \ref{DEC2SEMCOL}), as is the strong energy condition $SEC^{\phi}_3$ (Fig. \ref{SEC3SEMCOL}). Finally, for the dominant energy condition $DEC^{\phi}_{3}$, violation occurs for throat radii located outside the event horizon (Fig. \ref{DEC3ASEMCOL}) but not within the event horizon (Fig. \ref{DEC3BSEMCOL}).


\subsection{Case $\bar{n}=3$}\label{sec62}

\begin{figure}[htb!]
		\centering
		\subfigure[]
			{\label{cpforasemcolmod3}
				\includegraphics[scale=0.4]{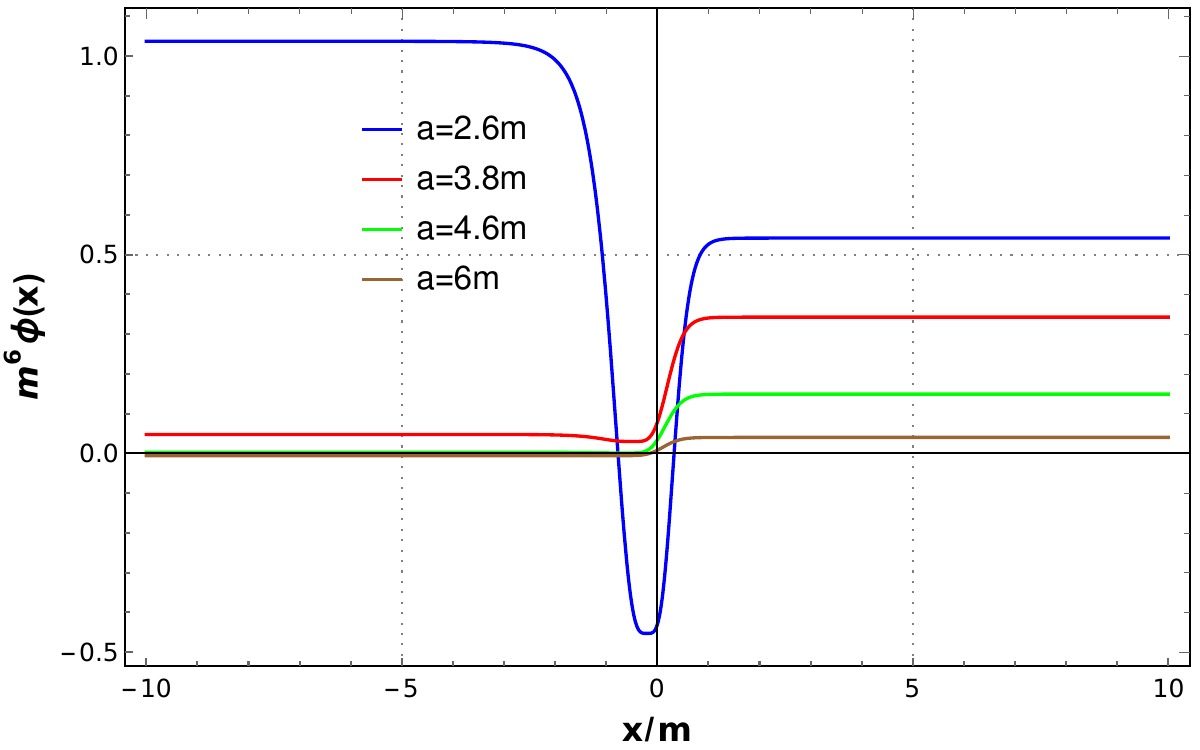}}\qquad
			\subfigure[]
			{\label{cpdentrosemcolmod3}
				\includegraphics[scale=0.4]{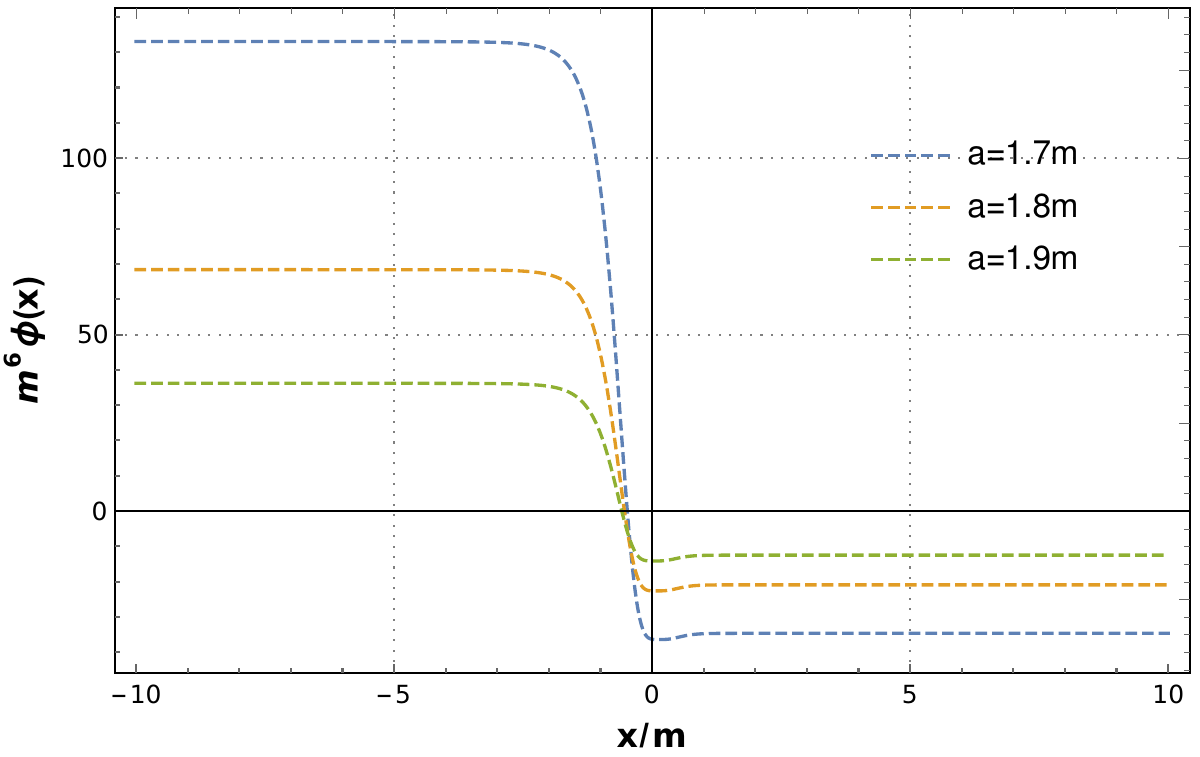}}
			\subfigure[]
			{\label{potforasemcolmod3}
				\includegraphics[scale=0.4]{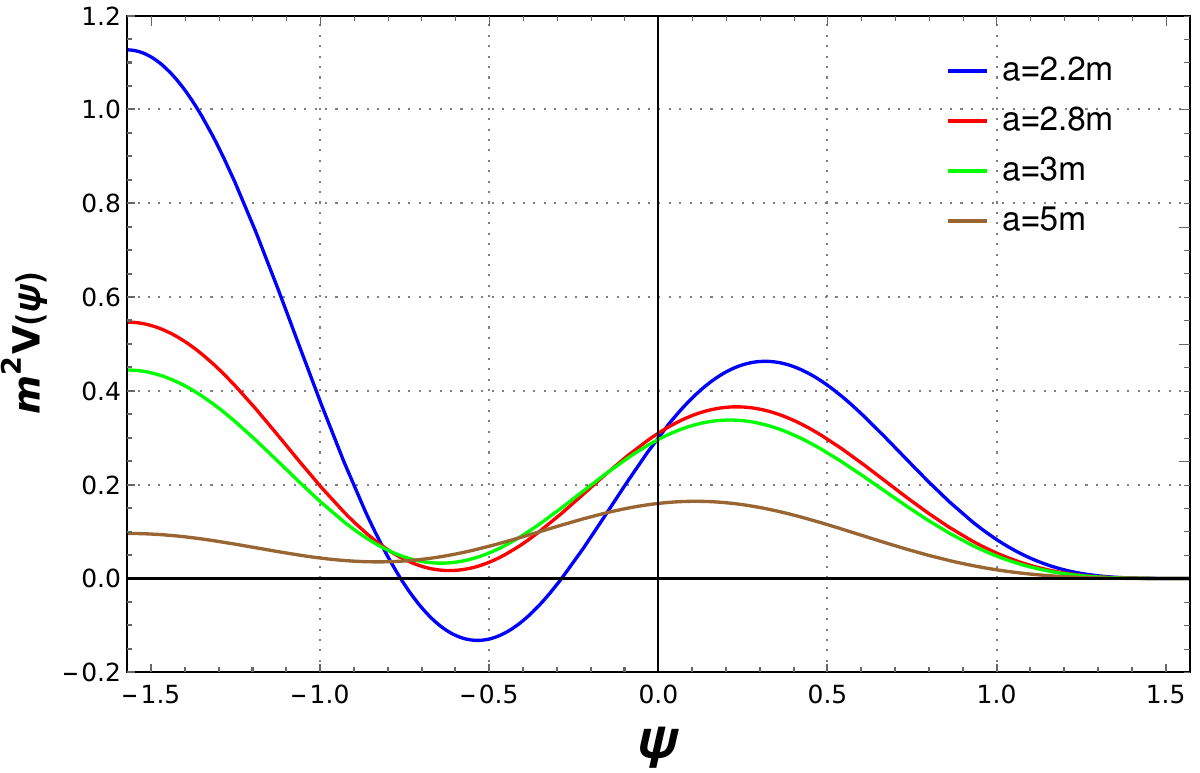}}\qquad
			\subfigure[]
			{\label{potdentrosemcolmod3}
				\includegraphics[scale=0.4]{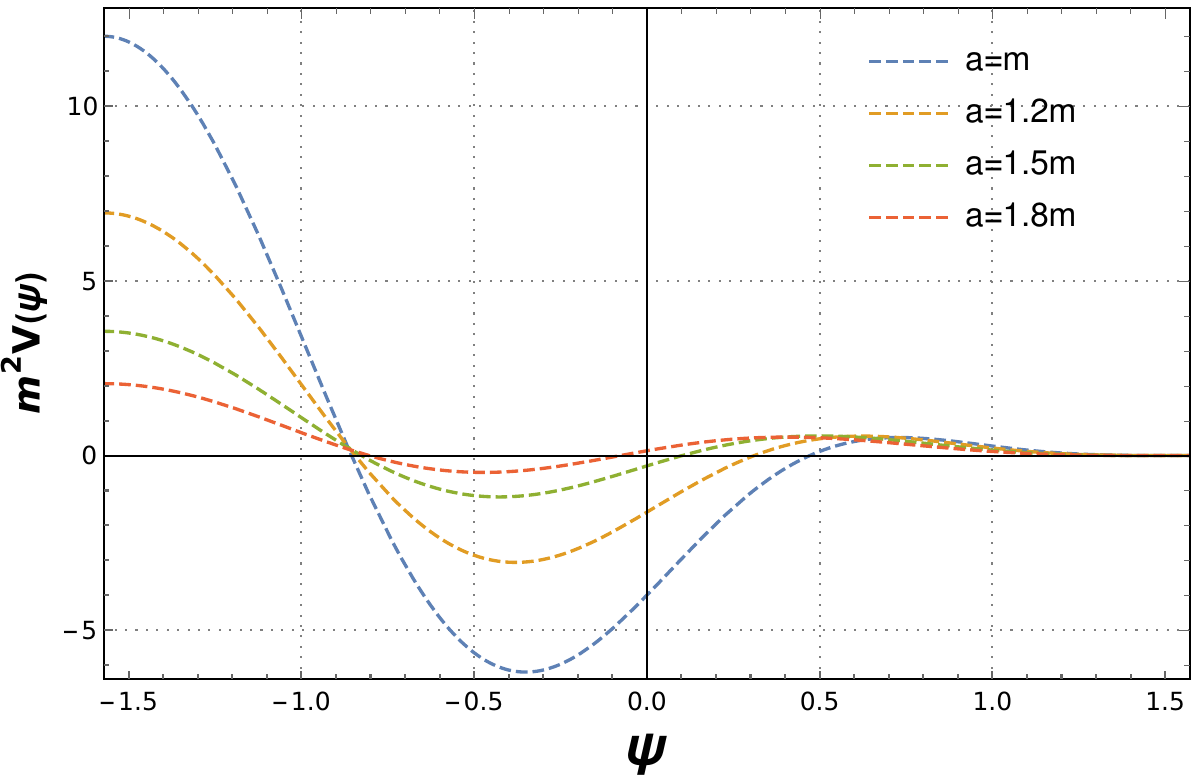}}

    \caption{Scalar and potential field for throat radius values inside and outside the event horizon. In both cases, the constant $F_0=1$ is fixed.}
\label{CASOIISEMCOLAGEM}
\end{figure}

As already mentioned in the previous section \ref{sec61} the behavior of the scalar and potential field is even described in the analysis of the section that it was separated by a region \ref{sec5} in the asymptotic limit in which $x\to\infty$. On the other hand, the behavior of these quantities changes in the asymptotic $x\to{-\infty}$ as can be seen in the panel \ref{CASOIISEMCOLAGEM}.

The asymptotic expression of the scalar field configuration is given below

\begin{eqnarray}\label{Assincaso2}
\phi\left(x\rightarrow{-\infty}\right)= -\frac{342\sqrt{14}\left[28 \left(4200 \pi ^2-3289\right) a m^2-480 \left(300 \pi ^2-539\right) m^3 + 315 \pi ^2 a^2 (11 a-108 m)\right]}{3840a^9\pi}.
\end{eqnarray}

For values increasingly within the event horizon, the scalar field exhibits a growing behavior, assuming positive values, as depicted in Fig. \ref{cpdentrosemcolmod3}. Conversely, for throat radii further outside the event horizon, the field tends towards zero (Fig. \ref{cpforasemcolmod3}). Regarding the potential, as in the previous model, it tends to assume increasingly larger positive values for increasing throat radii within the horizon (see Figs. \ref{potdentrosemcolmod3} and \ref{potforasemcolmod3}).

An intriguing observation is that in the analyses of configurations with match conducted in Sections \ref{sec4} and \ref{sec5}, as well as the case without matching investigated in Section \ref{sec61}, the emergence of a minimum point began to form for throat values already within the event horizon. In contrast, for this $\bar{n}=3$ configuration, the minimum tends to begin forming outside the event horizon and grows towards smaller throats or those on the order of $a \approx 2.66m$. This may suggest the possibility of normal or quasi-normal modes for scalar field perturbations for throat radii still outside the event horizon \cite{teller1,teller2,teller3,teller4,molina}.

Repeating the analysis of energy conditions is unnecessary, as they remain consistent with those analyzed in Section \ref{sec61}. The sections describing the metric by region demonstrated that these scalar field configurations do not alter the energy configurations.




\section{Conclusion}\label{sec7}

The possibility of introducing non-canonical scalar fields to investigate wormhole solutions within the Simpson-Visser Black-Bounce class was explored in Ref. \cite{matt}. Utilizing a characteristic area function $\Sigma^2=x^2 +a^2$, a novel region-defined metric function $A_{\pm}$ can be constructed \cite{CDJM}, along with a specific configuration of a ghost scalar field with $\bar{n}=1$. This combination, together with the corresponding potential, yields Black-Bounce solutions.

The present work verified that a class of new ghost scalar field configurations can be constructed using Eq. (\ref{eq16}). As the power values in the $k$-essence field increase, the form of the scalar field tends to grow with a positive power of the metric function, Eq. (\ref{eq15}), facilitating its explicit form through direct integration. The geometric properties of the metric function in Eq. (\ref{eq15}) have been previously explored in Ref. \cite{CDJM}, demonstrating the absence of divergences.

It was found that for the boundary conditions required to construct the function in Eq. (\ref{eq15}), at least qualitatively predicted results within the class of solutions obtained in Ref. \cite{PRL} can be recovered. Notably, the two scalar field configurations examined here, Eqs. (\ref{apen11}) and (\ref{apen21}), exhibit discontinuity at the origin of the coordinate system, precisely at the point where the metric function was mirrored in Ref. \cite{CDJM}. This suggests a dependence on the odd derivatives of the metric function $A_\pm$, inheriting its primary characteristic.

Finally, when analyzing the energy conditions for both scalar field configurations investigated in Sections \ref{sec41} and \ref{sec51}, it was found that for this class where the scalar field can assume different values and modify the expression for the potential associated with each of them, the energy conditions remain unchanged as a result. This further corroborates the solutions obtained in \cite{CDJM}. Intuitively, it is expected that this behavior will extend to configurations that were not explored here but that make up the same class of solutions $\bar{n}= 4,5,6,\dots$. Other ghost scalar field configurations can thus maintain the geometric structure of the solutions in Ref. \cite{CDJM}. However, no conclusions can be drawn about the energy conditions for other possible constructions of $k$-essence fields unless we have a complete analytical form for the polynomial form of the scalar field Eq. (\ref{eq16}) that provides an expression for all values of $\bar{n}$.

\begin{acknowledgments}
We thank CNPq, CAPES and FAPES for financial support.
\end{acknowledgments}


\appendix
\section{Field configurations}\label{appendix}

This appendix is intended to explicitly define the scalar field configurations for the cases investigated here.

\subsection{case $\bar{n}=2$}\label{apen1}

In this subsection we have the first case which consists of considering $\bar{n}=2$ which is equivalent to doing $n=\frac{1}{5}$. In this way, we have the general form of the scalar field defined by region

\begin{eqnarray}\label{apen11}
\phi_{\pm}\left(x\right)&=&\Omega_2\left(1-c\pi{a^2}+\frac{c^2\pi^2{a^4}}{2}\right)I_1 + \Omega_2\left(\pm{2ac}\mp{\pi{c^2a^3}}\right)I_2 + \Omega_2\left(a^2c^2-\pi{c}+\frac{\pi^2{c^2a^2}}{2}\right)I_3 \mp \Omega_2\left(\pi{ac^2}\right)I_4 + \nonumber \\
&+& \Omega_2\left(\frac{\pi^2{c^2}}{4}\right)I_5 + \Omega_2\left(\pm{2ca^2\mp{\pi{c^2a^4}}}\right)I_6 + \Omega_2\left(2c^2a^3\right)I_7 +\Omega_2\left(\pm{2c}\mp{2\pi{c^2a^2}}\right)I_8 + \Omega_2\left(2ac^2\right)I_9 + \nonumber \\
&\mp& \Omega_2\left(\pi{c^2}\right)I_{10} + \Omega_2\left(c^2a^4\right)I_{11} + \Omega_2\left(2a^2c^2\right)I_{12} + \Omega_2{c^2}I_{13},
\end{eqnarray} where the constants are defined by $c=\frac{4m}{\pi{a^3}}$ and $\Omega_2=\left(\frac{10a^2}{F_0}\right)^\frac{5}{2}$. The functions $I_j$ for $j=1,2,\dots,13$ are defined below

\begin{eqnarray}\label{apen12}
I_1&=&\frac{35\arctan\left(\frac{x}{a}\right)}{128a^9}+\frac{x\left[\frac{a^6}{\Sigma^6(x)}+\frac{7a^4}{6\Sigma^4\left(x\right)}+\frac{35a^2}{24\Sigma^2\left(x\right)}+\frac{35}{16}\right]}{8a^8\Sigma^2\left(x\right)}, \\\label{apen13}
I_2&=& -\frac{1}{8\Sigma^8\left(x\right)},  \\\label{apen14}
I_3&=&\frac{5\arctan\left(\frac{x}{a}\right)}{128a^7}+\frac{\left[\frac{8a^3x^3}{3\Sigma^6\left(x\right)}+\frac{ax\left(x^2-a^2\right)}{2\Sigma^4\left(x\right)}+\frac{ax^5\left(x^2-7 a^2\right)}{8\Sigma^8\left(x\right)}+\frac{a^5x\left(7x^2-a^2\right)}{8\Sigma^8\left(x\right)}\right]}{16a^7}, \\\label{apen15}
I_4&=& \frac{1}{2}\left[\frac{a^2}{4\Sigma^8\left(x\right)}-\frac{1}{3\Sigma^6\left(x\right)}\right], \\\label{apen16}
I_5&=& \frac{3\arctan\left(\frac{x}{a}\right)}{128a^5}+ \frac{\left[\frac{ax\left(x^2-a^2\right)}{2\Sigma^4\left(x\right)}+\frac{ax^5\left(7a^2-x^2\right)}{8\Sigma^8\left(x\right)}+\frac{a^5x\left(a^2-7x^2\right)}{8\Sigma^8\left(x\right)}\right]}{16a^5}, \\\label{apen17}
I_6&=&\frac{35\arctan^2\left(\frac{x}{a}\right)}{256a^9} +\frac{\left[-\frac{16a^3x^3}{3\Sigma^6\left(x\right)}+\frac{7ax\left(a^2-x^2\right)}{2\Sigma^4\left(x\right)}+\frac{ax^5\left(7a^2-x^2\right)}{8\Sigma^8\left(x\right)}+\frac{a^5x\left(a^2-7x^2\right)}{8\Sigma^8\left(x\right)}+\frac{8ax}{\Sigma^2\left(x\right)}\right]\arctan\left(\frac{x}{a}\right)}{16a^9} + \left(\frac{7175}{73728a^9}\right)  \nonumber \\
&+&\frac{\left[\frac{x^4\left(70a^4+x^4\right)}{512\Sigma^8\left(x\right)}-\frac{7a^2x^6}{128 \Sigma^8\left(x\right)}+\frac{7a^2\left(a^2-6x^2\right)}{32\Sigma^4\left(x\right)}+\frac{2\left(a^2-x^2\right)}{\Sigma^2\left(x\right)}+\frac{x^2-a^2}{3\Sigma^2\left(x\right)} +\frac{\left(a^2-x^2\right)^3}{9\Sigma^6\left(x\right)}+\frac{a^6\left(a^2-28x^2\right)}{512\Sigma^8\left(x\right)}+\frac{7x^4}{32\Sigma^4\left(x\right)}\right]}{16a^9},\\\label{apen18}
I_7&=&\frac{\left[\frac{35}{128}-\frac{a^8}{\Sigma^8\left(x\right)}\right] \arctan\left(\frac{x}{a}\right)}{8a^8}+\frac{x\left[\frac{a^6}{\Sigma^6\left(x\right)}+\frac{7a^4}{6\Sigma^4\left(x\right)}+\frac{35a^2}{24\Sigma^2\left(x\right)}+\frac{35}{16}\right]}{64a^7\Sigma^2\left(x\right)}, \\\label{apen19}
I_8&=& \frac{5\arctan^2\left(\frac{x}{a}\right)}{256a^7} + \left(\frac{665}{73728a^7}\right) + \frac{\left[-\frac{4ax\left(a^4+x^4\right)}{\Sigma^6\left(x\right)}+\frac{40a^3x^3}{3\Sigma^6\left(x\right)}+\frac{4ax\left(x^2-a^2\right)}{\Sigma^4\left(x\right)}+\frac{ax^5\left(x^2-7a^2\right)}{\Sigma^8\left(x\right)}+\frac{a^5x\left(7x^2-a^2\right)}{\Sigma^8\left(x\right)}+\frac{4ax}{\Sigma^2\left(x\right)}\right]\arctan\left(\frac{x}{a}\right)}{128a^7}  \nonumber  \\
&+&\frac{\left[-\frac{35a^4x^4}{32\Sigma^8\left(x\right)}+\frac{a^2-x^2}{\Sigma^2\left(x\right)}+\frac{x^6\left(28a^2-x^2\right)}{64 \Sigma^8\left(x\right)}+\frac{x^4\left(x^2-15a^2\right)}{9\Sigma^6\left(x\right)}+\frac{a^6\left(28x^2-a^2\right)}{64\Sigma^8\left(x\right)}+\frac{a^4\left(15x^2-a^2\right)}{9\Sigma^6\left(x\right)}+\frac{6a^2x^2-x^4-a^4}{4\Sigma^4\left(x\right)}\right]}{128a^7}, \\\label{apen110}
I_9&=& \frac{\left[\frac{35a^4x^4}{4\Sigma^8\left(x\right)}+\frac{a^2\left(6x^2-a^2\right)}{2 \Sigma^4\left(x\right)}+\frac{3\left(x^2-a^2\right)}{\Sigma^2\left(x\right)}+\frac{x^6 \left(x^2-28a^2\right)}{8\Sigma^8\left(x\right)}+\frac{x^4\left(15a^2-x^2\right)}{3\Sigma^6\left(x\right)}+\frac{a^6\left(a^2-28x^2\right)}{8\Sigma^8\left(x\right)}+\frac{a^4\left(a^2-15x^2\right)}{3\Sigma^6\left(x\right)}-\frac{x^4}{2\Sigma^4\left(x\right)}\right]\arctan\left(\frac{x}{a}\right)}{128a^6}  \nonumber \\
&+&\frac{\left[-\frac{2a^5x}{3\Sigma^6\left(x\right)}+\frac{ax\left(a^2-x^2\right)}{\Sigma^4\left(x\right)}+\frac{ax^5\left(x^2-7a^2\right)}{4\Sigma^8\left(x\right)}+\frac{2ax^3\left(10a^2-3x^2\right)}{9\Sigma^6\left(x\right)}+\frac{a^5x\left(7x^2-a^2\right)}{4\Sigma^8\left(x\right)}+\frac{6ax}{\Sigma^2\left(x\right)}\right]}{256a^6}, \\\label{apen111}
I_{10}&=& \frac{3\arctan^2\left(\frac{x}{a}\right)}{256a^5}+\frac{\left[\frac{4ax\left(x^2-a^2\right)}{\Sigma^4\left(x\right)}+\frac{ax^5\left(7a^2-x^2\right)}{\Sigma^8\left(x\right)}+\frac{a^5x\left(a^2-7x^2\right)}{\Sigma^8\left(x\right)}\right]\arctan\left(\frac{x}{a}\right)}{128a^5} + \left(\frac{15}{8192a^5}\right)  \nonumber \\
&+&\frac{\left[\frac{35a^4x^4}{8\Sigma^8\left(x\right)}+\frac{a^2\left(6x^2-a^2\right)}{\Sigma^4\left(x\right)}+\frac{x^6\left(x^2-28a^2\right)}{16\Sigma^8\left(x\right)}+\frac{a^6\left(a^2-28x^2\right)}{16\Sigma^8\left(x\right)}-\frac{x^4}{\Sigma^4\left(x\right)}\right]}{512a^5}, \\\label{apen112}
I_{11}&=& \frac{35\arctan^3\left(\frac{x}{a}\right)}{384a^9} + \frac{\left[\frac{28ax\left(a^2-x^2\right)}{\Sigma^4\left(x\right)}+\frac{ax^5\left(7a^2-x^2\right)}{\Sigma^8\left(x\right)}+\frac{a^5x\left(a^2-7x^2\right)}{\Sigma^8\left(x\right)}+\frac{8a^3x\left(3a^2-10x^2\right)}{3\Sigma^6\left(x\right)}+\frac{8ax^5}{\Sigma^6\left(x\right)}+\frac{56ax}{\Sigma^2\left(x\right)}\right] \arctan^2\left(\frac{x}{a}\right)}{128a^9}   \nonumber \\ 
&+& \frac{\left[\frac{35a^4x^4}{16\Sigma^8\left(x\right)}+\frac{7a^2\left(a^2-6x^2\right)}{2\Sigma^4\left(x\right)}+\frac{28\left(a^2-x^2\right)}{\Sigma^2\left(x\right)}+\frac{x^6\left(x^2-28a^2\right)}{32\Sigma^8\left(x\right)}+\frac{4x^4\left(15a^2-x^2\right)}{9\Sigma^6\left(x\right)}+\frac{a^6\left(a^2-28x^2\right)}{32\Sigma^8\left(x\right)}+\frac{4a^4\left(a^2-15x^2\right)}{9\Sigma^6\left(x\right)}+\frac{7x^4}{2\Sigma^4\left(x\right)}\right]\arctan\left(\frac{x}{a}\right)}{128a^9} \nonumber \\
&-&\frac{\left[\frac{7ax\left(a^2-x^2\right)}{2\Sigma^4\left(x\right)}+\frac{ax^5\left(7a^2-x^2\right)}{32\Sigma^8\left(x\right)}+\frac{a^5x\left(a^2-7x^2\right)}{32\Sigma^8\left(x\right)}+\frac{4a^3x\left(3a^2-10x^2\right)}{27\Sigma^6\left(x\right)}+\frac{4ax^5}{9\Sigma^6\left(x\right)}+\frac{28ax}{\Sigma^2\left(x\right)}\right]}{128a^9}, \\\label{apen113}
I_{12}&=& \frac{5\arctan^3\left(\frac{x}{a}\right)}{384a^7} + \frac{\left[\frac{4ax\left(x^2-a^2\right)}{\Sigma^4\left(x\right)}+\frac{ax^5\left(x^2-7a^2\right)}{\Sigma^8\left(x\right)}+\frac{a^5x\left(7x^2-a^2\right)}{\Sigma^8\left(x\right)}+\frac{4a^3x\left(10x^2-3a^2\right)}{3\Sigma^6\left(x\right)}-\frac{4ax^5}{\Sigma^6\left(x\right)}+\frac{4ax}{\Sigma^2\left(x\right)}\right]\arctan^2\left(\frac{x}{a}\right)}{128a^7}  \nonumber\\
&+& \frac{\left[-\frac{35a^4x^4}{16\Sigma^8\left(x\right)}+\frac{a^2\left(6x^2-a^2\right)}{2\Sigma^4\left(x\right)}+\frac{2\left(a^2-x^2\right)}{\Sigma^2\left(x\right)}+\frac{x^6\left(28a^2-x^2\right)}{32\Sigma^8\left(x\right)}+\frac{2x^4\left(x^2-15a^2\right)}{9\Sigma^6\left(x\right)}+\frac{a^6\left(28x^2-a^2\right)}{32\Sigma^8\left(x\right)}+\frac{2a^4\left(15x^2-a^2\right)}{9\Sigma^6\left(x\right)}-\frac{x^4}{2\Sigma^4\left(x\right)}\right]\arctan\left(\frac{x}{a}\right)}{128a^7}  \nonumber\\
&+& \frac{\left[\frac{a x \left(a^2-x^2\right)}{2\Sigma^4\left(x\right)}+\frac{ax^5\left(7a^2-x^2\right)}{32\Sigma^8\left(x\right)}+\frac{a^5x\left(a^2-7x^2\right)}{32\Sigma^8\left(x\right)}+\frac{2a^3x\left(3a^2-10x^2\right)}{27\Sigma^6\left(x\right)}+\frac{2ax^5}{9\Sigma^6\left(x\right)}-\frac{2ax}{\Sigma^2\left(x\right)}\right]}{128a^7}, \\\label{apen114}
I_{13}&=& \frac{\arctan^3\left(\frac{x}{a}\right)}{128a^5} + \frac{\left[\frac{4ax\left(x^2-a^2\right)}{\Sigma^4\left(x\right)}+\frac{ax^5\left(7a^2-x^2\right)}{\Sigma^8\left(x\right)}+\frac{a^5x\left(a^2-7x^2\right)}{\Sigma^8\left(x\right)}\right]\arctan^2\left(\frac{x}{a}\right)}{128a^5} + \frac{\left[\frac{4ax\left(a^2-x^2\right)}{\Sigma^4\left(x\right)}+\frac{ax^5\left(x^2-7a^2\right)}{4\Sigma^8\left(x\right)}+\frac{a^5x\left(7x^2-a^2\right)}{4\Sigma^8\left(x\right)}\right]}{1024a^5}  \nonumber \\
&+& \frac{\left[\frac{35a^4x^4}{8\Sigma^8\left(x\right)}+\frac{a^2\left(6x^2-a^2\right)}{\Sigma^4\left(x\right)}+\frac{x^6\left(x^2-28a^2\right)}{16\Sigma^8\left(x\right)}+\frac{a^6\left(a^2-28x^2\right)}{16\Sigma^8\left(x\right)}-\frac{x^4}{\Sigma^4\left(x\right)}\right]\arctan\left(\frac{x}{a}\right)}{256a^5}.
\end{eqnarray}

\subsection{case $\bar{n}=3$}\label{apen2}

In a similar way to what was presented above, we can express the scalar field defined by region, now for the configuration $\bar{n}=3$ which is equivalent to taking $n=\frac{1}{7}$. Therefore, we have to

\begin{eqnarray}\label{apen21}
\phi_{\pm}\left(x\right)&=&\Omega_3\left(1-\frac{\pi^3{a^6c^3}}{8} +\frac{3\pi^2{a^4c^2}}{4}-\frac{3\pi{a^2c}}{2}\right)II_1 + \Omega_3\left(\frac{3\pi^2{a^5c^3}}{4}-3\pi{a^3c^2} + 3ac\right)II_2 + \nonumber \\
&+& \Omega_3\left(3a^2c^2 -\frac{3\pi{a^4c^3}}{2} -\frac{3\pi^3{a^4c^3}}{8}  + \frac{3\pi^2{a^2c^2}}{2}-\frac{3c\pi}{2}\right)II_3 +   \Omega_3\left(a^3c^3 + \frac{3\pi^2{a^3c^3}}{2} -3\pi{ac^2}\right)II_4 + \nonumber \\
&+&   \Omega_3\left(\frac{3\pi^2{c^2}}{4}-\frac{3\pi{a^2c^3}}{2}-\frac{3\pi^3{a^2c^3}}{8} \right)II_5 +  \Omega_3\left(\frac{3\pi^2{ac^3}}{4} \right)II_6 - \Omega_3\left(\frac{c^3\pi^3}{8}\right)II_7 +  \Omega_3\left(3a^2c - 3\pi{a^4c^2} +\frac{3\pi^2{a^6c^3}}{4}   \right)II_8  \nonumber\\ &+&\Omega_3\left(6a^3c^2-3\pi{a^5c^3}\right)II_9 +  \Omega_3\left(3c + 3a^4c^3 -6\pi{a^2c^2} + \frac{9\pi^2{a^4c^3}}{4} \right)II_{10} +  \Omega_3\left(6ac^2-6\pi{a^3c^3}\right)II_{11} + \nonumber \\
&+&  \Omega_3\left(3a^2c^3 - 3\pi{c^2} + \frac{9\pi^2{a^2c^3}}{4} \right)II_{12} -  \Omega_3\left(3\pi{ac^3}\right)II_{13} +  \Omega_3\left(\frac{3\pi^2{c^3}}{4}\right)II_{14} +  \Omega_3\left(3a^4c^2 -\frac{3\pi{a^6 c^3}}{2} \right)II_{15} + \nonumber \\
&+& \Omega_3\left(3a^5c^3\right)II_{16} + \Omega_3\left(6a^2c^2 -\frac{9\pi{a^4c^3}}{2} \right)II_{17} + \Omega_3\left(6a^3c^3\right)II_{18} + \Omega_3\left(3c^2 -\frac{9\pi{a^2c^3}}{2} \right)II_{19} + \Omega_3\left(3ac^3\right)II_{20}  \nonumber \\
&-& \Omega_3\left(\frac{3\pi{c^3}}{2}\right)II_{21}  + \Omega_3\left(c^3a^6\right)II_{22} + \Omega_3\left(3c^3a^4\right)II_{23} + \Omega_3\left(3c^3a^2\right)II_{24} + \Omega_3\left(c^3\right)II_{25},
\end{eqnarray} where the constants $c=\frac{4m}{\pi{a^3}}$ and $\Omega_3=\left(\frac{14a^2}{F_0}\right)^\frac{7}{2 }$. The functions $II_i$ for $i=1,2,\dots,25$ are defined below

\begin{eqnarray}\label{apen22}
II_1&=& \frac{231\arctan\left(\frac{x}{a}\right)}{1024a^{13}} + \frac{\left[\frac{2a^{11}x}{\Sigma^{12}\left(x\right)}+\frac{11a^9x}{5\Sigma^{10}\left(x\right)}+\frac{99a^7x}{40\Sigma^8\left(x\right)}+\frac{231a^5x}{80\Sigma^6\left(x\right)}+\frac{231a^3x}{64\Sigma^4\left(x\right)}+\frac{693ax}{128\Sigma^2\left(x\right)}\right]}{24a^{13}}, \\\label{apen23}
II_2&=&-\frac{1}{12\Sigma^{12}\left(x\right)}, \\\label{apen24}
II_3&=& \frac{21\arctan\left(\frac{x}{a}\right)}{1024a^{11}} + \frac{\left[-\frac{1008a^5x^5}{5\Sigma^{10}\left(x\right)}-\frac{40ax\left(a^4+x^4\right)}{\Sigma^6\left(x\right)}+\frac{400a^3x^3}{3\Sigma^6\left(x\right)}+\frac{15ax\left(x^2-a^2\right)}{\Sigma^4\left(x\right)}+\frac{ax^9\left(3x^2-55a^2\right)}{3\Sigma^{12}\left(x\right)}+\frac{8ax^7\left(12a^2-x^2\right)}{\Sigma^{10}\left(x\right)}\right]}{2048a^{11}} + \nonumber \\
&+& \frac{\left[+\frac{26ax^5\left(x^2-7a^2\right)}{\Sigma^8\left(x\right)}+\frac{a^9x\left(55x^2-3a^2\right)}{3\Sigma^{12}\left(x\right)}+\frac{8a^7x\left(12x^2-a^2\right)}{\Sigma^{10}\left(x\right)}+\frac{26a^5x\left(7x^2-a^2\right)}{\Sigma^8\left(x\right)}+\frac{66a^5x^5\left(x^2-a^2\right)}{\Sigma^{12}\left(x\right)}+\frac{48ax}{\Sigma^2\left(x\right)}\right]}{2048a^{11}}, \\\label{apen25}
II_4&=& \frac{\left[\frac{a^2}{6\Sigma^2\left(x\right)}-\frac{1}{5}\right]}{2\Sigma^{10}\left(x\right)}, \\\label{apen26}
II_5&=& \frac{7\arctan\left(\frac{x}{a}\right)}{1024a^9} + \frac{\left[-\frac{12ax\left(a^4+x^4\right)}{\Sigma^6\left(x\right)}+\frac{40a^3x^3}{\Sigma^6\left(x\right)}+\frac{17ax \left(x^2-a^2\right)}{\Sigma^4\left(x\right)}+\frac{2ax^5\left(7a^2-x^2\right)}{\Sigma^8\left(x\right)}+\frac{2a^5x \left(a^2-7x^2\right)}{\Sigma^8\left(x\right)}+\frac{8ax}{\Sigma^2\left(x\right)}\right]}{2048a^9} \nonumber \\
&+& \frac{\left[\frac{4ax^5\left(126a^4 + 5x^4\right)}{5\Sigma^{10}\left(x\right)} -\frac{48a^3x^7}{\Sigma^{10}\left(x\right)}+\frac{ax^9\left(55a^2-3x^2\right)}{3\Sigma^{12}\left(x\right)}+\frac{a^9x\left(3a^2-55x^2\right)}{3\Sigma^{12}\left(x\right)}+ \frac{4a^7x\left(a^2-12x^2\right)}{\Sigma^{10}\left(x\right)}+\frac{66a^5x^5 \left(a^2-x^2\right)}{\Sigma^{12}\left(x\right)}\right]}{2048a^9}, \\\label{apen27}
II_6&=& \frac{\left[-\frac{a^4}{6\Sigma^4\left(x\right)}+\frac{2a^2}{5 \Sigma^2\left(x\right)}-\frac{1}{4}\right]}{2\Sigma^8\left(x\right)}, \\\label{apen28}
II_7&=& \frac{5\arctan\left(\frac{x}{a}\right)}{1024a^7} + \frac{\left[\frac{15ax\left(x^2-a^2\right)}{\Sigma^4\left(x\right)}+\frac{ax^9\left(3x^2-55a^2\right)}{3\Sigma^{12}\left(x\right)}+\frac{6ax^5\left(7a^2-x^2\right)}{\Sigma^8\left(x\right)}+\frac{a^9x\left(55x^2-3a^2\right)}{3\Sigma^{12}\left(x\right)}+\frac{6a^5x\left(a^2-7x^2\right)}{\Sigma^8\left(x\right)}+\frac{66a^5 x^5\left(x^2-a^2\right)}{\Sigma^{12}\left(x\right)}\right]}{2048a^7}, \nonumber \\\label{apen29}
II_8&=&  \frac{231\arctan^2\left(\frac{x}{a}\right)}{2048a^{13}} + \left(\frac{413413}{4915200a^{13}}\right) + \nonumber \\ 
&+&\frac{\arctan\left(\frac{x}{a}\right)\left[\frac{495ax\left(a^2-x^2\right)}{\Sigma^4\left(x\right)}+\frac{66ax^5\left(7a^2-x^2\right)}{\Sigma^8\left(x\right)}+\frac{12a^7x\left(a^2-12x^2\right)}{\Sigma^{10}\left(x\right)}+\frac{66a^5x\left(a^2-7x^2\right)}{\Sigma^8\left(x\right)}+\frac{220a^3x\left(3a^2-10x^2\right)}{3\Sigma^6\left(x\right)}+\frac{220ax^5}{\Sigma^6\left(x\right)}+\frac{792ax}{\Sigma^2\left(x\right)}\right]}{2048a^{13}}  \nonumber \\
&+& \frac{\arctan\left(\frac{x}{a}\right)\left[+\frac{1512a^5x^5}{5\Sigma^{10}\left(x\right)}+\frac{ax^9\left(55a^2-3x^2\right)}{3\Sigma^{12}\left(x\right)}+\frac{12ax^7\left(x^2-12a^2\right)}{\Sigma^{10}\left(x\right)}+\frac{a^9x\left(3a^2-55x^2\right)}{3\Sigma^{12}\left(x\right)}+\frac{66a^5x^5\left(a^2-x^2\right)}{\Sigma^{12}\left(x\right)}\right]}{2048a^{13}} + \nonumber \\
&+& \frac{\left[\frac{198\left(a^2-x^2\right)}{\Sigma^2\left(x\right)}+\frac{55x^4\left(15a^2-x^2\right)}{9\Sigma^6\left(x\right)}+\frac{33a^6\left(a^2-28x^2\right)}{32\Sigma^8\left(x\right)}+\frac{55a^4\left(a^2-15x^2\right)}{9\Sigma^6\left(x\right)}+\frac{495\left(a^4-6a^2x^2 + x^4\right)}{16\Sigma^4\left(x\right)}+\frac{33\left(70a^4x^4 - 28a^2x^6 + x^8\right)}{32\Sigma^8\left(x\right)}\right]}{2048a^{13}} + \nonumber \\
&+& \frac{\left[\frac{3x^8\left(45a^2-x^2\right)}{25\Sigma^{10}\left(x\right)}+\frac{a^{10} \left(a^2-66x^2\right)}{144\Sigma^{12}\left(x\right)}+\frac{3a^8\left(a^2-45x^2\right)}{25\Sigma^{10}\left(x\right)}+\frac{11a^6x^4\left(15a^2-28x^2\right)}{48\Sigma^{12}\left(x\right)}+\frac{126a^4x^4\left(a^2-x^2\right)}{5\Sigma^{10}\left(x\right)}+\frac{x^8 \left(495a^4-66a^2x^2 + x^4\right)}{144\Sigma^{12}\left(x\right)}\right]}{2048a^{13}},
\\\label{apen210}
II_9&=&  \frac{\arctan\left(\frac{x}{a}\right)\left[\frac{66\left(x^2-a^2\right)}{\Sigma^2\left(x\right)}+\frac{11x^6\left(28a^2-x^2\right)}{2\Sigma^8\left(x\right)}+\frac{55x^4\left(x^2-15a^2\right)}{3\Sigma^6\left(x\right)}+\frac{55a^4\left(15x^2-a^2\right)}{3\Sigma^6\left(x\right)}+\frac{165\left(6a^2x^2-a^4-x^4\right)}{4\Sigma^4\left(x\right)}\right]}{2048a^{12}}  \nonumber \\
&+&   \frac{\arctan\left(\frac{x}{a}\right)\left[\frac{x^{10} \left(66a^2-x^2\right)}{12\Sigma^{12}\left(x\right)}+\frac{11a^4x^6 \left(28a^2-15x^2\right)}{4\Sigma^{12}\left(x\right)}+\frac{x^6\left(210a^4 - 45a^2x^2 + x^4\right)}{\Sigma^{10}\left(x\right)}+\frac{a^8\left(66a^2x^2-a^4-495x^4\right)}{12\Sigma^{12}\left(x\right)}\right]}{2048a^{12}}   \nonumber \\
&+& \frac{\arctan\left(\frac{x}{a}\right)\left[\frac{11a^4\left(28a^2x^2- a^4 -70x^4\right)}{2\Sigma^8\left(x\right)} +\frac{a^6\left(45a^2x^2 - a^4 -210x^4\right)}{\Sigma^{10}\left(x\right)}\right]}{2048a^{12}} \nonumber \\
&+&  \frac{\left[\frac{165ax\left(a^2-x^2\right)}{4\Sigma^4\left(x\right)}+\frac{ax^7\left(x^2-12a^2\right)}{\Sigma^{10}\left(x\right)}+\frac{11ax^5\left(7a^2-x^2\right)}{2\Sigma^8\left(x\right)}+\frac{11a^5x\left(a^2-7x^2\right)}{2\Sigma^8\left(x\right)}+\frac{55a\left(3a^4x-10a^2x^3 +3x^5\right)}{9\Sigma^6\left(x\right)}+\frac{66ax}{\Sigma^2\left(x\right)}\right]}{2048a^{12}}  \nonumber \\
&+& \frac{\left[\frac{ax^9\left(55a^2-3x^2\right)}{36\Sigma^{12}\left(x\right)}+\frac{a^9x\left(3a^2-55x^2\right)}{36\Sigma^{12}\left(x\right)}+\frac{11a^5x^5\left(a^2-x^2\right)}{2\Sigma^{12}\left(x\right)}+\frac{a^5\left(5a^4x-60a^2x^3 + 126x^5\right)}{5\Sigma^{10}\left(x\right)}\right]}{2048a^{12}}, \\\label{apen211}
II_{10}&=&  \frac{21\arctan^2\left(\frac{x}{a}\right)}{2048a^{11}}+ \frac{\arctan\left(\frac{x}{a}\right)\left[\frac{15ax\left(x^2-a^2\right)}{\Sigma^4\left(x\right)}+\frac{26ax^5\left(x^2-7a^2\right)}{\Sigma^8\left(x\right)}+\frac{26a^5x\left(7x^2-a^2\right)}{\Sigma^8\left(x\right)}+\frac{40ax\left(10a^2x^2-3a^4-3x^4\right)}{3\Sigma^6\left(x\right)}+\frac{48ax}{\Sigma^2\left(x\right)}\right]}{2048a^{11}} \nonumber \\
&+&  \frac{\arctan\left(\frac{x}{a}\right)\left[\frac{ax^9\left(3x^2-55a^2\right)}{3\Sigma^{12}\left(x\right)}+\frac{a^9x\left(55x^2-3a^2\right)}{3\Sigma^{12}\left(x\right)}+\frac{8a^7x\left(12x^2-a^2\right)}{\Sigma^{10}\left(x\right)}+\frac{66 a^5x^5\left(x^2-a^2\right)}{\Sigma^{12}\left(x\right)}+\frac{8ax^5\left(60a^2x^2-126a^4-5x^4\right)}{5\Sigma^{10}\left(x\right)}\right]}{2048a^{11}} \nonumber \\
&+&  \frac{\left[\frac{12\left(a^2-x^2\right)}{\Sigma^2\left(x\right)}+\frac{13x^6\left(28a^2-x^2\right)}{32\Sigma^8\left(x\right)} +\frac{10x^4\left(x^2-15a^2\right)}{9\Sigma^6\left(x\right)}+\frac{13a^6\left(28x^2-a^2\right)}{32\Sigma^8\left(x\right)}+\frac{10a^4\left(15x^2-a^2\right)}{9\Sigma^6\left(x\right)}+\frac{15\left(6a^2x^2-a^4-x^4\right)}{16\Sigma^4\left(x\right)}-\frac{455a^4x^4}{16\Sigma^8\left(x\right)}\right]}{2048a^{11}} \nonumber \\
&+& \frac{\left[+\frac{2x^8\left(x^2-45a^2\right)}{25\Sigma^{10}\left(x\right)}+\frac{a^{10}\left(66x^2-a^2\right)}{144\Sigma^{12}\left(x\right)}+\frac{2a^8\left(45x^2-a^2\right)}{25\Sigma^{10}\left(x\right)}+\frac{11a^6x^4\left(28x^2-15a^2\right)}{48\Sigma^{12}\left(x\right)}+\frac{84a^4x^4\left(x^2-a^2\right)}{5\Sigma^{10}\left(x\right)}+\frac{x^8\left(66a^2x^2-x^4-495a^4\right)}{144\Sigma^{12}\left(x\right)}\right]}{2048a^{11}} \nonumber \\
&+& \left(\frac{29183}{4915200a^{11}}\right),  \\\label{apen212}
II_{11}&=& \frac{\arctan\left(\frac{x}{a}\right)\left[\frac{18 \left(x^2-a^2\right)}{\Sigma^2\left(x\right)}+\frac{3 x^6 \left(x^2-28 a^2\right)}{2 \Sigma^8\left(x\right)}+\frac{x^4 \left(15 a^2-x^2\right)}{3 \Sigma^6\left(x\right)}+\frac{a^4 \left(a^2-15 x^2\right)}{3 \Sigma^6\left(x\right)}+\frac{27 \left(-a^4+6 a^2 x^2-x^4\right)}{4 \Sigma^4\left(x\right)}\right]}{2048a^{10}} \nonumber \\
&+& \frac{\arctan\left(\frac{x}{a}\right)\left[+\frac{3 a^4 \left(a^4-28 a^2 x^2+70x^4\right)}{2\Sigma^8\left(x\right)}+\frac{3x^6\left(-210a^4 + 45a^2x^2- x^4\right)}{5\Sigma^{10}\left(x\right)} + \frac{3a^6\left(a^4 - 45a^2x^2 + 210x^4\right)}{5\Sigma^{10}\left(x\right)}\right]}{2048a^{10}} \nonumber\\
&+&\frac{\arctan\left(\frac{x}{a}\right)\left[-\frac{77a^6x^6}{\Sigma^{12}\left(x\right)} +\frac{x^8\left(495a^4 - 66a^2x^2 + x^4\right)}{12\Sigma^{12}\left(x\right)}+\frac{a^8\left(a^4-66a^2x^2 + 495x^4\right)}{12\Sigma^{12}\left(x\right)}\right]}{2048a^{10}} \nonumber \\
&+& \frac{\left[\frac{27 a x \left(a^2-x^2\right)}{4 \Sigma^4\left(x\right)}+\frac{3 a x^7 \left(12 a^2-x^2\right)}{5 \Sigma^{10}\left(x\right)}+\frac{3 a x^5 \left(x^2-7 a^2\right)}{2 \Sigma^8\left(x\right)}+\frac{3 a^5 x \left(7 x^2-a^2\right)}{2 \Sigma^8\left(x\right)}+\frac{a x \left(-3 a^4+10 a^2 x^2-3 x^4\right)}{9 \Sigma^6\left(x\right)}+\frac{18 a x}{\Sigma^2\left(x\right)}\right]}{2048a^{10}} \nonumber \\
&+& \frac{\left[\frac{a x^7 \left(198 a^4-55 a^2 x^2+3 x^4\right)}{36 \Sigma^{12}\left(x\right)}+\frac{a^7 x \left(-3 a^4+55 a^2 x^2-198 x^4\right)}{36 \Sigma^{12}\left(x\right)}++\frac{3 a^5 x \left(-5 a^4+60 a^2 x^2-126 x^4\right)}{25 \Sigma^{10}\left(x\right)}\right]}{2048a^{10}}, \\\label{apen213}
II_{12}&=&  \frac{7\arctan^2\left(\frac{x}{a}\right)}{2048 a^9} +  \frac{\arctan\left(\frac{x}{a}\right)\left[\frac{17 a x \left(x^2-a^2\right)}{\Sigma^4\left(x\right)}+\frac{2 a x^5 \left(7 a^2-x^2\right)}{\Sigma^8\left(x\right)}+\frac{2 a^5 x \left(a^2-7 x^2\right)}{\Sigma^8\left(x\right)}+\frac{4 a x \left(-3 a^4+10 a^2 x^2-3 x^4\right)}{\Sigma^6\left(x\right)}+\frac{8 a x}{\Sigma^2\left(x\right)}\right]}{2048 a^9} \nonumber\\
&+& \frac{\arctan\left(\frac{x}{a}\right)\left[\frac{4 a x^7 \left(x^2-12 a^2\right)}{\Sigma^{10}\left(x\right)}+\frac{a x^7 \left(-198 a^4+55 a^2 x^2-3 x^4\right)}{3 \Sigma^{12}\left(x\right)}+\frac{a^7 x \left(3 a^4-55 a^2 x^2+198 x^4\right)}{3 \Sigma^{12}\left(x\right)} +\frac{4 a^5 x \left(5 a^4-60 a^2 x^2+126 x^4\right)}{5 \Sigma^{10}\left(x\right)}\right]}{2048 a^9} \nonumber \\
&+& \frac{\left[\frac{2 \left(a^2-x^2\right)}{\Sigma^2\left(x\right)}+\frac{x^6 \left(x^2-28 a^2\right)}{32 \Sigma^8\left(x\right)}+\frac{x^4 \left(x^2-15 a^2\right)}{3 \Sigma^6\left(x\right)}+\frac{a^4 \left(15 x^2-a^2\right)}{3 \Sigma^6\left(x\right)}+\frac{17 \left(-a^4+6 a^2 x^2-x^4\right)}{16 \Sigma^4\left(x\right)}+\frac{a^4 \left(a^4-28 a^2 x^2+70 x^4\right)}{32 \Sigma^8\left(x\right)}\right]}{2048 a^9} + \left(\frac{19663}{14745600 a^9}\right)\nonumber \\
&+& \frac{\left[-\frac{77 a^6 x^6}{12 \Sigma^{12}\left(x\right)}+\frac{x^8 \left(495 a^4-66 a^2 x^2+x^4\right)}{144 \Sigma^{12}\left(x\right)} + \frac{x^6 \left(-210 a^4+45 a^2 x^2-x^4\right)}{25 \Sigma^{10}\left(x\right)} + \frac{a^8 \left(a^4-66 a^2 x^2+495 x^4\right)}{144 \Sigma^{12}\left(x\right)} +\frac{a^6 \left(a^4-45 a^2 x^2+210 x^4\right)}{25 \Sigma^{10}\left(x\right)}\right]}{2048 a^9}, \\\label{apen214}
II_{13}&=& \frac{\arctan\left(\frac{x}{a}\right)\left[\frac{10 \left(x^2-a^2\right)}{\Sigma^2\left(x\right)}+\frac{x^6 \left(x^2-28 a^2\right)}{2 \Sigma^8\left(x\right)}+\frac{5 x^4 \left(15 a^2-x^2\right)}{3 \Sigma^6\left(x\right)}+\frac{5 a^4 \left(a^2-15 x^2\right)}{3 \Sigma^6\left(x\right)}+\frac{5 \left(-a^4+6 a^2 x^2-x^4\right)}{4 \Sigma^4\left(x\right)}+\frac{a^4 \left(a^4-28 a^2 x^2+70 x^4\right)}{2 \Sigma^8\left(x\right)}\right]}{2048 a^8} \nonumber \\
&+&  \frac{\arctan\left(\frac{x}{a}\right)\left[\frac{77 a^6 x^6}{\Sigma^{12}\left(x\right)}+\frac{x^8 \left(-495 a^4+66 a^2 x^2-x^4\right)}{12 \Sigma^{12}\left(x\right)}+\frac{x^6 \left(210 a^4-45 a^2 x^2+x^4\right)}{5 \Sigma^{10}\left(x\right)}+\frac{a^8 \left(-a^4+66 a^2 x^2-495 x^4\right)}{12 \Sigma^{12}\left(x\right)}+\frac{a^6 \left(-a^4+45 a^2 x^2-210 x^4\right)}{5 \Sigma^{10}\left(x\right)}\right]}{2048 a^8} \nonumber \\
&+&  \frac{\left[\frac{5 a x \left(a^2-x^2\right)}{4 \Sigma^4\left(x\right)}+\frac{a x^5 \left(x^2-7 a^2\right)}{2 \Sigma^8\left(x\right)}+\frac{a^5 x \left(7 x^2-a^2\right)}{2 \Sigma^8\left(x\right)}+\frac{5 a x \left(-3 a^4+10 a^2 x^2-3 x^4\right)}{9 \Sigma^6\left(x\right)}+\frac{a^5 x \left(5 a^4-60 a^2 x^2+126 x^4\right)}{25 \Sigma^{10}\left(x\right)}+\frac{10 a x}{\Sigma^2\left(x\right)}\right]}{2048 a^8} \nonumber \\
&+&  \frac{\left[+\frac{a x^7 \left(x^2-12 a^2\right)}{5 \Sigma^{10}\left(x\right)}+\frac{a x^7 \left(-198 a^4+55 a^2 x^2-3 x^4\right)}{36 \Sigma^{12}\left(x\right)}+\frac{a^7 x \left(3 a^4-55 a^2 x^2+198 x^4\right)}{36 \Sigma^{12}\left(x\right)}\right]}{2048 a^8},\\\label{apen215}
II_{14}&=& \frac{5\arctan^2\left(\frac{x}{a}\right)}{2048 a^7} + \left(\frac{245}{589824 a^7}\right) \nonumber \\
&+& \frac{\arctan\left(\frac{x}{a}\right)\left[\frac{15 a x \left(x^2-a^2\right)}{\Sigma (x)^4}+\frac{6 a x^5 \left(7 a^2-x^2\right)}{\Sigma (x)^8}+\frac{6 a^5 x \left(a^2-7 x^2\right)}{\Sigma (x)^8}+\frac{a x^7 \left(198 a^4-55 a^2 x^2+3 x^4\right)}{3 \Sigma (x)^{12}}+\frac{a^7 x \left(-3 a^4+55 a^2 x^2-198 x^4\right)}{3 \Sigma (x)^{12}}\right]}{2048 a^7} \nonumber \\
&+& \frac{\left[\frac{77 a^6 x^6}{12 \Sigma (x)^{12}}+\frac{x^6 \left(3 x^2-84 a^2\right)}{32 \Sigma (x)^8}+\frac{3 a^4 \left(a^4-28 a^2 x^2+70 x^4\right)}{32 \Sigma (x)^8}+\frac{15 \left(-a^4+6 a^2 x^2-x^4\right)}{16 \Sigma (x)^4}\right]}{2048 a^7} \nonumber \\ 
&+& \frac{\left[+\frac{x^8 \left(-495 a^4+66 a^2 x^2-x^4\right)}{144 \Sigma (x)^{12}}+\frac{a^8 \left(-a^4+66 a^2 x^2-495 x^4\right)}{144 \Sigma (x)^{12}}\right]}{2048 a^7},\\\label{apen216}
II_{15}&=& \frac{77\arctan^3\left(\frac{x}{a}\right)}{1024 a^{13}} + \frac{\arctan^2\left(\frac{x}{a}\right)\left[\frac{495 a x \left(a^2-x^2\right)}{\Sigma^4\left(x\right)}+\frac{66 a x^5 \left(7 a^2-x^2\right)}{\Sigma^8\left(x\right)}+\frac{66 a^5 x \left(a^2-7 x^2\right)}{\Sigma^8\left(x\right)}+\frac{220 a x \left(3 a^4-10 a^2 x^2+3 x^4\right)}{3 \Sigma^6\left(x\right)}+\frac{792 a x}{\Sigma^2\left(x\right)}\right]}{2048 a^{13}} \nonumber \\
&+& \frac{\arctan^2\left(\frac{x}{a}\right)\left[\frac{12 a x^7 \left(x^2-12 a^2\right)}{\Sigma^{10}\left(x\right)}+\frac{a x^7 \left(-198 a^4+55 a^2 x^2-3 x^4\right)}{3 \Sigma^{12}\left(x\right)}+\frac{a^7 x \left(3 a^4-55 a^2 x^2+198 x^4\right)}{3 \Sigma^{12}\left(x\right)}+\frac{12 a^5 x \left(5 a^4-60 a^2 x^2+126 x^4\right)}{5 \Sigma^{10}\left(x\right)}\right]}{2048 a^{13}} \nonumber \\
&+& \frac{\arctan\left(\frac{x}{a}\right)\left[\frac{396 \left(a^2-x^2\right)}{\Sigma^2\left(x\right)}+\frac{110 x^4 \left(15 a^2-x^2\right)}{9 \Sigma^6\left(x\right)}+\frac{110 a^4 \left(a^2-15 x^2\right)}{9 \Sigma^6\left(x\right)}+\frac{33 a^4 \left(a^4-28 a^2 x^2+70 x^4\right)}{16 \Sigma^8\left(x\right)}+\frac{495 \left(a^4-6 a^2 x^2+x^4\right)}{8 \Sigma^4\left(x\right)}\right]}{2048 a^{13}} \nonumber \\
&+& \frac{\arctan\left(\frac{x}{a}\right)\left[+\frac{33 x^6 \left(x^2-28 a^2\right)}{16 \Sigma^8\left(x\right)}+\frac{6 x^6 \left(-210 a^4+45 a^2 x^2-x^4\right)}{25 \Sigma^{10}\left(x\right)}+\frac{6 a^6 \left(a^4-45 a^2 x^2+210 x^4\right)}{25 \Sigma^{10}\left(x\right)}\right]}{2048 a^{13}} \nonumber \\
&+& \frac{\arctan\left(\frac{x}{a}\right)\left[-\frac{77 a^6 x^6}{6 \Sigma^{12}\left(x\right)}+\frac{x^8 \left(495 a^4-66 a^2 x^2+x^4\right)}{72 \Sigma^{12}\left(x\right)}+\frac{a^8 \left(a^4-66 a^2 x^2+495 x^4\right)}{72 \Sigma^{12}\left(x\right)}\right]}{2048 a^{13}}  \nonumber \\
&+& \frac{\left[\frac{495 a x \left(x^2-a^2\right)}{8 \Sigma^4\left(x\right)}+\frac{33 a x^5 \left(x^2-7 a^2\right)}{16 \Sigma^8\left(x\right)}+\frac{33 a^5 x \left(7 x^2-a^2\right)}{16 \Sigma^8\left(x\right)}+\frac{110 a x \left(-3 a^4+10 a^2 x^2-3 x^4\right)}{27 \Sigma^6\left(x\right)}-\frac{396 a x}{\Sigma^2\left(x\right)}\right]}{2048 a^{13}} \nonumber \\
&+&  \frac{\left[\frac{6 a x^7 \left(12 a^2-x^2\right)}{25 \Sigma^{10}\left(x\right)}+\frac{a x^7 \left(198 a^4-55 a^2 x^2+3 x^4\right)}{216 \Sigma^{12}\left(x\right)}+\frac{a^7 x \left(-3 a^4+55 a^2 x^2-198 x^4\right)}{216 \Sigma^{12}\left(x\right)} +\frac{6 a^5 x \left(-5 a^4+60 a^2 x^2-126 x^4\right)}{125 \Sigma^{10}\left(x\right)}\right]}{2048 a^{13}}, 
\end{eqnarray}

\begin{eqnarray} \label{apen217}
II_{16}&=& \frac{\arctan^2\left(\frac{x}{a}\right)\left[\frac{66 \left(x^2-a^2\right)}{\Sigma^2\left(x\right)}+\frac{11 x^6 \left(28 a^2-x^2\right)}{2 \Sigma^8\left(x\right)}+\frac{55 x^4 \left(x^2-15 a^2\right)}{3 \Sigma^6\left(x\right)}+\frac{55 a^4 \left(15 x^2-a^2\right)}{3 \Sigma^6\left(x\right)}+\frac{165 \left(-a^4+6 a^2 x^2-x^4\right)}{4 \Sigma^4\left(x\right)}\right]}{2048 a^{12}} \nonumber \\ &+& \frac{\arctan^2\left(\frac{x}{a}\right)\left[+\frac{11 a^4 \left(-a^4+28 a^2 x^2-70 x^4\right)}{2 \Sigma^8\left(x\right)}+\frac{x^6 \left(210 a^4-45 a^2 x^2+x^4\right)}{\Sigma^{10}\left(x\right)}+\frac{a^6 \left(-a^4+45 a^2 x^2-210 x^4\right)}{\Sigma^{10}\left(x\right)}\right]}{2048 a^{12}} \nonumber \\
&+& \frac{\arctan^2\left(\frac{x}{a}\right)\left[\frac{77 a^6 x^6}{\Sigma^{12}\left(x\right)}+\frac{x^8 \left(-495 a^4+66 a^2 x^2-x^4\right)}{12 \Sigma^{12}\left(x\right)}+ \frac{a^8 \left(-a^4+66 a^2 x^2-495 x^4\right)}{12 \Sigma^{12}\left(x\right)}\right]}{2048 a^{12}} \nonumber \\
&+& \frac{\arctan\left(\frac{x}{a}\right)\left[\frac{165 a x \left(a^2-x^2\right)}{2 \Sigma^4\left(x\right)}+\frac{11 a x^5 \left(7 a^2-x^2\right)}{\Sigma^8\left(x\right)}+\frac{11 a^5 x \left(a^2-7 x^2\right)}{\Sigma^8\left(x\right)}+\frac{110 a x \left(3 a^4-10 a^2 x^2+3 x^4\right)}{9 \Sigma^6\left(x\right)}+\frac{132 a x}{\Sigma^2\left(x\right)}\right]}{2048 a^{12}} \nonumber \\
&+& \frac{\arctan\left(\frac{x}{a}\right)\left[\frac{2 a x^7 \left(x^2-12 a^2\right)}{\Sigma^{10}\left(x\right)}+\frac{a x^7 \left(-198 a^4+55 a^2 x^2-3 x^4\right)}{18 \Sigma^{12}\left(x\right)}+\frac{a^7 x \left(3 a^4-55 a^2 x^2+198 x^4\right)}{18 \Sigma^{12}\left(x\right)}+\frac{2 a^5 x \left(5 a^4-60 a^2 x^2+126 x^4\right)}{5 \Sigma^{10}\left(x\right)}\right]}{2048 a^{12}} \nonumber \\
&+& \frac{\left[\frac{33 \left(a^2-x^2\right)}{\Sigma^2\left(x\right)}+\frac{11 x^6 \left(x^2-28 a^2\right)}{64 \Sigma^8\left(x\right)}+\frac{55 x^4 \left(15 a^2-x^2\right)}{54 \Sigma^6\left(x\right)}+\frac{55 a^4 \left(a^2-15 x^2\right)}{54 \Sigma^6\left(x\right)}+\frac{165 \left(a^4-6 a^2 x^2+x^4\right)}{32 \Sigma^4\left(x\right)}+\frac{11 a^4 \left(a^4-28 a^2 x^2+70 x^4\right)}{64 \Sigma^8\left(x\right)}\right]}{2048 a^{12}} \nonumber \\
&+& \frac{\left[-\frac{77 a^6 x^6}{72 \Sigma^{12}\left(x\right)}+\frac{x^8 \left(495 a^4-66 a^2 x^2+x^4\right)}{864 \Sigma^{12}\left(x\right)}+\frac{x^6 \left(-210 a^4+45 a^2 x^2-x^4\right)}{50 \Sigma^{10}\left(x\right)}+\frac{a^8 \left(a^4-66 a^2 x^2+495 x^4\right)}{864 \Sigma^{12}\left(x\right)}+\frac{a^6 \left(a^4-45 a^2 x^2+210 x^4\right)}{50 \Sigma^{10}\left(x\right)}\right]}{2048 a^{12}} \nonumber \\
&+& \left(\frac{413413}{29491200 a^{12}}\right), \\\label{apen218}
II_{17}&=& \frac{7\arctan^3\left(\frac{x}{a}\right)}{1024 a^{11}} +\frac{\arctan^2\left(\frac{x}{a}\right)\left[\frac{15 a x \left(x^2-a^2\right)}{\Sigma^4\left(x\right)}+\frac{26 a^5 x \left(7 x^2-a^2\right)}{\Sigma^8\left(x\right)}+\frac{40 a x \left(-3 a^4+10 a^2 x^2-3 x^4\right)}{3 \Sigma^6\left(x\right)}+\frac{48 (a x)}{\Sigma^2\left(x\right)}\right]}{2048 a^{11}} \nonumber \\
&+& \frac{\arctan^2\left(\frac{x}{a}\right)\left[\frac{8 a x^7 \left(12 a^2-x^2\right)}{\Sigma^{10}\left(x\right)}+\frac{26 a x^5 \left(x^2-7 a^2\right)}{\Sigma^8\left(x\right)}+\frac{8 a^5 x \left(-5 a^4+60 a^2 x^2-126 x^4\right)}{5 \Sigma^{10}\left(x\right)}\right]}{2048 a^{11}} \nonumber \\
&+& \frac{\arctan^2\left(\frac{x}{a}\right)\left[\frac{a x^7 \left(198 a^4-55 a^2 x^2+3 x^4\right)}{3 \Sigma^{12}\left(x\right)}+\frac{a^7 x \left(-3 a^4+55 a^2 x^2-198 x^4\right)}{3 \Sigma^{12}\left(x\right)}\right]}{2048 a^{11}} \nonumber \\
&+& \frac{\arctan\left(\frac{x}{a}\right)\left[\frac{24 \left(a^2-x^2\right)}{\Sigma^2\left(x\right)}+\frac{20 x^4 \left(x^2-15 a^2\right)}{9 \Sigma^6\left(x\right)}+\frac{20 a^4 \left(15 x^2-a^2\right)}{9 \Sigma^6\left(x\right)}+\frac{15 \left(-a^4+6 a^2 x^2-x^4\right)}{8 \Sigma^4\left(x\right)}\right]}{2048 a^{11}} \nonumber \\
&+& \frac{\arctan\left(\frac{x}{a}\right)\left[\frac{77 a^6 x^6}{6 \Sigma^{12}\left(x\right)}+\frac{13 x^6 \left(28 a^2-x^2\right)}{16 \Sigma^8\left(x\right)}+\frac{13 a^4 \left(-a^4+28 a^2 x^2-70 x^4\right)}{16 \Sigma^8\left(x\right)}+\frac{4 a^6 \left(-a^4+45 a^2 x^2-210 x^4\right)}{25 \Sigma^{10}\left(x\right)}\right]}{2048 a^{11}} \nonumber \\
&+& \frac{\arctan\left(\frac{x}{a}\right)\left[\frac{x^8 \left(-495 a^4+66 a^2 x^2-x^4\right)}{72 \Sigma^{12}\left(x\right)} +\frac{4 x^6 \left(210 a^4-45 a^2 x^2+x^4\right)}{25 \Sigma^{10}\left(x\right)}+\frac{a^8 \left(-a^4+66 a^2 x^2-495 x^4\right)}{72 \Sigma^{12}\left(x\right)}\right]}{2048 a^{11}} \nonumber \\
&+& \frac{\left[\frac{15 a x \left(a^2-x^2\right)}{8 \Sigma^4\left(x\right)}+\frac{13 a x^5 \left(7 a^2-x^2\right)}{16 \Sigma^8\left(x\right)}+\frac{13 a^5 x \left(a^2-7 x^2\right)}{16 \Sigma^8\left(x\right)}+\frac{20 a x \left(3 a^4-10 a^2 x^2+3 x^4\right)}{27 \Sigma^6\left(x\right)}-\frac{24 (a x)}{\Sigma^2\left(x\right)}\right]}{2048 a^{11}} \nonumber \\
&+& \frac{\left[+\frac{4 a x^7 \left(x^2-12 a^2\right)}{25 \Sigma^{10}\left(x\right)}+\frac{a x^7 \left(-198 a^4+55 a^2 x^2-3 x^4\right)}{216 \Sigma^{12}\left(x\right)}+\frac{a^7 x \left(3 a^4-55 a^2 x^2+198 x^4\right)}{216 \Sigma^{12}\left(x\right)}+\frac{4 a^5 x \left(5 a^4-60 a^2 x^2+126 x^4\right)}{125 \Sigma^{10}\left(x\right)}\right]}{2048 a^{11}}, \\\label{apen219}
II_{18}&=& \frac{\arctan^2\left(\frac{x}{a}\right)\left[\frac{18 \left(x^2-a^2\right)}{\Sigma^2\left(x\right)}+\frac{x^4 \left(15 a^2-x^2\right)}{3 \Sigma^6\left(x\right)}+\frac{a^4 \left(a^2-15 x^2\right)}{3 \Sigma^6\left(x\right)}+\frac{3 a^4 \left(a^4-28 a^2 x^2+70 x^4\right)}{2 \Sigma^8\left(x\right)}+\frac{27 \left(-a^4+6 a^2 x^2-x^4\right)}{4 \Sigma^4\left(x\right)}\right]}{2048 a^{10}} \nonumber \\ &+& \frac{\arctan^2\left(\frac{x}{a}\right)\left[+\frac{x^6 \left(3 x^2-84 a^2\right)}{2 \Sigma^8\left(x\right)}+\frac{3 x^6 \left(-210 a^4+45 a^2 x^2-x^4\right)}{5 \Sigma^{10}\left(x\right)}+\frac{3 a^6 \left(a^4-45 a^2 x^2+210 x^4\right)}{5 \Sigma^{10}\left(x\right)}\right]}{2048 a^{10}} \nonumber \\
&+& \frac{\arctan^2\left(\frac{x}{a}\right)\left[-\frac{77 a^6 x^6}{\Sigma^{12}\left(x\right)}+\frac{x^8 \left(495 a^4-66 a^2 x^2+x^4\right)}{12 \Sigma^{12}\left(x\right)}+\frac{a^8 \left(a^4-66 a^2 x^2+495 x^4\right)}{12 \Sigma^{12}\left(x\right)}\right]}{2048 a^{10}} \nonumber \\
&+& \frac{\arctan\left(\frac{x}{a}\right)\left[\frac{27 a x \left(a^2-x^2\right)}{2 \Sigma^4\left(x\right)}+\frac{3 a x^5 \left(x^2-7 a^2\right)}{\Sigma^8\left(x\right)}+\frac{3 a^5 x \left(7 x^2-a^2\right)}{\Sigma^8\left(x\right)}+\frac{2 a x \left(-3 a^4+10 a^2 x^2-3 x^4\right)}{9 \Sigma^6\left(x\right)}+\frac{36 a x}{\Sigma^2\left(x\right)}\right]}{2048 a^{10}} \nonumber \\
&+& \frac{\arctan\left(\frac{x}{a}\right)\left[\frac{6 a x^7 \left(12 a^2-x^2\right)}{5 \Sigma^{10}\left(x\right)}+\frac{a x^7 \left(198 a^4-55 a^2 x^2+3 x^4\right)}{18 \Sigma^{12}\left(x\right)}+\frac{a^7 x \left(-3 a^4+55 a^2 x^2-198 x^4\right)}{18 \Sigma^{12}\left(x\right)}+\frac{6 a^5 x \left(-5 a^4+60 a^2 x^2-126 x^4\right)}{25 \Sigma^{10}\left(x\right)}\right]}{2048 a^{10}} \nonumber \\
&+& \frac{\left[\frac{9 \left(a^2-x^2\right)}{\Sigma^2\left(x\right)}+\frac{3 x^6 \left(28 a^2-x^2\right)}{64 \Sigma^8}+\frac{x^4 \left(x^2-15 a^2\right)}{54 \Sigma^6\left(x\right)}+\frac{a^4 \left(15 x^2-a^2\right)}{54 \Sigma^6\left(x\right)}+\frac{27 \left(a^4-6 a^2 x^2+x^4\right)}{32 \Sigma^4\left(x\right)}+\frac{3 a^4 \left(-a^4+28 a^2 x^2-70 x^4\right)}{64 \Sigma^8\left(x\right)}\right]}{2048 a^{10}} \nonumber\\
&+& \frac{\left[\frac{77 a^6 x^6}{72 \Sigma^{12}\left(x\right)}+\frac{x^8 \left(-495 a^4+66 a^2 x^2-x^4\right)}{864 \Sigma^{12}\left(x\right)}+\frac{3 x^6 \left(210 a^4-45 a^2 x^2+x^4\right)}{250 \Sigma^{10}\left(x\right)}+\frac{a^8 \left(-a^4+66 a^2 x^2-495 x^4\right)}{864 \Sigma^{12}\left(x\right)}+\frac{3 a^6 \left(-a^4+45 a^2 x^2-210 x^4\right)}{250 \Sigma^{10}\left(x\right)}\right]}{2048 a^{10}} \nonumber \\
&+& \left(\frac{588511}{147456000 a^{10}}\right), \\\label{apen220}
II_{19}&=&  \frac{7\arctan^3\left(\frac{x}{a}\right)}{3072 a^9} +\frac{\arctan^2\left(\frac{x}{a}\right)\left[\frac{17 a x \left(x^2-a^2\right)}{\Sigma^4\left(x\right)}+\frac{2 a x^5 \left(7 a^2-x^2\right)}{\Sigma^8\left(x\right)}+\frac{2 a^5 x \left(a^2-7 x^2\right)}{\Sigma^8\left(x\right)}+\frac{4 a x \left(-3 a^4+10 a^2 x^2-3 x^4\right)}{\Sigma^6\left(x\right)}+\frac{8 (a x)}{\Sigma^2\left(x\right)}\right]}{2048 a^9} \nonumber\\ 
&+& \frac{\arctan^2\left(\frac{x}{a}\right)\left[\frac{4 a x^7 \left(x^2-12 a^2\right)}{\Sigma^{10}\left(x\right)}+\frac{a x^7 \left(-198 a^4+55 a^2 x^2-3 x^4\right)}{3 \Sigma^{12}\left(x\right)}+\frac{a^7 x \left(3 a^4-55 a^2 x^2+198 x^4\right)}{3 \Sigma^{12}\left(x\right)}+\frac{4 a^5 x \left(5 a^4-60 a^2 x^2+126 x^4\right)}{5 \Sigma^{10}\left(x\right)}\right]}{2048 a^9} \nonumber \\
&+& \frac{\arctan\left(\frac{x}{a}\right)\left[\frac{4 \left(a^2-x^2\right)}{\Sigma^2\left(x\right)}+\frac{2 x^4 \left(x^2-15 a^2\right)}{3 \Sigma^6\left(x\right)}+\frac{2 a^4 \left(15 x^2-a^2\right)}{3 \Sigma^6\left(x\right)}+\frac{a^4 \left(a^4-28 a^2 x^2+70 x^4\right)}{16 \Sigma^8\left(x\right)}+\frac{17 \left(-a^4+6 a^2 x^2-x^4\right)}{8 \Sigma^4\left(x\right)}\right]}{2048 a^9} \nonumber \\
&+& \frac{\arctan\left(\frac{x}{a}\right)\left[+\frac{x^6 \left(x^2-28 a^2\right)}{16 \Sigma^8\left(x\right)}+\frac{2 x^6 \left(-210 a^4+45 a^2 x^2-x^4\right)}{25 \Sigma^{10}\left(x\right)}+\frac{2 a^6 \left(a^4-45 a^2 x^2+210 x^4\right)}{25 \Sigma^{10}\left(x\right)}\right]}{2048 a^9} \nonumber \\
&+& \frac{\arctan\left(\frac{x}{a}\right)\left[-\frac{77 a^6 x^6}{6 \Sigma^{12}\left(x\right)}+\frac{x^8 \left(495 a^4-66 a^2 x^2+x^4\right)}{72 \Sigma^{12}\left(x\right)}+\frac{a^8 \left(a^4-66 a^2 x^2+495 x^4\right)}{72 \Sigma^{12}\left(x\right)}\right]}{2048 a^9} \nonumber \\
&+& \frac{\left[\frac{17 a x \left(a^2-x^2\right)}{8 \Sigma^4\left(x\right)}+\frac{2 a x^7 \left(12 a^2-x^2\right)}{25 \Sigma^{10}\left(x\right)}+\frac{a x^5 \left(x^2-7 a^2\right)}{16 \Sigma^8\left(x\right)}+\frac{a^5 x \left(7 x^2-a^2\right)}{16 \Sigma^8\left(x\right)}+\frac{2 a x \left(3 a^4-10 a^2 x^2+3 x^4\right)}{9 \Sigma^6\left(x\right)}-\frac{4 (a x)}{\Sigma^2\left(x\right)}\right]}{2048 a^9} \nonumber \\
&+& \frac{\left[\frac{a x^7 \left(198 a^4-55 a^2 x^2+3 x^4\right)}{216 \Sigma^{12}\left(x\right)}+\frac{a^7 x \left(-3 a^4+55 a^2 x^2-198 x^4\right)}{216 \Sigma^{12}\left(x\right)}+\frac{2 a^5 x \left(-5 a^4+60 a^2 x^2-126 x^4\right)}{125 \Sigma^{10}\left(x\right)}\right]}{2048 a^9}, \\\label{apen221}
II_{20}&=&  \frac{\arctan^2\left(\frac{x}{a}\right)\left[\frac{10 \left(x^2-a^2\right)}{\Sigma^2\left(x\right)}+\frac{x^6 \left(x^2-28 a^2\right)}{2 \Sigma^8\left(x\right)}+\frac{5 x^4 \left(15 a^2-x^2\right)}{3 \Sigma^6\left(x\right)}+\frac{5 a^4 \left(a^2-15 x^2\right)}{3 \Sigma^6\left(x\right)}+\frac{5 \left(-a^4+6 a^2 x^2-x^4\right)}{4 \Sigma^4\left(x\right)}\right]}{2048 a^8} \nonumber \\ &+& \frac{\arctan^2\left(\frac{x}{a}\right)\left[+\frac{a^4 \left(a^4-28 a^2 x^2+70 x^4\right)}{2 \Sigma^8\left(x\right)}+\frac{x^6 \left(210 a^4-45 a^2 x^2+x^4\right)}{5 \Sigma^{10}\left(x\right)}+\frac{a^6 \left(-a^4+45 a^2 x^2-210 x^4\right)}{5 \Sigma^{10}\left(x\right)}\right]}{2048 a^8} \nonumber \\
&+& \frac{\arctan^2\left(\frac{x}{a}\right)\left[\frac{77 a^6 x^6}{\Sigma^{12}\left(x\right)}+\frac{x^8 \left(-495 a^4+66 a^2 x^2-x^4\right)}{12 \Sigma^{12}\left(x\right)}+\frac{a^8 \left(-a^4+66 a^2 x^2-495 x^4\right)}{12 \Sigma^{12}\left(x\right)}\right]}{2048 a^8} \nonumber \\
&+& \frac{\arctan\left(\frac{x}{a}\right)\left[\frac{5 a x \left(a^2-x^2\right)}{2 \Sigma^4\left(x\right)}+\frac{a x^5 \left(x^2-7 a^2\right)}{\Sigma^8\left(x\right)}+\frac{a^5 x \left(7 x^2-a^2\right)}{\Sigma^8\left(x\right)}+\frac{10 a x \left(-3 a^4+10 a^2 x^2-3 x^4\right)}{9 \Sigma^6\left(x\right)}+\frac{20 a x}{\Sigma^2\left(x\right)}\right]}{2048 a^8} \nonumber \\
&+& \frac{\arctan\left(\frac{x}{a}\right)\left[\frac{2 a x^7 \left(x^2-12 a^2\right)}{5 \Sigma^{10}\left(x\right)}+\frac{a x^7 \left(-198 a^4+55 a^2 x^2-3 x^4\right)}{18 \Sigma^{12}\left(x\right)}+\frac{a^7 x \left(3 a^4-55 a^2 x^2+198 x^4\right)}{18 \Sigma^{12}\left(x\right)}+\frac{2 a^5 x \left(5 a^4-60 a^2 x^2+126 x^4\right)}{25 \Sigma^{10}\left(x\right)}\right]}{2048 a^8} \nonumber \\
&+& \frac{\left[\frac{5 \left(a^2-x^2\right)}{\Sigma^2\left(x\right)}+\frac{x^6 \left(28 a^2-x^2\right)}{64 \Sigma^8\left(x\right)}+\frac{5 x^4 \left(x^2-15 a^2\right)}{54 \Sigma^6\left(x\right)}+\frac{5 a^4 \left(15 x^2-a^2\right)}{54 \Sigma^6\left(x\right)}+\frac{5 \left(a^4-6 a^2 x^2+x^4\right)}{32 \Sigma^4\left(x\right)}+\frac{a^4 \left(-a^4+28 a^2 x^2-70 x^4\right)}{64 \Sigma^8\left(x\right)}\right]}{2048 a^8} \nonumber \\
&+& \frac{\left[-\frac{77 a^6 x^6}{72 \Sigma^{12}\left(x\right)}+\frac{x^8 \left(495 a^4-66 a^2 x^2+x^4\right)}{864 \Sigma^{12}\left(x\right)}+\frac{x^6 \left(-210 a^4+45 a^2 x^2-x^4\right)}{250 \Sigma^{10}\left(x\right)}+\frac{a^8 \left(a^4-66 a^2 x^2+495 x^4\right)}{864 \Sigma^{12}\left(x\right)}+\frac{a^6 \left(a^4-45 a^2 x^2+210 x^4\right)}{250 \Sigma^{10}\left(x\right)}\right]}{2048 a^8}  \nonumber \\
&+& \left(\frac{38157}{16384000 a^8}\right), 
\end{eqnarray}

\begin{eqnarray}\\\label{apen222}
II_{21}&=& \frac{5\arctan^3\left(\frac{x}{a}\right)}{3072 a^7}+\frac{\arctan^2\left(\frac{x}{a}\right)\left[\frac{15 a x \left(x^2-a^2\right)}{\Sigma^4\left(x\right)}+\frac{6 a x^5 \left(7 a^2-x^2\right)}{\Sigma^8\left(x\right)}+\frac{6 a^5 x \left(a^2-7 x^2\right)}{\Sigma^8\left(x\right)}\right]}{2048 a^7} \nonumber \\
&+& \frac{\arctan^2\left(\frac{x}{a}\right)\left[\frac{a x^7 \left(198 a^4-55 a^2 x^2+3 x^4\right)}{3 \Sigma^{12}\left(x\right)}+\frac{a^7 x \left(-3 a^4+55 a^2 x^2-198 x^4\right)}{3 \Sigma^{12}\left(x\right)}\right]}{2048 a^7} \nonumber \\
&+&  \frac{\arctan\left(\frac{x}{a}\right)\left[\frac{3 x^6 \left(x^2-28 a^2\right)}{16 \Sigma^8\left(x\right)}+\frac{15 \left(-a^4+6 a^2 x^2-x^4\right)}{8 \Sigma^4\left(x\right)}+\frac{3 a^4 \left(a^4-28 a^2 x^2+70 x^4\right)}{16 \Sigma^8\left(x\right)}\right]}{2048 a^7} \nonumber \\
&+&  \frac{\arctan\left(\frac{x}{a}\right)\left[\frac{77 a^6 x^6}{6 \Sigma^{12}\left(x\right)}+\frac{x^8 \left(-495 a^4+66 a^2 x^2-x^4\right)}{72 \Sigma^{12}\left(x\right)}+\frac{a^8 \left(-a^4+66 a^2 x^2-495 x^4\right)}{72 \Sigma^{12}\left(x\right)}\right]}{2048 a^7} \nonumber \\
&+&  \frac{\left[\frac{15 a x \left(a^2-x^2\right)}{8 \Sigma^4\left(x\right)}+\frac{3 a x^5 \left(x^2-7 a^2\right)}{16 \Sigma^8\left(x\right)}+\frac{3 a^5 x \left(7 x^2-a^2\right)}{16 \Sigma^8\left(x\right)}+\frac{a x^7 \left(-198 a^4+55 a^2 x^2-3 x^4\right)}{216 \Sigma^{12}\left(x\right)}+\frac{a^7 x \left(3 a^4-55 a^2 x^2+198 x^4\right)}{216 \Sigma^{12}\left(x\right)}\right]}{2048 a^7}, \\\label{apen223}
II_{22}&=& \frac{231\arctan^4\left(\frac{x}{a}\right)}{4096 a^{13}} - \left(\frac{549748199}{3932160000 a^{13}}\right) \nonumber\\
&+& \frac{\arctan^3\left(\frac{x}{a}\right)\left[\frac{495 a x \left(a^2-x^2\right)}{\Sigma^4\left(x\right)}+\frac{66 a x^5 \left(7 a^2-x^2\right)}{\Sigma^8\left(x\right)}+\frac{66 a^5 x \left(a^2-7 x^2\right)}{\Sigma^8\left(x\right)}+\frac{220 a x \left(3 a^4-10 a^2 x^2+3 x^4\right)}{3 \Sigma^6\left(x\right)}+\frac{792 a x}{\Sigma^2\left(x\right)}\right]}{2048 a^{13}} \nonumber \\
&+& \frac{\arctan^3\left(\frac{x}{a}\right)\left[\frac{12 a x^7 \left(x^2-12 a^2\right)}{\Sigma^{10}\left(x\right)}+\frac{a x^7 \left(-198 a^4+55 a^2 x^2-3 x^4\right)}{3 \Sigma^{12}\left(x\right)}+\frac{a^7 x \left(3 a^4-55 a^2 x^2+198 x^4\right)}{3 \Sigma^{12}\left(x\right)}+\frac{12 a^5 x \left(5 a^4-60 a^2 x^2+126 x^4\right)}{5 \Sigma^{10}\left(x\right)}\right]}{2048 a^{13}}  \nonumber \\
&+& \frac{\arctan^2\left(\frac{x}{a}\right)\left[\frac{594 \left(a^2-x^2\right)}{\Sigma^2\left(x\right)}+\frac{99 x^6 \left(x^2-28 a^2\right)}{32 \Sigma^8\left(x\right)}+\frac{55 x^4 \left(15 a^2-x^2\right)}{3 \Sigma^6\left(x\right)}+\frac{55 a^4 \left(a^2-15 x^2\right)}{3 \Sigma^6\left(x\right)}+\frac{1485 \left(a^4-6 a^2 x^2+x^4\right)}{16 \Sigma^4\left(x\right)}\right]}{2048 a^{13}} \nonumber \\
&+& \frac{\arctan^2\left(\frac{x}{a}\right)\left[\frac{99 a^4 \left(a^4-28 a^2 x^2+70 x^4\right)}{32 \Sigma^8\left(x\right)}+\frac{9 x^6 \left(-210 a^4+45 a^2 x^2-x^4\right)}{25 \Sigma^{10}\left(x\right)}+\frac{9 a^6 \left(a^4-45 a^2 x^2+210 x^4\right)}{25 \Sigma^{10}\left(x\right)}\right]}{2048 a^{13}} \nonumber \\
&+& \frac{\arctan^2\left(\frac{x}{a}\right)\left[-\frac{231 a^6 x^6}{12 \Sigma^{12}\left(x\right)}+\frac{x^8 \left(495 a^4-66 a^2 x^2+x^4\right)}{48 \Sigma^{12}\left(x\right)}+\frac{a^8 \left(a^4-66 a^2 x^2+495 x^4\right)}{48 \Sigma^{12}\left(x\right)}\right]}{2048 a^{13}} \nonumber \\
&-& \frac{\arctan\left(\frac{x}{a}\right)\left[\frac{1485 a x \left(a^2-x^2\right)}{8 \Sigma^4\left(x\right)}+\frac{99 a^5 x \left(a^2-7 x^2\right)}{16 \Sigma^8\left(x\right)}+\frac{110 a x \left(3 a^4-10 a^2 x^2+3 x^4\right)}{9 \Sigma^6\left(x\right)}+\frac{1188 a x}{\Sigma^2\left(x\right)}\right]}{2048 a^{13}} \nonumber \\
&-& \frac{\arctan\left(\frac{x}{a}\right)\left[\frac{18 a x^7 \left(x^2-12 a^2\right)}{25 \Sigma^{10}\left(x\right)}+\frac{99 a x^5 \left(7 a^2-x^2\right)}{16 \Sigma^8\left(x\right)}+\frac{18 a^5 x \left(5 a^4-60 a^2 x^2+126 x^4\right)}{125 \Sigma^{10}\left(x\right)}\right]}{2048 a^{13}} \nonumber \\
&-& \frac{\arctan\left(\frac{x}{a}\right)\left[\frac{a x^7 \left(-198 a^4+55 a^2 x^2-3 x^4\right)}{72 \Sigma^{12}\left(x\right)}+\frac{a^7 x \left(3 a^4-55 a^2 x^2+198 x^4\right)}{72 \Sigma^{12}\left(x\right)}\right]}{2048 a^{13}} \nonumber \\
&-& \frac{\left[\frac{297 \left(a^2-x^2\right)}{\Sigma^2\left(x\right)}+\frac{99 x^6 \left(x^2-28 a^2\right)}{1024 \Sigma^8\left(x\right)}+\frac{55 x^4 \left(15 a^2-x^2\right)}{54 \Sigma^6\left(x\right)}+\frac{55 a^4 \left(a^2-15 x^2\right)}{54 \Sigma^6\left(x\right)}+\frac{1485 \left(a^4-6 a^2 x^2+x^4\right)}{128 \Sigma^4\left(x\right)}\right]}{2048 a^{13}} \nonumber \\
&-& \frac{\left[+\frac{99 a^4 \left(a^4-28 a^2 x^2+70 x^4\right)}{1024 \Sigma^8\left(x\right)}+\frac{9 x^6 \left(-210 a^4+45 a^2 x^2-x^4\right)}{1250 \Sigma^{10}\left(x\right)}+\frac{9 a^6 \left(a^4-45 a^2 x^2+210 x^4\right)}{1250 \Sigma^{10}\left(x\right)}\right]}{2048 a^{13}} \nonumber \\
&-& \frac{\left[-\frac{77 a^6 x^6}{288 \Sigma^{12}\left(x\right)}+\frac{x^8 \left(495 a^4-66 a^2 x^2+x^4\right)}{3456 \Sigma^{12}\left(x\right)}+\frac{a^8 \left(a^4-66 a^2 x^2+495 x^4\right)}{3456 \Sigma^{12}\left(x\right)}\right]}{2048 a^{13}},  \\\label{apen224}
II_{23}&=& \frac{21\arctan^4\left(\frac{x}{a}\right)}{4096 a^{11}} - \left(\frac{34943909}{3932160000 a^{11}}\right) \nonumber\\
&+&\frac{\arctan^3\left(\frac{x}{a}\right)\left[\frac{15 a x \left(x^2-a^2\right)}{\Sigma^4\left(x\right)}+\frac{26 a x^5 \left(x^2-7 a^2\right)}{\Sigma^8\left(x\right)}+\frac{26 a^5 x \left(7 x^2-a^2\right)}{\Sigma^8\left(x\right)}+\frac{40 a x \left(-3 a^4+10 a^2 x^2-3 x^4\right)}{3 \Sigma^6\left(x\right)}\right]}{2048 a^{11}} \nonumber \\
&+& \frac{\arctan^3\left(\frac{x}{a}\right)\left[\frac{8 a x^7 \left(12 a^2-x^2\right)}{\Sigma^{10}\left(x\right)}+\frac{a x^7 \left(198 a^4-55 a^2 x^2+3 x^4\right)}{3 \Sigma^{12}\left(x\right)}+\frac{a^7 x \left(-3 a^4+55 a^2 x^2-198 x^4\right)}{3 \Sigma^{12}\left(x\right)}\right]}{2048 a^{11}} \nonumber \\
&+& \frac{\arctan^3\left(\frac{x}{a}\right)\left[+\frac{48 a x}{\Sigma^2\left(x\right)}+\frac{8 a^5 x \left(-5 a^4+60 a^2 x^2-126 x^4\right)}{5 \Sigma^{10}\left(x\right)}\right]}{2048 a^{11}} \nonumber \\
&+& \frac{\arctan^2\left(\frac{x}{a}\right)\left[\frac{36 \left(a^2-x^2\right)}{\Sigma^2\left(x\right)}+\frac{10 x^4 \left(x^2-15 a^2\right)}{3 \Sigma^6\left(x\right)}+\frac{10 a^4 \left(15 x^2-a^2\right)}{3 \Sigma^6\left(x\right)}+\frac{45 \left(-a^4+6 a^2 x^2-x^4\right)}{16 \Sigma^4\left(x\right)}\right]}{2048 a^{11}} \nonumber \\
&+& \frac{\arctan^2\left(\frac{x}{a}\right)\left[\frac{77 a^6 x^6}{4 \Sigma^{12}\left(x\right)}+\frac{39 x^6 \left(28 a^2-x^2\right)}{32 \Sigma^8\left(x\right)}+\frac{39 a^4 \left(-a^4+28 a^2 x^2-70 x^4\right)}{32 \Sigma^8\left(x\right)}+\frac{6 a^6 \left(-a^4+45 a^2 x^2-210 x^4\right)}{25 \Sigma^{10}\left(x\right)}\right]}{2048 a^{11}} \nonumber \\
&+& \frac{\arctan^2\left(\frac{x}{a}\right)\left[\frac{x^8 \left(-495 a^4+66 a^2 x^2-x^4\right)}{48 \Sigma^{12}\left(x\right)}+\frac{6 x^6 \left(210 a^4-45 a^2 x^2+x^4\right)}{25 \Sigma^{10}\left(x\right)}+\frac{a^8 \left(-a^4+66 a^2 x^2-495 x^4\right)}{48 \Sigma^{12}\left(x\right)}\right]}{2048 a^{11}} \nonumber \\
&+& \frac{\arctan\left(\frac{x}{a}\right)\left[-\frac{72 a x}{\Sigma^2\left(x\right)}+\frac{45 a x \left(a^2-x^2\right)}{8 \Sigma^4\left(x\right)}+\frac{20 a x \left(3 a^4-10 a^2 x^2+3 x^4\right)}{9 \Sigma^6\left(x\right)}+\frac{39 a^5 x \left(a^2-7 x^2\right)}{16 \Sigma^8\left(x\right)}+\frac{39 a x^5 \left(7 a^2-x^2\right)}{16 \Sigma^8\left(x\right)}\right]}{2048 a^{11}} \nonumber \\
&+& \frac{\arctan\left(\frac{x}{a}\right)\left[\frac{12 a x^7 \left(x^2-12 a^2\right)}{25 \Sigma^{10}\left(x\right)}+\frac{a x^7 \left(-198 a^4+55 a^2 x^2-3 x^4\right)}{72 \Sigma^{12}\left(x\right)}+\frac{a^7 x \left(3 a^4-55 a^2 x^2+198 x^4\right)}{72 \Sigma^{12}\left(x\right)}+\frac{12 a^5 x \left(5 a^4-60 a^2 x^2+126 x^4\right)}{125 \Sigma^{10}\left(x\right)}\right]}{2048 a^{11}} \nonumber \\
&+& \frac{\left[\frac{18 \left(x^2-a^2\right)}{\Sigma^2\left(x\right)}+\frac{39 x^6 \left(x^2-28 a^2\right)}{1024 \Sigma^8\left(x\right)}+\frac{5 x^4 \left(15 a^2-x^2\right)}{27 \Sigma^6\left(x\right)}+\frac{5 a^4 \left(a^2-15 x^2\right)}{27 \Sigma^6\left(x\right)}+\frac{45 \left(a^4-6 a^2 x^2+x^4\right)}{128 \Sigma^4\left(x\right)}\right]}{2048 a^{11}} \nonumber  \\
&+& \frac{\left[+\frac{39 a^4 \left(a^4-28 a^2 x^2+70 x^4\right)}{1024 \Sigma^8\left(x\right)}+\frac{3 x^6 \left(-210 a^4+45 a^2 x^2-x^4\right)}{625 \Sigma^{10}\left(x\right)}+\frac{3 a^6 \left(a^4-45 a^2 x^2+210 x^4\right)}{625 \Sigma^{10}\left(x\right)}\right]}{2048 a^{11}} \nonumber \\
&+& \frac{\left[-\frac{77 a^6 x^6}{288 \Sigma^{12}\left(x\right)}+\frac{x^8 \left(495 a^4-66 a^2 x^2+x^4\right)}{3456 \Sigma^{12}\left(x\right)}+\frac{a^8 \left(a^4-66 a^2 x^2+495 x^4\right)}{3456 \Sigma^{12}\left(x\right)}\right]}{2048 a^{11}}, \\\label{apen225}
II_{24}&=& \frac{7\arctan^4\left(\frac{x}{a}\right)}{4096 a^9} -\left(\frac{57750847}{35389440000 a^9}\right)\nonumber \\
&+&\frac{\arctan^3\left(\frac{x}{a}\right)\left[\frac{17 a x \left(x^2-a^2\right)}{\Sigma^4\left(x\right)}+\frac{2 a x^5 \left(7 a^2-x^2\right)}{\Sigma^8\left(x\right)}+\frac{2 a^5 x \left(a^2-7 x^2\right)}{\Sigma^8\left(x\right)}+\frac{4 a x \left(-3 a^4+10 a^2 x^2-3 x^4\right)}{\Sigma^6\left(x\right)}+\frac{8 a x}{\Sigma^2\left(x\right)}\right]}{2048 a^9}   \nonumber \\
&+& \frac{\arctan^3\left(\frac{x}{a}\right)\left[\frac{4 a x^7 \left(x^2-12 a^2\right)}{\Sigma^{10}\left(x\right)}+\frac{a x^7 \left(-198 a^4+55 a^2 x^2-3 x^4\right)}{3 \Sigma^{12}\left(x\right)}+\frac{a^7 x \left(3 a^4-55 a^2 x^2+198 x^4\right)}{3 \Sigma^{12}\left(x\right)}+\frac{4 a^5 x \left(5 a^4-60 a^2 x^2+126 x^4\right)}{5 \Sigma^{10}\left(x\right)}\right]}{2048 a^9} \nonumber \\
&+& \frac{\arctan^2\left(\frac{x}{a}\right)\left[\frac{6 \left(a^2-x^2\right)}{\Sigma^2\left(x\right)}+\frac{3 x^6 \left(x^2-28 a^2\right)}{32 \Sigma^8\left(x\right)}+\frac{x^4 \left(x^2-15 a^2\right)}{\Sigma^6\left(x\right)}+\frac{a^4 \left(15 x^2-a^2\right)}{\Sigma^6\left(x\right)}+\frac{51 \left(-a^4+6 a^2 x^2-x^4\right)}{16 \Sigma^4\left(x\right)}\right]}{2048 a^9} \nonumber \\
&+& \frac{\arctan^2\left(\frac{x}{a}\right)\left[+\frac{3 a^4 \left(a^4-28 a^2 x^2+70 x^4\right)}{32 \Sigma^8\left(x\right)}+\frac{3 x^6 \left(-210 a^4+45 a^2 x^2-x^4\right)}{25 \Sigma^{10}\left(x\right)}+\frac{3 a^6 \left(a^4-45 a^2 x^2+210 x^4\right)}{25 \Sigma^{10}\left(x\right)}\right]}{2048 a^9} \nonumber \\
&+& \frac{\arctan^2\left(\frac{x}{a}\right)\left[-\frac{77 a^6 x^6}{4 \Sigma^{12}\left(x\right)}+\frac{x^8 \left(495 a^4-66 a^2 x^2+x^4\right)}{48 \Sigma^{12}\left(x\right)}+\frac{a^8 \left(a^4-66 a^2 x^2+495 x^4\right)}{48 \Sigma^{12}\left(x\right)}\right]}{2048 a^9} \nonumber \\
&+& \frac{\arctan\left(\frac{x}{a}\right)\left[\frac{51 a x \left(a^2-x^2\right)}{8 \Sigma^4\left(x\right)}+\frac{3 a x^5 \left(x^2-7 a^2\right)}{16 \Sigma^8\left(x\right)}+\frac{3 a^5 x \left(7 x^2-a^2\right)}{16 \Sigma^8\left(x\right)}+\frac{2 a x \left(3 a^4-10 a^2 x^2+3 x^4\right)}{3 \Sigma^6\left(x\right)}-\frac{12 a x}{\Sigma^2\left(x\right)}\right]}{2048 a^9} \nonumber \\
&+& \frac{\arctan\left(\frac{x}{a}\right)\left[\frac{6 a x^7 \left(12 a^2-x^2\right)}{25 \Sigma^{10}\left(x\right)}+\frac{a x^7 \left(198 a^4-55 a^2 x^2+3 x^4\right)}{72 \Sigma^{12}\left(x\right)}+\frac{a^7 x \left(-3 a^4+55 a^2 x^2-198 x^4\right)}{72 \Sigma^{12}\left(x\right)}+\frac{6 a^5 x \left(-5 a^4+60 a^2 x^2-126 x^4\right)}{125 \Sigma^{10}\left(x\right)}\right]}{2048 a^9} \nonumber \\
&+& \frac{\left[\frac{3 \left(x^2-a^2\right)}{\Sigma^2\left(x\right)}+\frac{3 x^6 \left(28 a^2-x^2\right)}{1024 \Sigma^8\left(x\right)}+\frac{3 x^4 \left(15 a^2-x^2\right)}{54 \Sigma^6\left(x\right)}+\frac{3 a^4 \left(a^2-15 x^2\right)}{54 \Sigma^6\left(x\right)}+\frac{51 \left(a^4-6 a^2 x^2+x^4\right)}{128 \Sigma^4\left(x\right)}\right]}{2048 a^9} \nonumber \\ 
&+& \frac{\left[\frac{3 a^4 \left(-a^4+28 a^2 x^2-70 x^4\right)}{1024 \Sigma^8\left(x\right)}+\frac{3 x^6 \left(210 a^4-45 a^2 x^2+x^4\right)}{1250 \Sigma^{10}\left(x\right)}+\frac{3 a^6 \left(-a^4+45 a^2 x^2-210 x^4\right)}{1250 \Sigma^{10}\left(x\right)}\right]}{2048 a^9} \nonumber \\
&+& \frac{\left[\frac{77 a^6 x^6}{288 \Sigma^{12}\left(x\right)}+\frac{x^8 \left(-495 a^4+66 a^2 x^2-x^4\right)}{3456 \Sigma^{12}\left(x\right)}+\frac{a^8 \left(-a^4+66 a^2 x^2-495 x^4\right)}{3456 \Sigma^{12}\left(x\right)}\right]}{2048 a^9}, \\\label{apen226}
II_{25}&=& \frac{5\arctan^4\left(\frac{x}{a}\right)}{4096 a^7} - \left(\frac{9485}{56623104 a^7}\right) \nonumber \\
&+& \frac{\arctan^3\left(\frac{x}{a}\right)\left[\frac{15 a x \left(x^2-a^2\right)}{\Sigma^4\left(x\right)}+\frac{6 a x^5 \left(7 a^2-x^2\right)}{\Sigma^8\left(x\right)}+\frac{6 a^5 x \left(a^2-7 x^2\right)}{\Sigma^8\left(x\right)}+\frac{a x^7 \left(198 a^4-55 a^2 x^2+3 x^4\right)}{3 \Sigma^{12}\left(x\right)}+\frac{a^7 x \left(-3 a^4+55 a^2 x^2-198 x^4\right)}{3 \Sigma^{12}\left(x\right)}\right]}{2048 a^7} \nonumber \\
&+& \frac{\arctan^2\left(\frac{x}{a}\right)\left[\frac{9 x^6 \left(x^2-28 a^2\right)}{32 \Sigma^8\left(x\right)}+\frac{45 \left(-a^4+6 a^2 x^2-x^4\right)}{16 \Sigma^4\left(x\right)}+\frac{9 a^4 \left(a^4-28 a^2 x^2+70 x^4\right)}{32 \Sigma^8\left(x\right)}\right]}{2048 a^7} \nonumber \\
&+& \frac{\arctan^2\left(\frac{x}{a}\right)\left[\frac{231 a^6 x^6}{12 \Sigma^{12}\left(x\right)}+\frac{x^8 \left(-495 a^4+66 a^2 x^2-x^4\right)}{48 \Sigma^{12}\left(x\right)}+\frac{a^8 \left(-a^4+66 a^2 x^2-495 x^4\right)}{48 \Sigma^{12}\left(x\right)}\right]}{2048 a^7} \nonumber \\
&+& \frac{\arctan\left(\frac{x}{a}\right)\left[\frac{45 a x \left(a^2-x^2\right)}{8 \Sigma^4\left(x\right)}+\frac{9 a x^5 \left(x^2-7 a^2\right)}{16 \Sigma^8\left(x\right)}+\frac{9 a^5 x \left(7 x^2-a^2\right)}{16 \Sigma^8\left(x\right)}+\frac{a x^7 \left(-198 a^4+55 a^2 x^2-3 x^4\right)}{72 \Sigma^{12}\left(x\right)}+\frac{a^7 x \left(3 a^4-55 a^2 x^2+198 x^4\right)}{72 \Sigma^{12}\left(x\right)}\right]}{2048 a^7} \nonumber \\
&+& \frac{\left[\frac{9 x^6 \left(28 a^2-x^2\right)}{1024 \Sigma^8\left(x\right)}+\frac{9 a^4 \left(-a^4+28 a^2 x^2-70 x^4\right)}{1024 \Sigma^8\left(x\right)}+\frac{45 \left(a^4-6 a^2 x^2+x^4\right)}{128 \Sigma^4\left(x\right)}+\frac{a^8 \left(a^4-66 a^2 x^2+495 x^4\right)}{3456 \Sigma^{12}\left(x\right)}\right]}{2048 a^7} \nonumber \\
&+& \frac{\left[+\frac{x^8 \left(495 a^4-66 a^2 x^2+x^4\right)}{3456 \Sigma^{12}\left(x\right)}-\frac{77 a^6 x^6}{288 \Sigma^{12}\left(x\right)}\right]}{2048 a^7}.
\end{eqnarray}

\clearpage

\nocite{*}

\begin{thebibliography}{99}

\bibitem{regular1} J. M. Bardeen, Proceedings of the International Conference GR5, Tbilisi, U.S.S.R. (1968).

\bibitem{regular2} E. Ayon-Beato and A. Garcia, Phys. Lett. B \textbf{493}, 149-152 (2000),[arXiv:gr-qc/0009077 [gr-qc]].

\bibitem{regular3} M. E. Rodrigues and M. V. d. Silva, JCAP \textbf{06}, 025 (2018).

\bibitem{regular4} M. E. Rodrigues, E. L. B. Junior and M. V. de S. Silva, JCAP \textbf{02}, 059 (2018).

\bibitem{regular5} E. Ayon-Beato and A. Garcia, Phys. Lett. B \textbf{464}, 25 (1999).

\bibitem{regular6} E. Ayon-Beato and A. Garcia, Phys. Rev. Lett. \textbf{80}, 5056-5059 (1998).

\bibitem{regular7} K. A. Bronnikov, Phys. Rev. Lett. \textbf{85}, 4641 (2000).

\bibitem{regular8} E. Ayon-Beato and A. Garcia, Gen. Rel. Grav. \textbf{37}, 635 (2005).

\bibitem{regular9} K. A. Bronnikov, Phys. Rev. D \textbf{63}, 044005 (2001).

\bibitem{regular10} I. Dymnikova, Class. Quant. Grav. \textbf{21}, 4417-4429 (2004).

\bibitem{regular11} L. Balart and E. C. Vagenas, Phys. Rev. D \textbf{90}, no.12, 124045 (2014).

\bibitem{regular12} L. Balart and E. C. Vagenas, Phys. Lett. B \textbf{730}, 14-17 (2014).

\bibitem{regular13} N. Uchikata, S. Yoshida and T. Futamase, Phys. Rev. D \textbf{86}, 084025 (2012).

\bibitem{regular14} J. Ponce de Leon, Phys. Rev. D \textbf{95}, no.12, 124015 (2017).

\bibitem{regular15} S. A. Hayward, Phys. Rev. Lett. \textbf{96}, 031103 (2006).

\bibitem{intro1} R. D'Inverno, Introducing Einstein's Relativity, Oxford University Press, New York (1998).

\bibitem{intro2} S. Weinberg; Gravitation and Cosmology: principles and applications of the general theory of relativity, 1972.

\bibitem{intro3} M. P. Hobson, G. P. Efstathiou e A. N. Lasenby, General Relativity-An Introduction for Physicists, Cambridge University Press, Nova York (2006).

\bibitem{intro4} S. Chandrasekhar, The mathematical theory of black holes, Oxford University Press, Nova York (2006).

\bibitem{intro5} M. Visser, Lorentzian Wormholes: from Einstein to Hawking, Springer-Verlag, Nova York (1996).

\bibitem{intro6} K. A. Bronnikov and R. K. Walia, Phys. Rev. D \textbf{105}, no.4, 044039 (2022).

\bibitem{intro7} P. Cañate, Phys. Rev. D \textbf{106}, no.2, 024031 (2022).


\bibitem{geo1}     A. Lima, G. Alencar, R. N. Costa Filho and R. R. Landim, General Relativity and Gravitation \textbf{55}, 108 (2023).

\bibitem{geo2} A. Lima, G Alencar, D. S. C. G\'omez, Physical Review D \textbf{109}, 064038  (2024).

\bibitem{geo3} A. M. Lima, G. M. de Alencar Filho and J. S. Furtado Neto, Symmetry \textbf{15}, 150 (2023).


\bibitem{intro8}  F. S. N. Lobo, M. E. Rodrigues, M. V. de S. Silva, A. Simpson and M. Visser, Phys. Rev. D \textbf{103}, 084052 (2021), arXiv:2009.12057.

\bibitem{intro9}  M. E. Rodrigues and M. V. de S. Silva, Class. Quantum Grav. \textbf{40}, 225011 (2023), arXiv:2204.11851.

\bibitem{intro10}  E. L. B. Junior and M. E. Rodrigues, Gen. Rel. Grav. \textbf{55}, 8 (2023), arXiv:2203.03629.

\bibitem{intro11}  J. C. Fabris, E. L. B. Junior and M. E. Rodrigues, Eur. Phys. J. C  \textbf{83}: 884 (2023).

\bibitem{intro12} E. Franzin, S. Liberati, J. Mazza, A. Simpson and M. Visser, JCAP \textbf{07}, 036 (2021).

\bibitem{intro13} J. Mazza, E. Franzin and S. Liberati, JCAP \textbf{04}, 082 (2021).

\bibitem{intro14}  J. R. Nascimento, A. Y. Petrov, P. J. Porfirio and A. R. Soares, Phys. Rev. D \textbf{102}, 044021 (2020), arXiv:2005.13096.

\bibitem{intro15}  Naoki Tsukamoto, Phys. Rev. D \textbf{104}, 064022 (2021).

\bibitem{intro16}  Naoki Tsukamoto, Phys. Rev. D \textbf{103}, 024033 (2021).

\bibitem{intro17}  S. Ghosh and A. Bhattacharyya, JCAP \textbf{11}, 006 (2022). 
 
\bibitem{intro18}  A. Chowdhuri, S. Ghosh, and A. Bhattacharyya, v. \textbf{11}, p. 164, (2023).

\bibitem{intro19}  H. Aounallah, A. R. Soares and R. L. L. Vit\'oria, Eur. Phys. J. C, \textbf{80} 447 (2020).

\bibitem{intro20}  A. R. Soares, R. L. L. Vit\'oria and H. Aounallah, Eur. Phys. J. Plus, \textbf{136} 966 (2021).

\bibitem{intro21}  C. F. S. Pereira, A. R. Soares, R. L. L. Vit\'oria, H. Belich, Eur. Phys. J. C \textbf{83}, 270 (2023).

\bibitem{intro22} C. F. S. Pereira,  R. L. L. Vit\'oria, A. R. Soares, H. Belich,  Modern Physics Letters A, \textbf{38}, 2350133 (2023). 

\bibitem{mukha1}
	C. Armendariz-Picon, V. Mukhanov and P. J. Steinhardt, Phys. Rev. D
	 {\bf 63}, 103510 (2001).

\bibitem{mukha2}
	C. Armendariz-Picon, T. Damour and V. Mukhanov, Phys. Lett. B {\bf
	 458}, 209 (1999).

\bibitem{mukha3}
	C. Armendariz-Picon, V. Mukhanov, and P. J. Steinhardt, Phys. Rev.
	 Lett. {\bf 85}, 4438 (2000).

\bibitem{dbi}
	R. Leigh, Mod. Phys. Lett. {\bf A4}, 2767 (1989).

\bibitem{bronnikov1}  K. A. Bronnikov, J. C. Fabris and D. C. Rodrigues, Gravit. Cosmol. \textbf{22}, 26-31 (2016).

\bibitem{simple}
    K. A. Bronnikov, V. A. G. Barcellos, L. P. de Carvalho and J. C. Fabris,
    Eur. Phys. J. C {\bf 81}, 395 (2021).
    
\bibitem{CDJM} C. F. S. Pereira, D. C. Rodrigues, J. C. Fabris and M. E. Rodrigues, Phys. Rev. D \textbf{109}, 044011 (2024). 

\bibitem{phan1} H.G. Ellis, J. Math. Phys. \textbf{14}, 104 (1973).

\bibitem{phan2} K.A. Bronnikov, Acta Phys. Pol. B \textbf{4}, 251 (1973).

\bibitem{phan3}  C. R. Almeida, J. C. Fabris, F. Sbisa and Y. Tavakoli, Quantum cosmology with k-Essence theory, arXiv:1604.00624. To appear in the proceedings of the 31st International Colloquium on Group Theoretical Methods in Physics.

\bibitem{phan4} K. A. Bronnikov, J. C. Fabris, O. F. Piattella, Denis C. Rodrigues and E. C. Santos, Duality between k-essence and Rastall gravity. Eur. Phys. J. C \textbf{77}, 409 (2017).

\bibitem{matt} A. Simpson and M. Visser, JCAP \textbf{02}, 042 (2019), arXiv:1812.07114.

\bibitem{PRL} K. A. Bronnikov and J. C. Fabris Phys. Rev. Lett. \textbf{96}, 251101 (2006).

\bibitem{KJD} K. A. Bronnikov, J. C. Fabris and D. C. Rodrigues, Gravit. Cosmol. \textbf{22}, 26-31 (2016).











    
\bibitem{teller1} U. C. da Silva, C. F. S. Pereira and A. A. Lima, Annals of Physics \textbf{460} 169549 (2024), arXiv:2308.04596.

\bibitem{teller2}  J. C. Fabris, M. G. Richarte, and A. Saa,  Phys. Rev. D \textbf{103}  no. 4, 045001 (2021).

\bibitem{teller3}  D.P. Du, B. Wang, and R.K. Su, Phys. Rev. D \textbf{70} 064024  (2004).

\bibitem{teller4} Y. Zhong, JHEP \textbf{09}  165 (2022).

\bibitem{molina}   C. Molina and J. C. S. Neves, Phys.Rev.D \textbf{86}, 024015 (2012).

\bibitem{desitter} De-Chang Dai, Djordje Minic and Dejan Stojkovic, Phys. Rev. D \textbf{98}, 124026 (2018).






\end{thebibliography}
		
\end{document}